\documentclass[lettersize, 12pt]{article}

\usepackage{graphics}
\usepackage{float}
\usepackage{theorem}
\usepackage{latexsym}
\usepackage{amsfonts}
\usepackage{amssymb}
\usepackage{amsmath}
\usepackage{graphicx}
\usepackage{natbib}
\usepackage{lscape}
\usepackage{booktabs}
\usepackage{threeparttable}
\usepackage{appendix}
\usepackage{subfig}
\usepackage[colorlinks=true,linkcolor=black,citecolor=black,urlcolor=black,pdfstartview=FitH,bookmarks=false]{hyperref}
\usepackage{setspace}
\usepackage{titlesec}
\usepackage{color}
\usepackage{natbib}
\usepackage{bm}
\usepackage{multicol}
\usepackage{soul}
\usepackage{tabularx} 

\newtheorem{result}{Result}[section]

\newtheorem{Remark}{Remark}
\newtheorem{Theorem}{Theorem}
\newtheorem{Corollary}{Corollary}

\newtheorem{Lemma}{Lemma}[section]
\newtheorem{Assumption}{Assumption}

\begin{document}
\title{\Large Partial Mean Processes with Generated Regressors: \\ Continuous Treatment Effects and Nonseparable Models}
\author{\normalsize
Ying-Ying Lee\footnote{\scriptsize The author is deeply grateful to Jack Porter and Bruce Hansen for invaluable guidance and encouragement.  
The author thanks Xiaoxia Shi, Chris Taber, and the seminar participants in UW-Madison for helpful comments  and discussion.
I also thank Debopam Bhattacharya, Juan Carlo Escanciano, Antonio Galvao, Yu-Chin Hsu, Martin Huber, David Jacho-Ch\'{a}vez, Arthur Lebwel, Enno Mammen, Whitney Newey, Christoph Rothe, Anne Vanhems, Ingrid van Keilegom, Melanie Schienle, Sami Stouli, Kyungchul Song, as well as the seminar participants in Academia Sinica, UCL, University of Bristol, University of Mannheim, University of Oxford, TSE, 2012/2013 Midwest Econometrics Group Meeting, 2012 Info-Metrics Nonparametric Conference, 2013 SETA, and 2014 Cambridge Nonparametric and Semiparametric Methods Conference, 2018 California Econometrics Conference.} 
\\
\it{\normalsize University of California Irvine}\footnote{Department of economics, 3151 Social Science Plaza, University of California Irvine, Irvine, CA 92697.
E-mail: \href{yingying.lee@uci.edu}{yingying.lee@uci.edu}}
}
\date{\normalsize October 2018}
\maketitle
\begin{center}
{\bf \small Abstract}
\end{center}

\vspace{-5pt}

Partial mean with generated regressors arises in several econometric problems, such as the distribution of potential outcomes with continuous treatments and the quantile structural
function in a nonseparable triangular model. This paper proposes a nonparametric estimator for the partial mean process, where the second step consists of a kernel regression on regressors that are estimated in the first step.  The main contribution is a uniform expansion that characterizes in detail how the estimation error associated with the generated regressor affects the limiting distribution of the marginal integration estimator. The general results are illustrated with two examples: the generalized propensity score for a continuous treatment \citep*{HI04} and control variables in triangular models \citep*{NPV99ETA, IN09ETA}.
An empirical application to the Job Corps program evaluation demonstrates the usefulness of the method.
%

\section{Introduction}
This paper studies nonparametric estimation of continuous treatment effects. 
Endogeneity is allowed in the continuous treatment variable and is corrected by including a control variable as an additional regressor.\footnote{According to \cite{M07}, ``A {\it control variable} is a function of observable variables such that conditioning on its value purges any statistical dependence that may exist between the observable and unobservable explanatory variables in an original model."}
This approach leads to identification of the unconditional distribution of the potential outcome with continuous treatments by a partial mean, where a partial mean is defined as a marginal integration of the conditional outcome distribution over the control variable with the continuous treatments held fixed. 
We allow the control variable to be estimated in a preliminary step and enter the partial mean as a generated regressor. 
A generated regressor can be viewed as a possibly infinite-dimensional nuisance parameter in many economics examples: the control variables in triangular models \citep*{NPV99ETA, IN09ETA}, the generalized propensity score for continuous treatment \citep*{HI04}, or the propensity score for sample selection \citep*{DNV03}.
The main contribution of this paper is to characterize how the estimation error of general generated regressors affects the limit properties of partial mean estimation and to provide uniform inference for the entire potential outcome distribution of continuous treatments.

Our results provide new insights on nonparametric estimation of continuous treatment effect models. 
Two key features of these models are non-separability in the unobservables and heterogeneity in treatment intensity effects.
The proposed methods capture heterogeneous treatment intensity effects by estimating an array of  distributional structural features that can be applied to a variety of economics questions.
For example, when evaluating a social program, researchers might be interested in how the length of exposure to the program affects the wage distribution.
The proposed method includes inference on smooth functionals of the outcome distribution process.
In this example, a researcher could consider how inequality responds to the length of exposure to an anti-poverty program by tracing out the Gini coefficient of the wage distribution by time in the program.
In demand analysis, we can estimate the Engel curve that describes how the distribution of household food consumption responds to an exogenous change in total expenditure.
Analyzing these questions often involves generated regressors to account for the endogeneity of the continuous explanatory variable.  
Understanding the estimation error of the generated regressors is fundamental to perform correct inference.


We define the weighted {\it partial mean process} indexed by $(y,t)$ by
\begin{align}
 \mathbb{E}\big[ F_{Y|TV}(y|t, V)  W\big] = \int\ \mathbb{E}\big[ {\bf 1}{\{ Y \leq y \}} \big| T=t, V = v \big] w\ \ dF_{V W}(v, w). 
\label{PMGR}
\end{align}
\footnote{The notation ${\bf 1}{\{A\}}$ is an indicator function for the event $A$.}
The conditional expectation $F_{Y|TV}(y|t,V)$ is the conditional cumulative distribution function (CDF) of the outcome $Y$ given the continuous explanatory variables $T$ and control variables $V$. 
The regressor $V = v_0(\cdot)$ is a function of observables, which can be estimated in a first step as a {\it generated regressor}.
In the second step, the conditional CDF $F_{Y|TV}(y|t,V)$ is estimated nonparametrically by a kernel regression using the first-step generated regressor.
The third step is sample analogue of a marginal integration that averages out the regressor~$V$ and the weight $W$, but fixes the value of the continuous treatment $T$ at $t$. 
Because the regression function $F_{Y|TV}(y|t,V)$ contains more arguments than being averaged over in the third step, this is known as a {\it partial mean} in the terminology of \cite*{Newey94ET}. 
As $T$ is continuous, the partial mean is an infinite-dimensional nonparametric object.
We analyze the three-step estimator for the partial mean of the weighted conditional CDF in Eq.$\!$~(\ref{PMGR}) that builds on and extends the partial mean literature. 
Our main contribution is a stochastic expansion that is uniform over $(y,t)$ and accounts for the estimation error of the generated regressors.
The influence of the generated regressors is characterized by an  integral operator on the estimation error $\hat v(\cdot) - v_0(\cdot)$. 
The estimator $\hat v$ can be (semi)parametric or nonparametric for a general function $v_0$, such as mean regression, density, or quantile regression.


Another set of our results is weak convergence of the partial mean process indexed by the threshold value $y$, for each $t$.
By extending the results to the Hadamard-differentiable functionals of the partial mean process, we provide the limiting distribution and uniform inference method for estimating various distributional structural features and common inequality measures, such as the Gini coefficient.
The mean is the {\it average structural function} or the {\it dose response function} \citep*{BP03,Flores07}.
The quantile is the {\it quantile structural function} \citep*{IN09ETA}. 
The difference between two treatment levels is the quantile treatment effect.
We show that a multiplier bootstrap method is valid for uniform inference over $y$, which enables functional hypotheses tests for the whole distribution, such as tests for stochastic dominance.

\medskip

We illustrate the usefulness of our general results by two examples.
The first is the generalized propensity score (GPS), defined as the conditional density function of treatment given observable characteristics $X$.  
Under the unconfoundedness assumption, the GPS is known to reduce dimensionality in the second-step regression \citep*{HI04}.
Using the GPS as a generated regressor $V = f_{T|X}(t|X)$ or weight is now a common practice in  program evaluation where the length of participation is often taken as the continuous treatment \citep*{ACW10, FFGN12ReStat, KSUZ12, GW, HHLP}.
This paper is the first presentation of a complete limit theory of nonparametric kernel regression on the estimated GPS.
Another causal object of interest is the treatment effect {\it for the treated}, where {\it the treated} subpopulation is defined by individuals who are currently choosing treatment value~$\bar t$.
The local average response and the marginal treatment effect for the treated are based on the conditional average structural function given $T = \bar t$ \citep*{AM05ETA, FHMV08ETA}.
We add to the literature a new estimator of the effect for the treated that regresses on two GPSs, one for the counterfactual treatment value $t$ and one for the treated value $\bar t$, i.e., $V = (f_{T|X}(t|X), f_{T|X}(\bar t|X))'$.
We illustrate the effectiveness of our methods by analyzing the Job Corps program.
We estimate the causal effects of the length of participation on employment 
that uncovers heterogeneities in the effects of dosages of academic and vocational training.
For example, we find that extending the length in Job Corps from one month to six months increases the proportion of weeks employed in the second year by 2.5\% on average.

\medskip

The second example is the control variable as in the triangular simultaneous equations models in \cite*{NPV99ETA} and \cite*{IN09ETA}. 
Our limit theory applies to nonparametrically estimate the bounds of the average and quantile structural functions.
Moreover the stochastic expansion, which is uniform over $t$ and $y$, is useful when a partial mean is an intermediate step in a multi-step estimation a semiparametric setting.
In such cases, the estimation error from the control variable might not be ignorable.
This extension to functionals of the partial mean is a direct calculation from our stochastic expansion.  
For example, the compensating variation in welfare analysis studied by \cite*{DB} can be expressed as functionals of the average structural function.


\medskip

To our best knowledge, this is the first paper that analyzes a marginal integration estimator for partial means of a kernel regression estimator, accounting for general generated regressors.
There are two infinite-dimensional parameters in the partial mean: the generated regressor $V=v_0(\cdot)$ and the regression function $F_{Y|TV}(y|t, \cdot )$.
The generated regressor enters the partial mean estimation through two channels: first, it is an {\it argument} of the regression function that directly enters the partial mean; second, it is a {\it regressor} that determines the functional form of the regression $F_{Y|TV}(y|t, \cdot)$.
Our new uniform expansion distinguishes the {\it argument} and {\it regressor} roles of the generated regressor and contains three important {\it elements} --- (i) partial mean structure,
(ii) generated regressor as an index function of observables, and (iii) projection of the weight $\mathbb{E}[W|V]$.
This finding appears to be of significant practical importance by providing conditions under which the estimation error is negligible. 
These {\it elements} are generic for marginal integration estimators that involve a kernel regression with generated regressors, in both nonparametric or semiparametric models.

\medskip

The rest of the paper is organized as follows:
Section~\ref{SecLit} discusses our contribution and the related literature.
Section~\ref{SPO} introduces the setup and causal parameters of interest.
We outline three-step nonparametric kernel estimation for the weighted partial mean process with generated regressors defined in Eq.$\!$~(\ref{PMGR}).
Section~\ref{SknownW} presents the main asymptotic theorems.
Section~\ref{SecEx} illustrates the usefulness of our results by economic examples.
Section~\ref{SecInf} presents the limit theories for treatment effects for the Hadamard-differentiable policy functionals
and specifically for the quantile processes.
A multiplier bootstrap method enables uniform inference.
Section~\ref{SecEm} is the empirical application on the Job Corps program. 
Section~\ref{SecCon} concludes this paper.
The proofs are in the Appendix.

Throughout this paper, let an uppercase letter $V$ be a generic $d_v$-dimensional random vector with support $\mathcal{V} = Supp(V)\subseteq \mathbb{R}^{d_v}$, where $d_v$ denotes the dimension of $V$.
Denote the interior support of $V$ to be $\mathcal{V}_0$.
The realized value of $V$ is denoted by a lowercase letter $v$.

\subsection{Related Literature}
\label{SecLit}


The {\it argument} role of the generated regressor is played by an unknown function affecting the sampling variation of the final estimator directly and has been studied extensively in the literature; for example, \cite*{PP89ETA}, \cite*{Andrews94}, \cite*{Sherman94},  \cite*{CLK03ETA}, \cite*{IL10}, among many others.
The challenge lies in the {\it regressor} role of the generated regressor that indirectly affects the sampling variation of the final estimator through the second-step nonparametric regression.
There is a recent literature on nonparametric regression with generated regressors, for example, \cite*{Song08ET,Song14JoE}, \cite*{Sperlich}, \cite*{MRS12A, MRS12B},  \cite*{HR13ETA, HR16}, \cite*{HahnLiaoRidder}, \cite*{EJL}, among others.
Most of the previous work focuses on the impact of generated regressors on the nonparametric regression and its application in semiparametric models.
In contrast to the semiparametric models where the estimator integrates the regression function over {\it all} continuous regressors and hence is a full mean,
our partial mean is a marginal integration over the generated regressor but evaluates the continuous regressor $T$ at a fixed value $t$.

\cite*{HR13ETA} are among the first to characterize the asymptotic variance of a semiparametric estimator contributed by the dual role of the generated regressors by using \cite*{Newey94ETA} path-derivative method.
\cite{HR16} further consider the control variable estimator with non-separable errors and endogenous regressors.
In contrast, we use stochastic expansion that gives sufficient conditions under which the asymptotically linear representation of the kernel-based estimator is valid.
The empirical process approach allows us to express the impact of the generated regressor on the final estimator by an integral operator on the estimation error $\hat v(\cdot) - v_0(\cdot)$.
So our result can be applied to general estimators for the generated regressor, such as regression or density.\footnote{
\cite*{HahnLiaoRidder} study two-step sieve M estimation, where the sieve estimation may involve generated regressors.
Their parameters of interest are {\it known} functionals of the unknown functions estimated in both steps that do not include the partial mean/full mean.
\cite*{EJL} use a stochastic equicontinuity argument on a full mean based on a general class of conditional mean functions.
\cite*{MRS12B} study a class of semiparametric optimization estimators that involves nonparametric regression with generated regressors and verify sufficient conditions for the asymptotic normality developed in \cite*{CLK03ETA}.
Besides the apparent difference to the aforementioned papers that our partial mean is infinite-dimensional, we 
 give conditions directly on the kernel regression estimator and generated regressor. 
We do not assume the high-level smoothness assumption that the regression function is very smooth with respect to the regressor. 
We discuss more technical detail in Section~\ref{SecTR} in the Appendix.}

Our empirical process approach builds on the stochastic equicontinuity argument for the kernel estimator developed in \cite*{MRS12A}.
While \cite*{MRS12A} contribute a detailed characterization of how the generated regressor affects the regression estimator, we focus on its impact on the partial mean of such regression.
The partial mean is a  function of the continuous variables $t$ and hence is estimated nonparametrically at a convergence rate slower than regular root-$n$.
Deriving the weak convergence and multiplier method demands more involved technical arguments than applying the standard Donsker's Theorem as in \cite*{Rothe10JoE, FP11, DHB, CFM12}, for example.



\section{Setup and Estimation}
\label{SPO}
This section introduces the causal objects of interest, which are identified by functionals of the partial means in Eq.$\!$~(\ref{PMGR}). 
The partial mean is also a statistical object with broad applications, such as additive nonparametric models and differential equation solution introduced in \cite*{Newey94ET}.
We use the potential outcome framework that is convenient for interpretation and simplifies notation.
The treatment effect model is known to be equivalent to a nonseparable outcome with a general disturbance, where the outcome equation is  $Y = \phi(T, \epsilon)$, e.g.,  \cite{IN09ETA}, \cite{WC13ER}.
The error $\epsilon$ represents unobservable individual heterogeneity.
No functional form assumption is imposed on the general disturbances $\epsilon$, like monotonicity, dimensionality, or separability. 
Let $Y(t) = \phi(t,\epsilon)$ denote the potential outcome corresponding to the treatment level $t$, where the randomness comes from the unobserved disturbances $\epsilon$.
If the treatment value $\bar t$ is chosen by an individual $i$, then $T_i = \bar t$ and the observed outcome $Y_i =Y_i(\bar t)$ is one of the potential outcomes $\{Y_i(t) = \phi(t, \epsilon_i) \}_{t \in {\mathcal{T}}}$.
For example, we observe the quantity demanded at the observed price, but cannot observe what the demand would have been given other prices.
Regarding program evaluation, we observe the wage of a participant after one year in a job training program, but we wish to learn what the wage would have been if the participant had stayed in the program for two years.

The key causal object of interest is the unconditional distribution of $Y(t)$ by averaging out other covariates and unobservable heterogeneity.
The cumulative distribution function (CDF) of the potential outcome  $F_{Y(t)}(y) = \mathbb{E}[{\bf 1}{\{\phi(t, \epsilon) \leq y\}}] = \int {\bf 1}{\{\phi(t, \epsilon) \leq y\}} f_\epsilon(\epsilon) d\epsilon$ is the outcome distribution when the value of the treatment $T$ is fixed at $t$ and the expectation is taken with respect to the marginal density of $\epsilon$, $f_\epsilon$.
In other words, it is the unconditional outcome distribution if, hypothetically, the individual had been assigned to the treatment level $t$.
An array of estimands are often of interest based on these causal outcome distributions.
We consider a general class of functionals $\Gamma$ on $F_{Y(t)}(\cdot)$.
For example, if interest centers on the quantile treatment effect, we let $\Gamma$ be the quantile operator $\Gamma(F_{Y(t)}) = Q_\tau(Y(t)) = \inf\{y: F_{Y(t)}(y) \geq \tau\}$.
The quantile treatment effect corresponding to a change from $\bar t$ to $t$ is $\Gamma(F_{Y(t)}) - \Gamma(F_{Y(\bar t)})$.
When the outcome structural function $\phi$ is monotone in the error $\epsilon$, the {\it quantile structural function} $Q_\tau(Y(t))$ equals the structural function evaluated at $(t, \tau)$, $\phi(t,\tau)$, by a normalization.  
If interest is on the mean treatment effect, then let $\Gamma$ be the mean operator.
When the outcome is separable $Y = \phi(T) + \epsilon$, the {\it average structural function} $\mathbb{E}[Y(t)]$ equals $\phi(t)$ up to an additive constant.  
Other inequality measures are also applicable, such as the coefficient of variation, the interquantile range, the Theil index, the Gini coefficient, the Lorenz curve.  

We use a conditional independence assumption to show that the partial mean process in Eq.$\!$~(\ref{PMGR}) identifies the causal objects of interest.

\begin{Assumption}[Conditional Independence Assumption]
There exists an observable or identifiable control variable $V$ such that $T$ and $\epsilon$ are independent conditional on $V$.
An equivalent expression in the potential outcome framework: for any $t$, the potential outcome $Y(t) = \phi(t,\epsilon)$ is independent of the treatment $T$, given $V$.
\label{ACIA}
\end{Assumption}

Under Assumption~\ref{ACIA}, 
\begin{align}
\mathbb{E}[Y(t)|V = v] &= \int_{Supp(\epsilon|V=v)} \phi(t, \epsilon) f_{\epsilon|V}(\epsilon|v) d\epsilon \notag \\
&=  \int_{Supp(\epsilon|T=t, V=v)} \phi(t, \epsilon) f_{\epsilon|TV}(\epsilon|t, v) d\epsilon \notag \\
&= \mathbb{E}[Y |T=t, V = v], 
\label{IDCIA}
\end{align}
for any $v \in Supp(V|T=t)$. 
Let $\mathcal{V}^*$ denote a  subset of the support of $V$ conditional on $T=t$.
We will give examples of particular $\mathcal{V}^*$ in the following sections.
Then for any $v \in \mathcal{V}^*$, $\mathbb{E}[Y|T=t, V = v]$ is well-defined and hence $\mathbb{E}[Y(t)|V = v]$ is identified from Eq.$\!$ (\ref{IDCIA}).
In words, the conditional mean of the potential outcome given the control variable is recovered by the mean of the observed outcome of those who had received $t$ and with the same value of  the control variable.
Therefore, we can identify the unconditional mean of $Y(t)$ defined on $\mathcal{V}^*$,
\begin{align}
\mathbb{E}^*[Y(t)] &\equiv 
\int_{\mathcal{V}^*} \mathbb{E}[Y(t)|V = v] f_V(v) dv 
= \mathbb{E}\big[  \mathbb{E}\big[ Y\big| T=t, V \big] \pi(V) \big],
\label{DRFlb}
\end{align}
where $\pi(v) = {\bf 1}\{v \in \mathcal{V}^\ast\}$.
This is the {\it dose response function} or {\it average structural function} for the subpopulation whose control variable takes values in $\mathcal{V}^*$.

When the common support $\mathcal{V}^*$ equals the support of $V$, i.e., $\pi(v) = 1$, the population average structural function $\mathbb{E}^*[Y(t)] = \mathbb{E}[Y(t)]$ is therefore identified.
This is known as the {\it common support} or {\it overlapping assumption} in the literature, i.e., the support of $V$ conditional on $T=t$ equals to the support of $V$.
The common support assumption requires everyone in the population to be able to find a match who shares the same value of the control variable $V$ and has received the counterfactual value $t$.

Nevertheless, the common support assumption is known to be strong in many applications. 
We do not impose this common support assumption, but estimate the causal object defined by the common support $\mathcal{V}^*$:
\begin{align}
F_{Y(t)}\big(y;V, W\pi \big) &\equiv \int \mathbb{E}\big[ {\bf 1}{\{ Y \leq y \}} \big| T=t, V = v \big]\ w \ \pi(v)\ dF_{V W}(v, w), \label{PMGRest}
\end{align}
where 
the trimming function $\pi(v) \equiv {\bf 1}{\left\{\inf_{t\in \mathcal{T}^*} f_{TV}(t, v) \geq c\right\}} = {\bf 1}{\{v \in \mathcal{V^*}\}}$ with a positive constant $c$ and a range of treatment values of interest $\mathcal{T}^*$ that is an interior subsupport of $T$. 
The trimming function selects a common support $\mathcal{V}^*$, which is a subset of the intersection of the supports of $V$ conditional on $T=t$ for $t \in \mathcal{T}^*$,
\begin{align}
\mathcal{V}^* \equiv \Big\{v: \inf_{t\in \mathcal{T}^*} f_{TV}(t, v) \geq c \Big\}\ \subseteq\ \bigcap_{t\in \mathcal{T}^*} Supp(V|T=t). \label{CSdef}
\end{align}
An advantage of the trimming approach is that the identification set is valid for any positive trimming parameter $c$.
Another purpose of trimming is technical: 
The trimming function allows us to work on a compact interior subsupport where the density functions are bounded away from zero, satisfying the smoothness Assumption~\ref{Asm}.\footnote{We can then use uniform convergence of the first- and second-step estimators over the range of integration that suffices for deriving the properties of the final estimator.
The edge effect and the boundary problem of the local constant estimator are avoided.
\cite{IN09ETA} refer the edge effect to the case when the convergence rate of the estimator depends on how fast the joint density goes to zero on the boundary.
As the joint density of the endogenous variable $T$ and the control variable $V$ often goes to zero at the boundary of the support of the control variable in the triangular model, averaging over the control variable over the entire support upweights the tails relative to the joint distribution. 
} 
The trimming approach has been used in similar contexts, for example, \cite{Newey94ET}, \cite{IN09ETA}, \cite{IchimuraTodd}, \cite*{FFGN12ReStat}, to simplify the theoretical analysis and practical application. 

The weight $W$ is the additional observed variable that is not involved in the regression function $F_{Y|TV}$.
We will discuss specific estimators for the economic examples in the later sections.
We find that the estimation approach outlined below has different properties depending on the details of implementation corresponding to each estimand.
As a result, different asymptotic distribution results are proceeded for the different versions of this general estimation approach described below.
The estimation procedure involves three steps:
\begin{changemargin}{1cm}{0.3cm} 
\begin{enumerate}
\item[Step 1.]
(Generated Regressor)\ 
The regressor $V = v_0(S_v)$ is a vector of measurable functions of observables $S_v$.
The estimator $\hat V_i = \hat v(S_{vi})$ satisfies certain smoothness and uniform convergence conditions, specified in Assumption \ref{Acom2}.


\item[Step 2.]
(Regression)\ A local polynomial estimator of the regression function of the indicator ${\bf 1}{\{Y \leq y\}}$ on $(T, \hat V)$ from Step 1, evaluated at $(t, v)$;
 for example, a local constant estimator is
$\hat F_{Y|T\hat V}(y|t,v) = n^{-1}\sum_{j=1}^n {\bf 1}{\{ Y_j\leq y \}} K_h(T_j-t) K_h\big(\hat v(S_{vj}) - v\big)\big/\hat f_{T\hat V}(t, v)$,
where $\hat f_{T\hat V}(t, v)$ is the standard kernel density estimator.
The product kernel is $K_h(u) \equiv h^{-d_u} \Pi_{l=1}^{d_u} k\big(u_l/h\big)$,
where $h$ is the bandwidth assumed the same for all the elements of the vector $u$ for simplicity and $k$ is the $r$-order kernel function satisfying Assumption~\ref{AkernelEJL}.
Let the dimension of the regressors at this step be $d = d_t + d_v$.

\item[Step 3.]
(Partial Mean)\ A marginal integration of the regression function from Step 2 averages over the generated regressor $\hat V$ from Step 1 and fixing the treatment variable $T$ at level $t \in \mathcal{T}^*$,
\begin{align*}
&\hat F_{Y(t)}\big(y;\hat V,W\hat \pi \big) = \frac{1}{n}\sum_{i=1}^n \hat F_{Y|T\hat V}\big(y|t, \hat v(S_{vi})\big)  W_i \hat \pi(\hat V_i), 
\end{align*}
where $\hat \pi(\hat V_i) = {\bf 1}{\{ \inf_{t \in \mathcal{T}^*}\hat f_{T\hat V}(y|t, \hat V_i) \geq c \}}$.\footnote{To choose the trimming parameter $c$, we could construct $L$ grid points $\{ t_{(1)}, t_{(2)},..,t_{(L)}\} \subset \mathcal{T}^*$ to be quantiles of $T$ over the range $\mathcal{T}^*$.
For each grid point $l \in \{1,2,.., L\}$, let $c_l$ be the sample $2.5\%$-quantile of the positive estimated densities $\{ z: z = \hat f_{T\hat V}(t_{(l)}, v) >  0, v\in \hat{\mathcal{V}}\}$, where $\hat{\mathcal{V}} = \{\hat V_1, \hat V_2, ..., \hat V_n\}$ estimates the support of $V$.
Let the trimming parameter be $c = \max\{c_1, c_2,..., c_L\}$. 
Alternative approaches to select the common support or the trimming function could be found in \cite{FFGN12ReStat} and \cite{KSUZ12}, for example.
\label{ftrim}}


\end{enumerate}
\end{changemargin}

From the above procedure, we see that the generated regressors $\{\hat V_i\}_{i=1}^n$ in Step 1 are used as {\it regressors} in Step 2 and used again as {\it arguments} of the regression function in Step~3.
The treatments $T$ or covariates $V$ could contain discrete variables and the kernel is replaced by an indicator function, known as the frequency method.
For notational convenience, discrete covariates are not allowed for.
The smoothness Assumption~\ref{Asm} (ii) requires that  the treatment variables cannot have point masses, i.e., ${\rm Pr}(T=t) = 0$ for $t \in \mathcal{T}^\ast$. 

\section{Asymptotic Results}
\label{SknownW} 
We first present the limit theory for estimating the partial mean process when all the regressors are observed.
This is for the case to identify the causal distribution of the potential outcome under unconfoundedness, where the Conditional Independence Assumption~\ref{ACIA} is satisfied by $V = X$ observable characteristics.
This limiting distribution is also for the case when the Step 1 estimation error of the generated regressor is asymptotically ignorable.
We show the partial mean estimator weakly converges to a Gaussian process indexed by $y$ for each $t$, which serves as a building block for analyzing the estimator involving generated regressors in Section~\ref{SecGR}.
Our main result is a uniform stochastic expansion of the three-step estimator, revealing the influence of estimating the generated regressors of general function form on the final estimator.  
The stochastic expansion is uniform over the treatment value of interest $t$ and $y$ for the CDF.
We characterize the leading bias that can be useful to compute the asymptotically mean squared error optimal bandwidth and to conduct robust inference; see recent development by \cite*{CCF} and \cite{AK}, for example.
The asymptotic theorems focus on the local constant estimator.
Then we present the results for the local polynomial estimators.


\subsection{Observable regressors}
\label{SOB}
Following the procedure described in Section~\ref{SPO},
we estimate the partial mean process with observable regressors $V$ in Eq.$\!$ (\ref{PMGRest}) by 
$\hat F_{Y(t)}(y;V,W\hat \pi) = n^{-1} \sum_{i=1}^n \hat F_{Y|TV}(y|t, V_i) \ W_i\hat \pi(V_i)$,
where the trimming function $\hat \pi(V_i) = {\bf 1}{\{\inf_{t\in\mathcal{T}^*}\hat f_{TV}(t, V_i) \geq c\}}$.
\begin{Theorem}[Weak Convergence]
Suppose Assumptions~\ref{ACIA}, \ref{Asm}-\ref{Abw} in the Appendix hold with $\Delta_k \geq a$ and $\Delta \geq a + r$.
Assume $W$ is uniformly bounded.
Assume the derivatives of $\mathbb{E}[W|V]$ up to order $r$ exist and are uniformly bounded and continuous.  
Define 
$\Sigma_{y_1y_2}(V) \equiv \mathbb{E}\Big[ \Big( F_{Y|TV}(\min\{y_1, y_2\} | t,V) - F_{Y|TV}(y_1|t,V) F_{Y|TV}(y_2|t,V) \Big) \mathbb{E}[W| V]^2 \pi(V) /f_{T|V}(t|V) \Big] \int k^2(u) du$,
for any $y_1, y_2 \in \mathcal{Y}$.
Then 
letting $nh^{2r+d_t} \rightarrow C_B \in [0, \infty)$,
for any $t \in \mathcal{T}^*$ and $y \in \mathcal{Y}$,
$\sqrt{nh^{d_t}} \big( \hat F_{Y(t)}(y;V,W\hat \pi) - F_{Y(t)}(y;V,W\pi) - h^r \mathfrak{B}_{LC}  \big) \rightarrow \mathcal{N}(0, \Sigma_{yy}(V))$, where 
$\mathfrak{B}_{LC} \equiv \mathbb{E}\big[ \mathsf{B}_{LC}(t,V) \mathbb{E}[W|V] \pi(V) \big]$ and $\mathsf{B}_{LC}$ is the leading bias of the second-step local constant estimator given in Eq.$\!$ (\ref{eBiasLC}) in the Appendix.

Letting $nh^{2r+d_t} \rightarrow 0$, for any fixed $t \in \mathcal{T}^*$,
$\sqrt{nh^{d_t}} \big( \hat F_{Y(t)}(\cdot;V,W\hat \pi) - F_{Y(t)}(\cdot;V,W\pi)  \big) \Longrightarrow \mathbb{G}_t(\cdot;V,W\pi).$ 
The empirical process converges weakly to a Gaussian process $\mathbb{G}_t(\cdot;V,W\pi)$ with mean zero and the covariance kernel
$\Sigma_{y_1y_2}(V) = Cov \big(\mathbb{G}_t(y_1;V,W\pi), \mathbb{G}_t(y_2;V,W\pi)\big)$.
\label{TGaussian}
\end{Theorem}

The convergence rate of the partial mean estimator is $\sqrt{nh^{d_t}}$ depending on $d_t$, the dimension of the continuous conditioning variables that are fixed in the partial mean.
It is known that when more conditioning variables are averaged out in the partial mean, the nonparametric estimator converges at a faster rate \citep*{Newey94ET}.
A {\it full mean} estimator is the case when all arguments of the regressors are averaged out.
That is the case when the regressor $T$ is dropped or $T$ is discrete ($d_t = 0$) in our estimation procedure and limit theory.  
So the converge rate is root-$n$; for example, the average derivative in \cite*{PSS89ETA} and the propensity score regression estimator for multivalued treatment effects in \cite{Lee14DTT}.

\subsection{Generated regressors}
\label{SecGR}
This section presents the asymptotic theory for nonparametric estimation of the partial mean process  in Eq.$\!$ (\ref{PMGRest}) when the regressors $V$ are unobserved and estimated at the first step.
In particular, we consider $V = v_0(T,S)$, where $S \subseteq (X,Z)$.



To distinguish the {\it regressor} and {\it argument} roles of the generated regressor in its impact on the final estimator, we introduce some notation:
\begin{align*}
REG_{yt}\big(s,v\big) \equiv &\ \frac{f_{T|S}(t|s)}{ f_{T|V}(t| v)} \bigg\{
- \nabla_v F_{Y|TV}(y|t,v) \mathbb{E}\big[W\big|V=v\big] \\[0pt]
&+  \left( F_{Y|TV}(y|t,v) - F_{Y|TS}(y|t,s)\right)\!\!\left( \frac{\nabla_v f_{T|V}(t| v)}{f_{T| V}(t|v)}  \mathbb{E}\big[W\big|V\!=\!v \big]\!-\!\nabla_v \mathbb{E}\big[W\big|V\!=\!v\big] \right)\!\!\bigg\} \\
ARG_{yt}\big(w, v\big) \equiv&\  \nabla_v F_{Y|TV}(y|t,v)\ w, 
\end{align*} 
where $\nabla_v$ denotes the vector differential operator with respective to $v$.
$REG_{yt}$ is for using $\hat V$ as the {\it regressor} to estimate the regression function, and $ARG_{yt}$ is for using $\hat V$ as the {\it argument} of the regression function.
Let $\mathcal{T}^\dagger \times \mathcal{S}^\dagger$ be an interior subsupport of $(T, S)$, defined by the trimming function  $\pi_i = {\bf 1}{\{\inf_{t \in \mathcal{T}^*} f_{TV}(t, V_i) \geq c\}} = {\bf 1}{\{V_i=v_0(T_{i}, S_{i}) \in \mathcal{V}^*\}} =  {\bf 1}{\{(T_{i}, S_{i}) \in \mathcal{T}^\dagger \times\mathcal{S}^\dagger\}}$.
The following theorem states the main result.


\begin{Theorem}[Stochastic Expansion]
Suppose the conditions in Theorem~\ref{TGaussian} and Assumption~\ref{Acom2} hold.
Then uniformly in $y \in \mathcal{Y}$ and $t \in \mathcal{T}^*$,
\begin{align}
& \sqrt{nh^{d_t}} \Big( \hat F_{Y(t)}\big(y; \hat V,W\hat \pi\big) - F_{Y(t)}\big(y; V,W \pi\big) \Big) \notag \\
& = \frac{1}{\sqrt{n}} \sum_{i=1}^n \psi_{tin}\big(y; V,W\pi \big) \notag
+ \sqrt{nh^{d_t}} \Delta_{yt}(\hat v(T,S)) \notag \\
&\ \ \ +\sqrt{h^{d_t}} \frac{1}{\sqrt{n}} \sum_{i=1}^n \Big( F_{Y|T V}(y|t,V_i) W_i \pi_i - \mathbb{E}\big[ F_{Y|T V}(y|t,V) W \pi\big]\Big) + \sqrt{nh^{d_t}}R_n(y,t), \label{so0}
\end{align}
where the influence of estimating the Step 2 regression is
\begin{align*}
\psi_{tin}(y;V, W\pi) &\equiv \ \frac{\sqrt{h^{d_t}} K_{h}\big(T_i-t\big)}{f_{T|V}(t|V_i)}  \Big( {\bf 1}{\{Y_i \leq y\}} -  F_{Y|TV}(y|t,V_i) \Big)  \mathbb{E}\big[W\big| V = V_i \big] \pi(V_i),
\end{align*}
the influence of estimating the Step 1 generated regressor is
\begin{align*}
\Delta_{yt}(\hat v(T,S)) \equiv&\  \int_{\mathcal{S}^\dagger} \big(\hat v(t,s) - v_0(t,s)\big)' REG_{yt}\big(s,v_0(t,s)\big)\ dF_{S}(s) \\[5pt]
&+
\int_{\mathcal{W}} \!\int_{\mathcal{S}^\dagger}\!  \int_{\mathcal{T}^\dagger}  
\big(\hat v(\iota,s) - v_0(\iota,s)\big)' ARG_{yt}\big(w, v_0(\iota,s)\big)  \ dF_{TSW}(\iota,s,w),
\end{align*}
and the remainder term $R_n(y,t)$ satisfies $\sup_{y \in \mathcal{Y}, t \in \mathcal{T}^*}|R_n(y,t)| = O_p(n^{-\kappa_1} + n^{-\kappa_2})$ with 
$0 < \kappa_1 <   \min\big\{ (1-d_v \eta)/2 + \delta-\eta, (1 - d_t\eta)/2 + \delta,  (1-d\eta)/2 + 2(\delta - \eta) \big\} - (\delta + \xi^*)d_1/(2\alpha)$ and $0 < \kappa_2 < \min\big\{1- d\eta , 2(\delta-\eta), r\eta + \delta \big\}$.
\label{TGR}
\end{Theorem}

The impact of the generated regressor $\Delta_{yt}(\hat v)$ is characterized by an integral operator on the estimation error $\hat v(\cdot) - v_0(\cdot)$.
The Step 1 estimator $\hat v(\cdot)$ is taken as a fixed function given the sample and the integration is taken over the underlying random variables $(T, S, W)$.
The integral operator enables a further derivation by plugging in a linear representation of the estimation error $\hat v - v_0$ of a specific estimator.
We illustrate the stochastic expansion by kernel estimators and parametric estimators $\hat v$ in the examples in Section~\ref{SecEx}.  

We emphasize three {\it elements} of our stochastic expansion that summarize the important insight of the impact of generated regressors: (i) the partial/full mean structure of $\Delta_{yt}(\hat v)$,
(ii) {\it index bias} $F_{Y|TV} - F_{Y|TS}$ and {\it index ratio} $f_{T|S}/f_{T|V}$, and (iii) the {\it projection of the weight} $\mathbb{E}[W|V]$.
These {\it elements} are generic for the marginal integration estimators that involve a kernel regression with generated regressors.
We discuss these elements in more detail below and illustrate by the examples in Section~\ref{SecEx}.


\begin{itemize}
\item[(i)]
The partial/full mean structure of $\Delta_{yt}(\hat v)$ depends on how $T$ enters the generated regressor $v_0(T,S)$ as a regressor or in the dependent variable.
The partial/full mean suggests a convergence rate, as discussed in Section~\ref{SOB}.
As a result, we do not assume the sampling variation from Step~1 nonparametric estimation to be of smaller order by restricting the tuning parameter.

\item[(ii)]
The generated regressor can be viewed as an index function of observables $S$.
The index ratio $f_{T|S}/f_{T|V}$ is the impact of $S$ on the treatment density that is not captured by the ``index" $v_0(t, S)$.
The index ratio is a special feature of the Step 3 partial mean where the treatment is fixed.
The index bias $F_{Y|TV} - F_{Y|TS}$ is the impact of the underlying variable $S$ on the outcome distribution that is not captured by $v_0(t, S)$.
The index bias and ratio come from the Step 2 regression that averages over the underlying variable $(T, S)$ of the generated regressor. 
It results in the distribution functions conditional on $S$, rather than $V$ that is taken as a given index function $v_0$.
The existing literature has known the index bias from the second-step nonparametric regression; for example, the semiparametric models in \cite*{HR13ETA}, \cite*{EJL}, and \cite*{MRS12B}.

\item[(iii)]
The {\it projection of the weight} $\mathbb{E}[W|V]$ comes from the Step 3 partial mean that averages over the arguments $V$ and the additional variable $W$.
Step~3 can be viewed as regressing $W$ on the arguments $V$ in the second-step kernel regression estimator.

\end{itemize}

The last two terms (\ref{so0}) in the stochastic expansion from the estimation error of the Step~3 summation  diminish at a root-$n$ rate.
We maintain the smaller-order terms in the expansion for two reasons: 
First, when the partial mean is an intermediate step in a multi-step estimation procedure, our uniform expansion can be used to analyze the asymptotic property of the final estimator. 
In the case when the final estimator is $\sqrt{n}$-consistent, the impact of Step 3 partial mean and the impact of Step~1 generated regressors could be of first order.   
The second reason is to improve the asymptotic approximation in finite sample. 
The asymptotic variance could be estimated by the second moment of the estimated influence function 
in the stochastic expansion, 
including the terms that are asymptotically negligible. 

\begin{Remark}[Transformed  outcome]
{\rm Theorem~\ref{TGR} is readily extended to a transformed dependent variable $A_y(Y)$ replacing the dependent variable ${\bf 1}{\{Y \leq y\}}$; the Proof of Lemma~\ref{LMRS3}~(I\!I) in the Appendix.
Consider the parametric family $\{A_y: y \in \mathcal{A}\}$ composed of uniformly bounded functions $A_y$ indexed by a finite-dimensional parameter $y \in \mathcal{A}$, a compact set in $\mathbb{R}$, and $A_y$ is Lipschitz continuous in $\alpha$. 
Theorem~\ref{TGR} provides a stochastic expansion of the partial mean with generated regressors $n^{-1}\sum_{i=1}^n \hat{\mathbb{E}}\big[A_y(Y)|T=t, \hat V = \hat V_i\big] W_i\hat \pi_i$ uniformly over $y \in \mathcal{A}$ and $t\in \mathcal{T}^*$.
This result can be applied to various contexts, for example, semiparametric transformation model in \cite*{VV13}.
}
\label{RVV}
\end{Remark}

Next we consider the Step~2 regression to be the $p$th-order local polynomial estimator.
Theorem~\ref{TLP} shows that the first-order asymptotic properties are the same as described in Theorems~\ref{TGaussian} and \ref{TGR} for the local constant estimator, under an undersmoothing condition such that the leading bias is of smaller order.
We apply Theorem 2 in \cite*{MRS12B} for bounds on the supremum norm of the approximation error and modify the corresponding regularity conditions;
see the detailed discussion in Section~\ref{SecTR} in the Appendix.

\begin{Theorem}[Local polynomial estimator]
Let Assumptions~\ref{ACIA}, \ref{Asm}-\ref{AsmSE} hold with $\Delta = p+1$ and $\Delta_k = 2$.
Let Assumption~\ref{Abw} hold with (i) $nh^{2(p+1)+d_t} \rightarrow C_B \in [0, \infty)$.
Then the results in Theorem~\ref{TGaussian} hold: 
$\sqrt{nh^{d_t}} \big( \hat F_{Y(t)}(y;V,W\hat \pi) - F_{Y(t)}(y;V,W\pi) - h^{p+1} \mathfrak{B}_{LP}  \big) \rightarrow \mathcal{N}(0, \Sigma_{yy}(V))$, where 
$\mathfrak{B}_{LP} \equiv \mathbb{E}\big[ \mathsf{B}_{LP}(t,V) \mathbb{E}[W|V] \pi(V) \big]$ and $\mathsf{B}_{LP}$ is the leading bias of the Step~2 local polynomial estimator.\footnote{We focus on the local polynomial estimator with an odd order $p$.   
It is straightforward to show the result for an even order $p$, where the bias is of order $h^{p+2}$.}

Further let Assumption~\ref{Acom2} hold with (ii) $\delta > \eta$ and
assume Assumption~\ref{Acon}.
Then the results in Theorem~\ref{TGR} hold with
$\sup_{y\in\mathcal{Y}, t \in \mathcal{T}^\ast}|R_n(y,t)| = O_p( n^{-\kappa_{LP}})$, where
$0 < \kappa_{LP} < \min\big\{(1-d\eta)/2 + \delta - \eta -(\delta + \xi^*)d_1/(2\alpha), p \eta + \delta, 2\delta - \eta, r\eta + \delta\big\}$.
\label{TLP}
\end{Theorem}

\section{Examples}
\label{SecEx}

The usefulness of the general estimation procedure and asymptotic theory is illustrated by, but not limited to, the economic examples in this section.
The three {\it elements} --- partial mean, index, and projection of the weight --- critically determine the influence of estimating the generated regressors on the partial mean estimator.
We allow the generated regressors to be general functions; for the examples considered in this section, the generated regressors are conditional density functions, conditional CDF, or residual from a regression.
The Step~1 estimator of the generated regressors can be (semi)parametric or nonparametric, satisfying Assumption~\ref{Acom2} and with an asymptotic linear representation.
We illustrate with kernel estimators and parametric estimators in these examples.
For the parametric estimator, the weak convergence of the partial mean estimator is asymptotically equivalent to the Gaussian process as if the true generated regressor was observed. 
For the nonparametric estimator, we provide conditions under which the Step 1 estimation error is asymptotically ignorable. 
We follow the setup in Section~\ref{SPO} and consider a single continuous treatment variable ($d_t =1$)
and a set of treatment values of interest $\mathcal{T}^*$.

\subsection{Continuous treatment effects}
\label{SGRGPS}
Under unconfoundedness or selection on observables assumption, the observable characteristics $X$ is a valid control variable satisfying Assumption~\ref{ACIA}.
\cite*{HI04} show that regressing on the generalized propensity score (GPS) is sufficient for estimating continuous treatment effects {\it for the population}, i.e., $\mathbb{E}[Y(t) - Y(\bar t)]$ for $(t, \bar t) \in \mathcal{T}^\ast$.
The key identifying assumption is that for each $t$, $Y(t)$ is independent of $T$ given $V = v_0(X) = f_{T|X}(t|X)$.
We extend the results for the effects for the population in \cite{HI04} to the effects {\it for the treated} that is $\mathbb{E}[Y(t) - Y(\bar t)|T=\bar t]$ by switching the treatment value from $\bar t$ to $t$ {\it for the subpopulation with treatment value} $\bar t$.
We find that to identify $\mathbb{E}[Y(t)|T=\bar t]$, we need to adjust for two generalized propensity scores --- one for the potential treatment value $t$ and one for the treated value $\bar t$.
Thus the generated regressor is $V= v_0(X) = (f_{T|X}(t|X), f_{T|X}(\bar t|X))'$.
We extend the kernel-based propensity score regression estimator in \cite*{HIT98RES} to a continuous treatment case and provide a limit theory.

\bigskip

We first discuss what we learn from the three elements of the stochastic expansion $\Delta_{yt}$ in Theorem~\ref{TGR}.
(i) 
The GPS $f_{T|X}(t|X)$ has two regressors $(T, X)$, but $T$ is evaluated at $t$ in the estimation procedure, i.e., $f_{T|X}(t|X)$ is not a function of $T$.
As a result, $\Delta_{yt}$ is a partial mean in the sense that only $X$ is integrated out and $T$ is fixed at $t$. 
It follows that when the GPS is estimated nonparametrically, $\Delta_{yt}$ converges at a nonparametric rate.
(ii) 
The index ratio $f_{T|X}(t|X)/f_{T|V}(t|v_0(X))=1$ and $\nabla_v f_{T|V}(t|v)|_{v=V} = 1$ for $V = f_{T|X}(t|X)$.
(iii) 
For the population treatment effect, $W=1$ and hence $\nabla_v \mathbb{E}[W|V=v] = 0$. 
For the treatment effect for the treated where $V= (f_{T|X}(t|X), f_{T|X}(\bar t|X))'$, $\mathbb{E}[W|V] = f_{T|X}(\bar t|X)/f_T(\bar t)$ and hence $\nabla_v \mathbb{E}[W|V=v] = (0,1/f_T(\bar t))'$.

\bigskip

Now we present the detail of the estimator for the effects for the population and its asymptotic properties. 
We estimate the CDF of $Y(t)$ for the subpopulation whose observable characteristics take values on a common support $\mathcal{X}^*$, using the GPS as the generated regressor $V=v_0(X) = f_{T|X}(t|X)$ for each $t \in \mathcal{T}^*$, i.e., as defined in Eq. $\!$(\ref{PMGRest}), 
\begin{align*}
F_{Y(t)}^*(y) &= \mathbb{E}[F_{Y(t)|V}(y|V) \pi(V)] = \int F_{Y|TV}(y|t, v) \pi(v) dF_V(v) = F_{Y(t)}(y; V, \pi), 
\end{align*}
where the trimming function $\pi(V)  = {\bf 1}{\{ \inf_{t\in\mathcal{T}^*} f_{T|X}(t|X) \geq c \}} = {\bf 1}{\{ \inf_{t\in\mathcal{T}^*} V \geq c \}} = {\bf 1}{\{ X \in \mathcal{X}^*\}}$.\footnote{The support of $V$ conditional on $T=t$ equals the support of $X$ conditional on $T=t$ by definition.  So the common support $\mathcal{V^*} = \mathcal{X}^*$ is defined the same as the case using $X$ as the control variable in Eq.$\!$ (\ref{CSdef}).
}

The stochastic expansion in Theorem~\ref{TGR} implies the influence from estimating the GPS is expressed by an integral operator on $\hat f_{T|X}(t|\cdot) - f_{T|X}(t|\cdot)$:
\begin{align}
\Delta_{yt}(\hat f_{T|X}(t|X)) &= \int_{\mathcal{X}^*} \left(\hat f_{T|X}(t|x) - f_{T|X}(t|x)\right)' \Lambda_{yt}(x)\ dF_X(x), \ \mbox{where} \label{IFGPS}
\\
\Lambda_{yt}(x) &\equiv \left( F_{Y|TV}(y|t, f_{T|X}(t|x)) - F_{Y|TX}(y|t,x)\right) \big/ f_{T|X}(t|x). \notag
\end{align}
We see that the influence $\Delta_{yt}$ is determined by the index bias $F_{Y|TV} - F_{Y|TX}$.

\bigskip

We consider two estimators for the GPS in Step~1.
Step 2 and Step 3 follow the procedure described in Section~\ref{SPO}.
We can choose the bandwidth $h \sim n^{-0.22}$ for Step~2 local linear estimator or local constant estimator. 
\begin{changemargin}{1cm}{0.6cm} 
\begin{enumerate}
\item[Step 1.](GPS)\ \
\begin{enumerate}
\item (Kernel)
$\hat V_i = \hat f_{T|X}(t|X_i)$ is the standard conditional kernel density estimator, using a $r_1$-order kernel function with bandwidth $h_1$, satisfying Assumption~\ref{AkernelEJL}.

\item (MLE)\ \ 
Assume the conditional distribution of $T$ given $X$ is normal $\mathcal{N}(X'\beta_0, \sigma_0^2)$.
The probit maximum likelihood estimator is
$\hat V_i = \exp\big({-(t-X_i'\hat \beta)^2/(2\hat \sigma^2})\big)/\sqrt{2\pi\hat \sigma^2} \equiv \zeta(t, X_i'\hat\beta, \hat \sigma^2)$. 
\end{enumerate}
\end{enumerate}
\end{changemargin}

In the following Corollary~\ref{GRGPS}(I), we present the influence from estimating the GPS by the two estimators described above.
For a nonparametric kernel estimator in Step~1(a), the influence $\Delta_{yt} = O_p((nh_1)^{-1/2})$.
We also characterize the leading bias from Step~1(a).
For the commonly used probit estimator in Step~1(b), the influence $\Delta_{yt} = O_p(n^{-1/2})$.
Corollary~\ref{GRGPS}(I\!I) presents the limiting property when the nonparametric estimation of the GPS is not first-order ignorable.
Finally in Corollary~\ref{GRGPS}(I\!I\!I) when the GPS is estimated parametrically or nonparametrically with larger bandwidth such that $h/h_1 \rightarrow 0$, the first-order asymptotic property is the same as if the true GPS was observed.
In this case, the influence functions provided in Corollary~\ref{GRGPS}(I) could be useful 
for estimating functionals of the partial means or for improving asymptotic approximation in finite sample.

\begin{Corollary}[Generalized Propensity Score]
Let the conditions in Theorem~\ref{TGR} hold.
The asymptotic variance $\Sigma_{yy}$ and the Gaussian process $\mathbb{G}_t(\cdot;V,\pi)$ are given in Theorem~\ref{TGaussian}.
\begin{itemize}
\item[(I)]  In the stochastic expansion in Theorem~\ref{TGR}, for each $t \in \mathcal{T}^*$,\footnote{In this case of using the GPS as the generated regressor $v_0(X) = f_{T|X}(t|X)$, the expansion implied by Theorem~\ref{TGR} is uniform over $y$, but point-wise in $t$.  
We lose the uniformity in $t$ because now the generated regressor is also indexed by $t$. 
And we do not impose the high-level assumption that the CDF $F_{Y|Tv_0(X)}$ is smooth with respect to the regressor $v_0(\cdot)$.
} 
the influences from estimating the GPS by {\rm Step~1(a) and (b)} are, respectively,
\begin{enumerate}
\item[(a)] {\rm (Kernel)}
\begin{align*}
\Delta_{yt}(\hat v(X))  &=  \frac{1}{n}\sum_{i=1}^n \left( K_{h_1}(T_i - t) - f_{T|X}(t|X_i) \right) 
\Lambda_{yt}(X_i)\pi_i + O_p\left(h_1^{r_1} +  \frac{\log n}{nh_1^{d_x+d_t/2}}  \right). 
\end{align*}

\item[(b)] {\rm (MLE)}
\begin{align*}
\Delta_{yt}(\hat v(X)) =&\ \Big((\hat \beta - \beta_0)', \hat \sigma^2 - \sigma_0^2 \Big) 
\\&
\times \int_{\mathcal{X}^*} \left( x \frac{\partial \zeta(t, \mu, \sigma_0^2)}{\partial \mu} \Big|_{\mu=x'\beta_0},\frac{\partial  \zeta(t, x'\beta_0, \sigma_0^2)}{\partial \sigma^2_0} \right)'  \Lambda_{yt}(x) dF_X(x) + O_p\left(n^{-1}\right).
\end{align*}
\end{enumerate}

\item[(I\!I)]
Consider the nonparametric kernel estimator in {\rm Step~1(a)}  satisfying Assumption~\ref{Abwgps} (ii).  
Then 
for $t \in \mathcal{T}^*$,
$\sqrt{nh} \Big( \hat F_{Y(t)}\big(\cdot;\hat V, \hat \pi \big) - F_{Y(t)}\big(\cdot;V, \pi \big) \Big) \Longrightarrow \mathbb{G}_t(\cdot;X,\pi)$,
a Gaussian process with the covariance kernel $\Sigma_{y_1y_2}(X)$.

\item[(I\!I\!I)]
Consider the two estimators in {\rm Step 1}.
Let the nonparametric kernel estimator {\rm (a)} satisfy $h = o(h_1)$ and Assumption~\ref{Abwgps} (i).
Then 
for $t \in \mathcal{T}^*$ and $y \in \mathcal{Y}$,
$\sqrt{nh} \Big( \hat F_{Y(t)}\big(y;\hat V, \hat \pi \big) - F_{Y(t)}\big(y;V, \pi \big) - \big( h^r \mathfrak{B}_{LC} + h_1^{r_1}\mathfrak{B}_{GPS} \big) \Big) \rightarrow \mathcal{N}(0, \Sigma_{yy}(V))$,
where 
$\mathfrak{B}_{GPS}$ is given in Eq.$\!$ (\ref{EBiasGPS}) in the Appendix for Step~1(a){\rm (Kernel)}
and $\mathfrak{B}_{GPS}=0$ for Step~1(b){\rm (MLE)}.

Assuming $\sqrt{nh} (h^r + h_1^{r_1}) \rightarrow 0$, for $t \in \mathcal{T}^*$,
$\sqrt{nh} \Big( \hat F_{Y(t)}\big(\cdot;\hat V, \hat \pi\big) - F_{Y(t)}(\cdot; V, \pi) \Big) \Longrightarrow \mathbb{G}_t(\cdot;V,\pi)$,
a Gaussian process with the covariance kernel $\Sigma_{y_1y_2}(V)$.

\end{itemize}
\label{GRGPS}
\end{Corollary}

Next we estimate the treatment effects for the treated.
Let the generated regressor $V = v_0(X) = (f_{T|X}(t|X), f_{T|X}(\bar t|X))'$.
Theorem~\ref{CIAGPS} below extends Theorems 2.1 and 3.1 in \cite{HI04}.
\begin{Theorem}
Let Assumption~\ref{ACIA} hold. 
For $s \in \{t, \bar t\} \subseteq \mathcal{T}^\ast$ and for any $x \in \{x: f_{T|X}(t|x)f_{T|X}(\bar t|x) > 0\}$, let $(v, \bar v) \equiv (f_{T|X}(t|x), f_{T|X}(\bar t|x))$.
Then $f_{T|Y(t) V}((s|y, v, \bar v) = f_{T|V}(s|v,\bar v)$ and $f_{Y(t)|TV}(y|s, v, \bar v) = f_{Y(t)|V}(y|v, \bar v)$.
\label{CIAGPS}
\end{Theorem}

By the same arguments in Eq.$\!$ (\ref{IDCIA}), Theorem~\ref{CIAGPS} implies that 
$\mathbb{E}[Y(t)|T=\bar t, V = (v, \bar v)]  = \mathbb{E}[Y(t)|T=t, V = (v, \bar v)] = \mathbb{E}[Y |T=t, V = (v, \bar v)]$
for any $(v, \bar v) \in \mathcal{V}^* \subseteq Supp(V|T=t)\cap  Supp(V|T=\bar t)$.
Then we can identify the conditional mean of $Y(t)$ given $T=\bar t$, defined on $\mathcal{V}^*$,
by the following,
\begin{align*}
\mathbb{E}^*\big[Y(t)\big|T=\bar t\big] &\equiv 
 \int_{\mathcal{V}^*} \mathbb{E}\big[Y(t)\big|T=\bar t, V = (v, \bar v)\big] dF_{V|T}(v,\bar v|\bar t)  \\
&=  \int_{\mathcal{V}^*} \mathbb{E}\big[Y\big|T=t, V = (v, \bar v)\big] dF_{V|T}(v,\bar v|\bar t)  \\
&=\mathbb{E}\big[  \mathbb{E}\big[ Y\big|T=t, V\big] \pi(V) \big|T = \bar t\big],
\end{align*}
where the trimming function $\pi(V) \equiv {\bf 1}\{ V \in \mathcal{V}^\ast\}$.
This is the {\it conditional average structural function} given $T=\bar t$ 
or the {\it average dose response function for the treated} $\bar t$ subpopulation whose control variable takes values in $\mathcal{V}^*$.
Therefore we can identify the corresponding CDF, $F^\ast_{Y(t)|T}(y|\bar t) \equiv \mathbb{E}^*\big[{\bf 1}\{Y(t) \leq y\}\big|T=\bar t\big] 
= F_{Y(t)}(y; V, W\pi)$, where $W = f_{T|X}(\bar t|X)/f_T(\bar t)$.
We propose two partial mean estimators following the procedure in Section~\ref{SPO} with an estimated weight: 
\begin{enumerate}
\item[(a)] (Kernel)\ \ $\hat V_i = (\hat f_{T|X}(t|X_i), \hat f_{T|X}(\bar t|X_i))'$ is the standard conditional kernel density estimator, using a $r_1$-order kernel function with bandwidth $h$, satisfying Assumption~\ref{AkernelEJL}.
\begin{align*}
&\hat F_{Y(t)}(y; \hat V, \hat W_1\hat \pi) = \frac{1}{n}\sum_{i=1}^n \hat F_{Y|T\hat V}(y|t,\hat V_i )  \hat W_{1i} \hat \pi(\hat V_i), \mbox{\ where\ } \hat W_{1i} = \frac{K_h(T-\bar t)}{ n^{-1}\sum_{i=1}^n K_h(T_i - \bar t)}.
\end{align*}

\item[(b)] (MLE)\ \ 
Assume the conditional distribution of $T$ given $X$ is normal $\mathcal{N}(X'\beta_0, \sigma_0^2)$.
$\hat V_i = (\hat f_{T|X}(t|X_i), \hat f_{T|X}(\bar t|X_i))'$ is the probit MLE and
\begin{align*}
&\hat F_{Y(t)}(y; \hat V, \hat W_2\hat \pi) = \frac{1}{n}\sum_{i=1}^n \hat F_{Y|T\hat V}(y|t,\hat V_i )  \hat W_{2i} \hat \pi(\hat V_i), \mbox{\ where\ } \hat W_{2i} = \frac{\hat f_{T|X}(\bar t|X_i)}{n^{-1}\sum_{i=1}^n \hat f_{T|X}(\bar t|X_i)}.
\end{align*}
\end{enumerate}

\begin{Theorem}[Treatment effects for the treated]
Suppose the conditions in Corollary~\ref{GRGPS} hold.  
Then for each $t \in \mathcal{T}^\ast$, uniformly in $y \in \mathcal{Y}$, 
\begin{itemize}
\item[(a)] {\rm (Kernel)}
\begin{align*}
&\sqrt{nh}\big(\hat F_{Y(t)}(y; \hat V, \hat W_1\hat \pi) - F^\ast_{Y(t)|T}(y|\bar t)\big) 
\\
=\ & \sqrt{nh}\big(\hat F_{Y(t)}(y; X, \hat W_1\hat \pi) - F^\ast_{Y(t)|T}(y|\bar t)\big) + o_p(1)\\
=\ & \sqrt{\frac{h}{n}}\sum_{i=1}^n  \frac{K_h(T_i - t)}{f_{T|X}(t|X_i)}\big(  {\bf 1}{\{Y_i \leq y\}} - F_{Y|TX}(y|t, X_i) \big) \frac{f_{T|X}(\bar t|X_i)}{f_T(\bar t)} \\
&+ \frac{K_{h}(T_i - \bar t)}{f_T(\bar t)}  \big( F_{Y|TX}(y|t, X_i) - F_{Y(t)|T}(y|\bar t)\big)
   + o_p(1).  
\end{align*}

\item[(b)] {\rm (MLE)}
\begin{align*}
&\sqrt{nh}\big(\hat F_{Y(t)}(y; \hat V, \hat W_2\hat \pi) - F^\ast_{Y(t)|T}(y|\bar t)\big) 
\\&
= \sqrt{\frac{h}{n}} \sum_{i=1}^n
\frac{K_h(T_i - t)}{f_{T|X}(t|X_i)}\big(  {\bf 1}{\{Y_i \leq y\}} - F_{Y|TV}(y|t, V_i) \big) \frac{f_{T|X}(\bar t|X_i)}{f_T(\bar t)}.
\end{align*}
\end{itemize}
Asymptotic normality follows.
\label{TTT}
\end{Theorem}

Corollary~\ref{GRGPS} and Theorem~\ref{TTT} suggest that regression on the nonparametrically estimated GPS is first-order asymptotically equivalent to regressing on $X$, using the same bandwidth and kernel function for both  the Step 1 and Step 2 estimators.
So there is no efficiency gain in using the GPS.
These results parallel the findings for the binary treatment in \cite*{HIT98RES}, \cite{HR13ETA},  \cite*{MRS12B}, and the multivalued treatment effects for the treated in \cite{Lee14DTT}.
This is because the whole set of observables $X$ provides finer conditioning variables than its index $f_{T|X}(t|X)$ or $f_{T|X}(\bar t|X)$.  
When the index bias is not zero, i.e., there exists $x$ such that $F_{Y|TV}(y|t,v_{0}(x)) \neq F_{Y|TX}(y|t,x)$, the inequality between the corresponding asymptotic variances is strict.
Lemma~\ref{LvarGPS} in the Appendix provides a formal and general result comparing expectations of the conditional variances given the whole set of observables $X$ and given the index $v_0(X)$, respectively.

\subsection{Control variables in the triangular models}
\label{SecCV}
When unconfoundedness is violated, one approach to satisfy Assumption~\ref{ACIA} is through control variables as in the triangular simultaneous equations model.
For example, \cite*{IN09ETA} show the conditional distribution function of the endogenous variable given the instrumental variables is a valid control variable $V = F_{T|Z}(T|Z)$, where $Z$ is an exogenous instrumental variable.
Let the bounded support of $Y$ be $[y_l, y_u]$ and the trimmed proportion $P^* \equiv {\rm Pr}(V \notin \mathcal{V}^*)$.
\cite{IN09ETA} identify the bounds for the average structural function $\mathbb{E}[Y(t)]$ by
$\mathbb{E}^*[Y(t)] +\  y_l  P^* \leq \mathbb{E}[Y(t)] \leq \mathbb{E}^*[Y(t)] +\ y_u P^*.$
For a smaller trimming threshold $c$, $P^*$ is smaller and hence the identified set is smaller.
By replacing the dependent variable $Y(t)$ with ${\bf 1}{\{Y(t) \leq y\}}$, the CDF of the outcome at a hypothetical value $t$ is bounded by the following
\begin{align*}
&{F^*_{Y(t)}}(y) \leq {F_{Y(t)}}(y) \leq {F^*_{Y(t)}}(y) + P^*, \mbox{where}\  
{F^*_{Y(t)}}(y) \equiv \int_{\mathcal{V}^*} \mathbb{E}[{\bf 1}{\{Y \leq y\}}|T=t, V=v] \ dF_V(v)
\end{align*}
as defined in Eq. $\!$(\ref{PMGRest}).
It follows that the lower bound of the $\tau$-quantile structural function is $y_l$ if $\tau \leq P^*$ and is ${F^{*-1}_{Y(t)}}(\tau - P^*)$ if $\tau > P^*$. 
The upper bound of the $\tau$-quantile structural function is ${F^{*-1}_{Y(t)}}(\tau)$ if $\tau < 1-P^*$ and is $y_u$ if $\tau \geq 1-P^*$. 
The partial mean estimator  for the bounds of the average and quantile structural functions is described in Section~\ref{SPO} with $W=1$.
And $\hat P^* = 1-n^{-1}\sum_{i=1}^n \hat \pi(\hat V_i)$.

We first discuss what we learn from the three elements of the stochastic expansion in Theorem~\ref{TGR}.
The exclusion assumption of the instrumental variable implies (ii) the index bias is zero $F_{Y|TZ}(y|t,Z) = F_{Y|TV}(y|t,v_0(t, Z))$.
So Theorem~\ref{TGR} implies the influence of the generated regressors for the {\it regressor} role is simplified to 
$REG_{yt}(z,v) = - \frac{\partial}{\partial v} F_{Y|TV}(y|t, v) \frac{f_{T|Z}(t|z)}{ f_{T|V}(t| v)}$. 
The integral operator in $\Delta_{yt}(\hat v)$ gives (i) a full mean structure, where the regressor $Z$ is integrated out and $T$ is in the dependent variable.
It implies that the influence of the nonparametrically estimated control variable converges at a root-$n$ rate.
Our stochastic expansion characterizes the terms that are asymptotically negligible for the partial mean estimator and could be used to improve finite-sample inference.
Moreover, it is the building block when the average structural function is an intermediate ingredient of a more complicated estimation procedure, such as \cite{VV13}, \cite{BL14}, and \cite{HR16}.
In such cases, the final estimator are functionals of the partial mean and the influence of the estimated control variable might not be asymptotic ignorable.

In an earlier working paper of \cite{Lee}, we derive a complete inference theory that accounts for the estimation error of the control variables.  
To conserve space in this paper, we present the detailed results in a separate paper. 
In particular, we consider three estimators of the control variable in Step~1.
(I) a local constant estimator of the CDF $F_{T|Z}(T|Z)$ in \cite{IN09ETA} and the residual $T-\mathbb{E}[T|Z]$ in \cite{NPV99ETA}, where $\mathbb{E}[T|Z]$ is estimated by (I\!I) a local constant estimator and (I\!I\!I) an ordinary least squares estimator.  
Our results contribute to the literature on inference in triangular models, for example, \cite*{NPV99ETA}, \cite*{SuUllah},  \cite{IN09ETA}, \cite*{MRS12A}, \cite*{Stouli}, among others.\footnote{
\cite{IN09ETA} obtain a convergence rate of a power series estimator for the average structural function.
\cite{NPV99ETA} use a two-step series estimator to exploit their additive structural function and characterize the estimation error of the control variable in the asymptotic variance.
For estimating the triangular model in \cite{NPV99ETA}, \cite{SuUllah} and \cite{MRS12A} control the sampling variation of the reduced form residual to be first-order ignorable by choosing the tuning parameters. 
In contrast, the full mean structure of our stochastic expansion implies the sampling variation from Step 1 nonparametric estimation is of order $n^{-1/2}$
that is first-order ignorable without further restricting the tuning parameters.
\cite{Stouli} estimate semiparametric nonseparable triangular models. }

\section{Inference for the Treatment Effects}
\label{SecInf}
Often the objects of ultimate interest are policy effects or inequality measures. 
Such objects can be expressed as functionals of the potential outcome distributions identified by the partial mean $F_{Y(t)}(y;V, W\pi)$ and estimated in previous sections.
The key to the distribution theory for a class of smooth functionals is the functional delta method for Hadamard-differentiable functionals.
The results are illustrated by the mean and quantile operators. 
In this section, we let $\theta_t(y) = F_{Y(t)}(y;V, W\pi)$ by suppressing $(V, W\pi)$ in the notation for brevity.
The corresponding asymptotic theorem derived in previous sections provides the influence function and weak convergence: denoting as $\sqrt{nh^{d_t}}\big( \hat \theta_t - \theta_t \big) = n^{-1/2} \sum_{i=1}^n \psi_{tin} + o_p(1)$ and converges weakly to a Gaussian process $\mathbb{G}_t$.

\begin{Assumption}[Hadamard]
The functional $\Gamma$: $\mathcal{F}_\Gamma \mapsto \mathbb{R}$ defined over the marginal distribution function of the potential outcome $Y(t)$ is Hadamard differentiable at $F_{Y(t)}(; V, W\pi)$ tangentially to the subset $\mathcal{F}_{\Gamma;V, W\pi}$.\footnote{See, for example, \cite*{V00} for definition:
let $\Gamma$ be a Hadamard-differentiable functional mapping from $\mathcal{F}$ to some normed space $\mathcal{E}$, with derivative $\Gamma'_f$, a continuous linear map $\mathcal{F} \mapsto \mathcal{E}$. 
For every $h_n \rightarrow h$ and $f \in \mathcal{F}$,
$\lim_{u \rightarrow 0} \big(\Gamma(f + uh_n) - \Gamma(f) \big)/u = \Gamma'_f(h)$.
}
\label{Adef}
\end{Assumption}
These Hadamard-differentiable functionals can be highly nonlinear functionals of the CDF, but admit a linear functional derivative.
Weak convergence of the estimators will be implied by the functional delta method in empirical process theory.
Assumption~\ref{Adef} is a high-level assumption that could impose restrictions or smoothness on the distribution functions of potential outcomes.
In particular, when $\Gamma$ is the $\tau$-quantile operator on $\theta_t(y) = F_{Y(t)}(y;V, W\pi)$, $\Gamma$ is a generalized inverse $\theta_t^{-1}: (0,1) \rightarrow \mathcal{Y}$ given by $\theta_t^{-1}(\tau)= \inf\{y: \theta_t(y) \geq \tau\}$.
Then Assumption~\ref{Adef} means $\theta_t(y)$ is continuously differentiable at the $\tau$-quantile, with the derivative being strictly positive and bounded over a compact neighborhood.
Additional assumptions might be needed for different policy functionals.
For example, \cite*{B07JoE} gives regularity conditions for Hadamard-differentiability of Lorenz and Gini functionals. 


\begin{Theorem}
Assume the conditions in the asymptotic theorem for $\hat \theta_t$ hold.
Consider the parameter $\theta$ as an element of a parameter space $D_\theta \subset l^\infty(\mathcal{Y})$ with $D_\theta$ containing the true value $\theta_t$. 
Suppose a functional $\Gamma(\theta)$ mapping $D_\theta$ to $l^\infty(\mathcal{U})$ is Hadamard differentiable in $\theta$ at $\theta_t$ with derivative $\Gamma'_\theta$
and Assumption~\ref{Adef} holds.
Then 
\begin{enumerate}
\item (Functional Delta Method)
\begin{align*}
\Big| &\sqrt{nh^{d_t}}\big( \Gamma(\hat \theta_t)(u) - \Gamma(\theta_t)(u) \big) 
- \frac{1}{\sqrt{n}} \sum_{i=1}^n \Gamma_\theta'(\psi_{tin})(u) \Big| = o_p(1)\\
&\sqrt{nh^{d_t}}\big( \Gamma(\hat \theta_t)(u) - \Gamma(\theta_t)(u) \big) \Longrightarrow \Gamma_\theta'(\mathbb{G}_t)(u) \equiv G(u),
\end{align*}
where $G$ is a Gaussian process indexed by $u\in \mathcal{U}$ in $l^\infty(\mathcal{U})$, with mean zero and covariance kernel defined by the limit of the second moment of $\Gamma_\theta'(\psi_{tin})$.

\item (Causal effects)
\begin{align*}
\sqrt{nh^{d_t}} \left( 
\begin{array}{c}
\hat \theta_t(\cdot) - \theta_t(\cdot)\\
\hat \theta_{\bar t}(\cdot) -\theta_{\bar t}(\cdot)
\end{array}
 \right) 
= \frac{1}{\sqrt n}\sum_{i=1}^n 
\left( \begin{array}{c}
\psi_{tin}(\cdot) \\
\psi_{\bar tin}(\cdot)
\end{array} \right) + o_p(1)
 \Longrightarrow \mathbb{G}_{t\bar t}(\cdot)
\end{align*}
that is a Gaussian process with zero mean.
The diagonal elements of the covariance matrix are the covariance matrix of $\mathbb{G}_t$ and $\mathbb{G}_{\bar t}$.
And the off-diagonal terms are zero.  
Theorem~\ref{TFDM} implies
\begin{align*}
\sqrt{nh^{d_t}} \Big( \Gamma(\hat \theta_t) -  \Gamma(\hat \theta_{\bar t}) - \big( \Gamma( \theta_t) - \Gamma(\theta_{\bar t}) \big) \Big) 
= \frac{1}{\sqrt{n}} \sum_{i=1}^n \Big( \Gamma'_\theta(\psi_{tin}) - \Gamma'_\theta(\psi_{\bar tin}) \Big) + o_p(1) 
&\Longrightarrow \mathbb{G}^\Gamma_{t\bar t}
\end{align*}
that is a mean-zero Gaussian process whose covariance kernel is the summation of the covariance of $\Gamma'_\theta(\mathbb{G}_t)$ and the covariance of $\Gamma'_\theta(\mathbb{G}_{\bar t})$.
\end{enumerate}
\label{TFDM}
\end{Theorem}
%

The above result gives the policy/inequality treatment effects of shifting the treatment from $\bar t$ to $t$, $\Gamma\big(\theta_t\big) - \Gamma\big(\theta_{\bar t}\big)$.
The estimators of the distributional features at different treatment levels $t$ and $\bar t$, $\Gamma\big(\theta_t\big)$ and $ \Gamma\big(\theta_{\bar t}\big)$, are asymptotically uncorrelated.

\subsection{Mean and quantile}
\label{SHDF}
The mean for the CDF $\theta_t$ is $\Gamma(\theta_t) = \int_{\mathcal Y} u d\theta_t(u)$, which has the Hadamard derivative $\Gamma'(\theta) = \int u d\theta(u)$.
Then the estimator is $\int y\ d\hat \theta_t(y)$ by replacing the dependent variable ${\bf 1}{\{Y \leq y\}}$ with $Y$ in the estimation procedure described in Section~\ref{SPO}.
Alternatively, we can use the transformed outcome in Remark \ref{RVV}.
Theorems \ref{TGaussian} and \ref{TGR} provide the asymptotic theory of estimating the mean $\mathbb{E}^*[Y(t)]$ by simply replacing $F_{Y|TV}(y|t, V_i)$ with $\mathbb{E}[Y|T=t, V= V_i]$ in the influence function $\psi_{tin}(y;V, W\pi)$, $ARG$, and $REG$ given in Theorem~\ref{TGR}.

The unconditional quantile function is inverted directly from the unconditional CDF.
For the quantile process $\{Q_\tau: \tau \in (0,1)\}$ of the CDF $\theta_t$, $Q_\tau \equiv \inf \{y : \theta_t(y) \geq \tau\}$.
The following corollary gives the asymptotic theory of estimating 
unconditional quantile function of $Y(t)$ for the whole population assuming unconfoundedness and using control variables.

\begin{Corollary}[Quantile Process]
Suppose the conditions in Theorem~\ref{TFDM} hold.
Suppose $\sqrt{nh^{d_t}}\big( \hat \theta_t(\cdot) - \theta_t(\cdot) \big)  = n^{-1/2} \sum_{i=1}^n \psi_{tin}(\cdot) + o_p(1) \Longrightarrow \mathbb{G}_t(\cdot)$.
Assume $\theta_t$ is continuously differentiable with strictly positive derivative $\nabla_y \theta_t(y)\big|_{y=Q_\tau} \equiv \theta_t'(Q_\tau)$.
Then the influence function for estimating the quantile process is 
$\psi^Q_{tin}(\tau)\equiv -\psi_{tin}\big(Q_\tau \big) \big/  \theta_t'(Q_\tau)$. 
Therefore, 
\begin{align*}
\sqrt{nh^{d_t}}\big( \hat Q_\cdot - Q_\cdot\ \big) 
= \frac{1}{\sqrt{n}} \sum_{i=1}^n \psi_{tin}^Q(\cdot)  + o_p(1) \Longrightarrow - \mathbb{G}_t \big(Q_\cdot \big) \Big/  \theta_t'(Q_\cdot) \equiv \mathbb{G}^Q_t(\cdot),
\end{align*}
where $\mathbb{G}^Q_t$ is a Gaussian process indexed by $\tau \in [a,b] \subset (0,1)$ in the metric space $l^\infty([a,b])$.  
The Gaussian process $\mathbb{G}_t^Q$ has zero mean and covariance kernel, for any $\tau_1 < \tau_2 \in [a,b]$, 
$Cov\big(\mathbb{G}^Q_t(\tau_1), \mathbb{G}^Q_t(\tau_2)\big) = \lim_{n \rightarrow \infty} \mathbb{E}\big[ \psi^Q_{tin}(\tau_1) \psi^Q_{tin}(\tau_2) \big]$.
\label{CQP}
\end{Corollary}


\subsection{Inference}
\label{SecMCLT}
The asymptotic theorems in the previous sections can be used to calculate point-wise confidence intervals. 
The influence function can be estimated by replacing unknown functions with consistent estimators.
Then the covariance matrix can be estimated by the sample variance of the estimated influence functions.
Our expansions in Theorem~\ref{TGR} characterize the second-order terms, which could be included in the estimated influence function to improve asymptotic approximation.  
Alternatively, the covariance matrix can be estimated by a plug-in method that is a sample analogue with consistently estimated unknown functions.  
The procedure is standard and omitted for brevity.

Bandwidth selection is essential to implement the nonparametric estimation procedure.
\cite*{SuUllah} propose a plug-in method by minimizing the asymptotic integrated mean squared error.  
The stochastic expansion in \cite*{EJL} is uniform over the bandwidth which allows the use of a data-driven bandwidth choice procedure.  
Although a bandwidth selection procedure is beyond the scope of this paper, a rule of thumb method satisfying the bandwidth Assumptions in this paper can be implemented in practice and our empirical application in Section~\ref{SecEm}.

Besides point-wise inference, we might be interested in testing a hypothesis involving a policy on the whole distribution: constant effect or stochastic dominance.
We suggest using a multiplier method to simulate the empirical processes defined in Theorem~\ref{TGaussian}.
The multiplier method has been used in \cite*{DHB} and \cite*{DH14JoE} to simulate a distribution process.
It is easy to perform asymptotically valid inference on distributional features defined by the Hadamard-differentiable functionals. 
Let $\{U_i\}_{i=1}^n$ be a sequence of $i.i.d.$ random variables with mean zero and variance one, for example,  $\mathcal{N}(0,1)$, independent of the data.
We assume that  there exists a uniformly consistent estimator  $\hat \psi_{tin}$ of the influence function $\psi_{tin}$  that is monotone in $y$ by using second-order kernel or a monotone transformation.\footnote{
This low-level assumption on $\hat \psi_{tin}$ is sufficient for the general high-level assumptions for the validity of the multiplier bootstrap in \cite{HsuMB}, for example.
}
The following theorem shows that $n^{-1/2} \sum_{i=1}^n U_i \hat \psi_{tin}(\cdot)$ simulates the asymptotic distribution of the estimator.

\begin{Theorem}[Multiplier method]
Assume the conditions in Theorem~\ref{TGaussian} that gives 
$ \sqrt{nh^{d_t}}\big( \hat \theta_t(\cdot) - \theta_t(\cdot) \big)  = n^{-1/2} \sum_{i=1}^n \psi_{tin}(\cdot)  + o_p(1) \Longrightarrow \mathbb{G}_t(\cdot)$.
Then for any $t \in \mathcal{T}^*$,
$ \mathbb{G}^M_{tin}(\cdot) \equiv n^{-1/2} \sum_{i=1}^n U_i \hat \psi_{tin}(\cdot) \Longrightarrow \mathbb{G}_t(\cdot)$,
conditional on sample path $\{(Y_i, T_i, X_i, Z_i, W_i): 1 \leq i \leq n\}$ w.p.a.1.  
For the Hadamard-differentiable functional $\Gamma$, 
$\Gamma'\Big( \mathbb{G}^M_{tin}(\cdot)  \Big) = n^{-1/2} \sum_{i=1}^n U_i \Gamma'\big( \hat \psi_{tin}(\cdot) \big)\Longrightarrow \Gamma'\big( \mathbb{G}_t(\cdot)\big)$.
\label{TMCLT}
\end{Theorem}

\section{Empirical application}
\label{SecEm}
We illustrate our method by evaluating the Job Corps program in the United States conducted in the mid-1990s.
The largest publicly founded job training program targets disadvantaged youth.
The participants are exposed to different numbers of actual hours of academic and vocational training.
The participants' labor market outcomes may differ if they accumulate different amounts of human capital acquired through different lengths of exposure.
We estimate the average/quantile dose response functions to investigate the relationship between employment and the length of exposure to academic and vocational training.
We also estimate the dose response functions for the treated~$\bar t$, i.e., for the participants who have received $\bar t$ hours of training.
Therefore, we are able to estimate the quantile treatment effect for the treated and contribute to the literature that have been focusing on the binary treatment effects of Job Corps by employing a binary indicator of participation and
the average continuous treatment effects for the population.
As our analysis builds on \cite*{FFGN12ReStat} and \cite*{HHLP}, we refer the readers to the reference therein for the detail of Job Corps.


\paragraph{Data}
We use the same dataset in \cite*{HHLP}.
We consider the outcome variable ($Y$) to be the proportion of weeks employed in the second year following the program assignment.
The continuous treatment variable ($T$) is the total hours spent in academic and vocational training in the first year.
The Conditional Independence Assumption~\ref{ACIA} means that selection into different levels of the treatment is random, conditional on a rich set of observed covariates, denoted by $X$.
The identifying Assumption~\ref{ACIA} is indirectly assessed in \cite*{FFGN12ReStat}.
In addition to socio-economic characteristics at base line, 
$X$ includes expectations about Job Corps and interaction with the recruiters that are predictive for the duration in Job Corps.
Our sample consists of 4,024 individuals who completed at least 40 hours (one week) of academic and vocational training.
Table~\ref{Tss} provides descriptive statistics.

 \begin{table}[!htbp] \centering 
  \caption{Descriptive statistics} 
  \label{Tss} 
\resizebox{\textwidth}{!}{%
\begin{tabular}{@{\extracolsep{5pt}}lcccccc} 
\\[-1.8ex]\hline 
\hline \\[-1.8ex] 
Variable  & \multicolumn{1}{c}{Mean} & \multicolumn{1}{c}{Median} & \multicolumn{1}{c}{StdDev} & \multicolumn{1}{c}{Min} & \multicolumn{1}{c}{Max} & \multicolumn{1}{c}{Non-missing} \\ 
\hline \\[-1.8ex] 
  share of weeks employed in 2nd year ($Y$)                     &44  &40.38  &37.89  &0  &100  &4024    \\
   total hours spent in 1st-year training ($T$)                    &1219.8  &992.86  &961.74  &40  &6188.57  &4024    \\
 female              &0.43  &0  &0.5  &0  &1  &4024    \\
 age                 &18.33  &18  &2.15  &16  &24  &4024    \\
 white          &0.25  &0  &0.43  &0  &1  &4024    \\
 black          &0.5  &0.5  &0.5  &0  &1  &4024    \\
 hispanic  &0.17  &0  &0.38  &0  &1  &4024    \\
   years of education                &10.05  &10  &1.54  &0  &20  &3965    \\
   GED diploma &0.04  &0  &0.2  &0  &1  &4004    \\
  high school diploma  &0.18  &0  &0.39  &0  &1  &4004    \\
  native English           &0.85  &1  &0.35  &0  &1  &3970    \\
divorced   &0.01  &0  &0.09  &0  &1  &3973    \\
separated  &0.01  &0  &0.11  &0  &1  &3973    \\
cohabiting  &0.03  &0  &0.18  &0  &1  &3973    \\
married  &0.02  &0  &0.13  &0  &1  &3973    \\
had children            &0.18  &0  &0.38  &0  &1  &4004    \\
ever workded             &0.4  &0  &0.49  &0  &1  &1415    \\
average weekly gross earnings               &19.72  &0  &99.07  &0  &2000  &4023    \\
    is household head                  &0.1  &0  &0.31  &0  &1  &3952    \\
household size  &3.54  &3  &2.01  &0  &15  &3963    \\
designated for nonresidential slot              &0.17  &0  &0.38  &0  &1  &4024    \\
  dad did not work when 14   &0.06  &0  &0.23  &0  &1  &3604    \\
  received food stamps                  &0.45  &0  &0.5  &0  &1  &3856    \\
  welfare receipt during childhood   &2.06  &2  &1.19  &1  &4  &3744    \\
  poor/fair general health status &0.13  &0  &0.33  &0  &1  &3973    \\
   physical/emotional problems                  &0.04  &0  &0.2  &0  &1  &3970    \\
extent of marijuana use                 &2.53  &3  &1.55  &0  &4  &1480    \\
  extent of smoking                  &1.53  &1  &0.98  &0  &4  &2085    \\
   extent of alcohol consumption                  &3.14  &4  &1.21  &0  &4  &2310    \\
ever arrested                  &0.25  &0  &0.43  &0  &1  &3970    \\
number of times in prison                 &0.07  &0  &0.35  &0  &5  &4024    \\
  time by recruiter speaking of Job Corps                 &2.07  &2  &0.95  &1  &4  &3941    \\
 extent of recruiter support                 &1.59  &1  &1.06  &1  &5  &3931    \\
  idea about wished training                 &0.85  &1  &0.35  &0  &1  &3964    \\
  expected improvement in maths                 &1.32  &1  &0.53  &1  &3  &3937    \\
   expected improvement in reading skills                 &1.53  &1  &0.65  &1  &3  &3953    \\
   expected improvement in social skills                 &1.48  &1  &0.68  &1  &3  &3951    \\
  expected to be training for a job                 &1.04  &1  &0.23  &1  &3  &3942    \\
  worried about Job Corps                 &0.37  &0  &0.48  &0  &1  &3964    \\
  1st contact with recruiter by           &0.41  &0  &0.49  &0  &1  &3973    \\
  expected stay in Job Corps                &6.64  &0  &9.78  &0  &36  &4024    \\
\hline \\[-1.8ex] 
\end{tabular} 
}\begin{tablenotes}
      \footnotesize
      \item Notes: Summary statistics for 4,024 individuals who completed at least 40 hours of academic and vocational training.
The set of covariates includes 40 variables measured at the baseline survey and missing dummies to account for non-response in discrete variables.  
    \end{tablenotes}
\end{table}

\paragraph{Calculations}
To calculate the average and quantile dose response functions (DRF) for the population and for the treated $\bar t$, we follow the semiparametric method described in Section~\ref{SecEx}. 
The details are as follows.
Construct grid points of hours in training $\mathcal{T}^\dagger \equiv \{80, 120,..., 2000\}$ and consider $t, \bar t \in \mathcal{T}^\dagger$.
\begin{changemargin}{0.6cm}{0.3cm} 
\begin{enumerate}
\item[Step 1.](GPS-MLE)\ 
Assume the conditional distribution of $T$ given $X$ to be $Lognormal(X'\beta_0, \sigma_0^2)$.
The probit maximum likelihood estimator of the generalized propensity score is
$\hat f_{T|X}(t|X_i) = \exp\big({-(\log(t)-X_i'\hat \beta)^2/(2\hat \sigma^2})\big)\big/\big(t \sqrt{2\pi\hat \sigma^2} \big)$. 

\item[Step 2.](Regression) \ 
For the DRF for the population, $\hat V_i = \hat f_{T|X}(t|X_i)$.
For the DRF for the treated, $\hat V_i = \big(\hat f_{T|X}(t|X_i), \hat f_{T|X}(\bar t|X_i)\big)^\prime$.
$\hat{\mathbb{E}}[Y|T=t, \hat V = v]$ is a local linear estimate of the regression function of $Y$  on $(T, \hat V)$, evaluated at $(t, v)$, 
with bandwidth $h = C\times StdDev(Z) \times n^{-0.22}$ for $Z = T, \hat V$, respectively.  
We choose $C=5$ and vary the constant C over a range from 2 to 8 for robustness check. 

\item[Step 3.](Partial mean)\ \ To select the common support, we trim the observation $i$ if there exists $t \in \mathcal{T}^\dagger$ such that $\hat f_{T|X}(t|X_i)$ is smaller than its 2.5\%-quantile, as described in footnote~\ref{ftrim}.
The trimmed sample has $3,822$ observations that is 95\% of our original sample.
\begin{enumerate}
\item Average DRF for the population: $\hat{\mathbb{E}}[Y(t)] = n^{-1}\sum_{i=1}^n \hat{\mathbb{E}}[Y|T=t, \hat V = \hat V_i]$, where $\hat V_i = \hat f_{T|X}(t|X_i)$. 

\item Average DRF for the treated $\bar t$: $\hat{\mathbb{E}}[Y(t)|T=\bar t] = n^{-1}\sum_{i=1}^n \hat{\mathbb{E}}[Y|T=t, \hat V = \hat V_i]\hat W_i$, where $\hat W_i = \hat f_{T|X}(\bar t|X_i)\big/ \big( n^{-1}\sum_{i=1}^n \hat f_{T|X}(\bar t|X_i) \big)$ and $\hat V_i = \big(\hat f_{T|X}(t|X_i), \hat f_{T|X}(\bar t|X_i)\big)^\prime$.
\end{enumerate}

\item[Step 4.] (Confidence interval)\ 
The $95\%$ point-wise confidence intervals are obtained by the covariance matrix derived in Corollary~\ref{GRGPS}, Theorem~\ref{TTT}, and Theorem~\ref{TFDM}
and estimated by the sample variance of the estimated influence function.

\end{enumerate}
\end{changemargin}

To estimate the $\tau$-quantile  DRF, replace the dependent variable with ${\bf 1}{\{Y \leq y\}}$ in the above procedure and solve the inverse function of the CDF estimate.
Specifically, the $\tau$-quantile DRF for the population is estimated by $\hat Q_\tau(Y(t)) = \hat F^{-1}_{Y(t)}(\tau)$, where $\hat F_{Y(t)}(y) = n^{-1}\sum_{i=1}^n \hat{\mathbb{E}}[{\bf 1}{\{Y \leq y\}}|T=t, \hat V = \hat V_i]$ in Step~3(a). 
The $\tau$-quantile DRF for the treated $\bar t$ is estimated by $\hat Q_\tau(Y(t)|T=\bar t) = \hat F^{-1}_{Y(t)|T}(\tau|\bar t)$, where $\hat F_{Y(t)|T}(y|\bar t) = n^{-1}\sum_{i=1}^n \hat{\mathbb{E}}[{\bf 1}{\{Y \leq y\}}|T=t, \hat V = \hat V_i] \hat W_i$ in Step~3(b).
In Step~4, the derivative $f_{Y(t)}(y)$ in the influence function derived in Corollary~\ref{CQP} is estimated using the same procedure for $\hat F_{Y(t)}(y)$ by changing the Step~2 local linear regression to a kernel conditional density estimation $\hat f_{Y|T\hat V}(y|t, v)$.

\begin{figure}[!htp]
\centering

\caption{Average dose response functions
\label{FADRF}}
\includegraphics[width=0.32\textwidth]{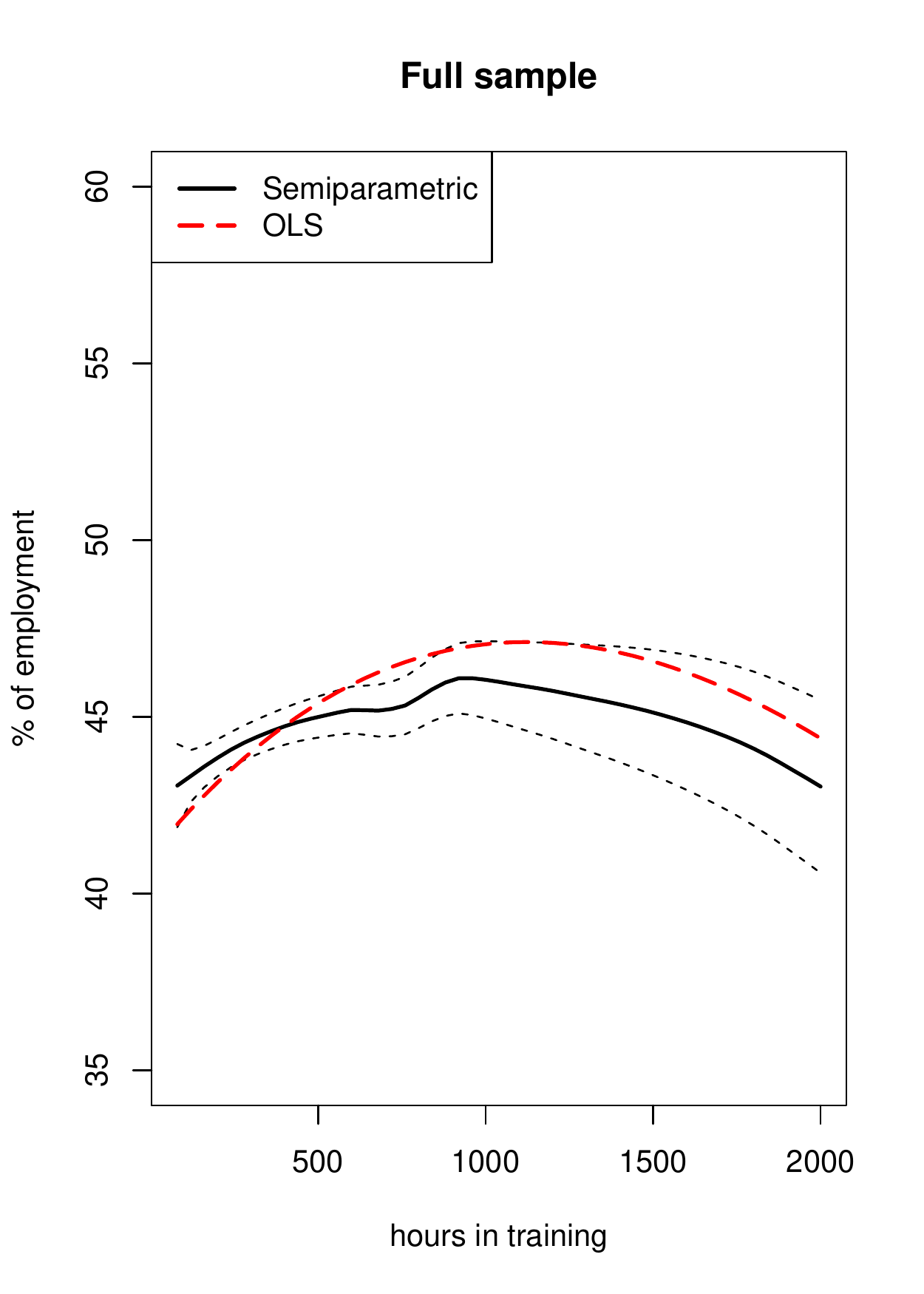}
\includegraphics[width=0.32\textwidth]{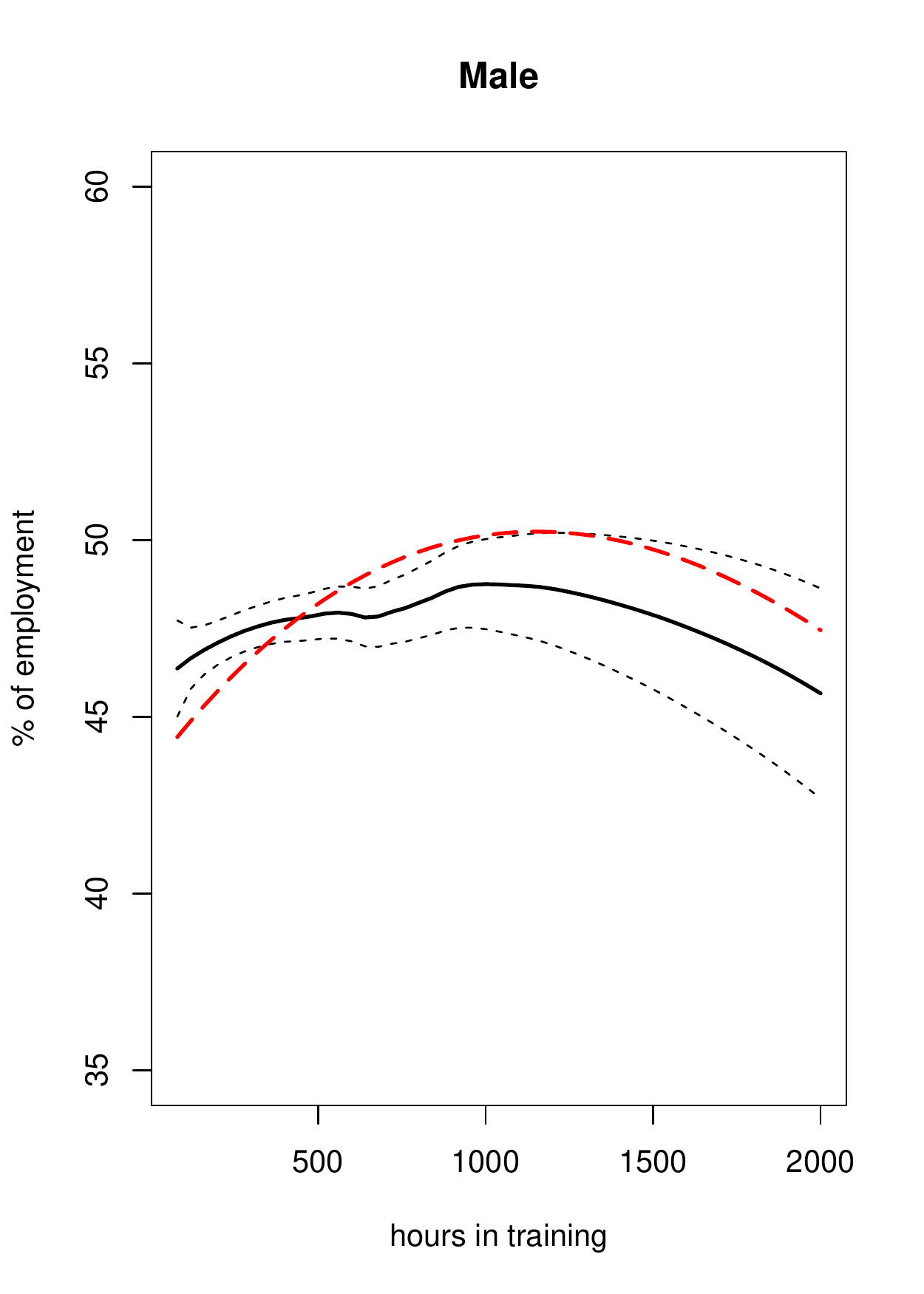}
\includegraphics[width=0.32\textwidth]{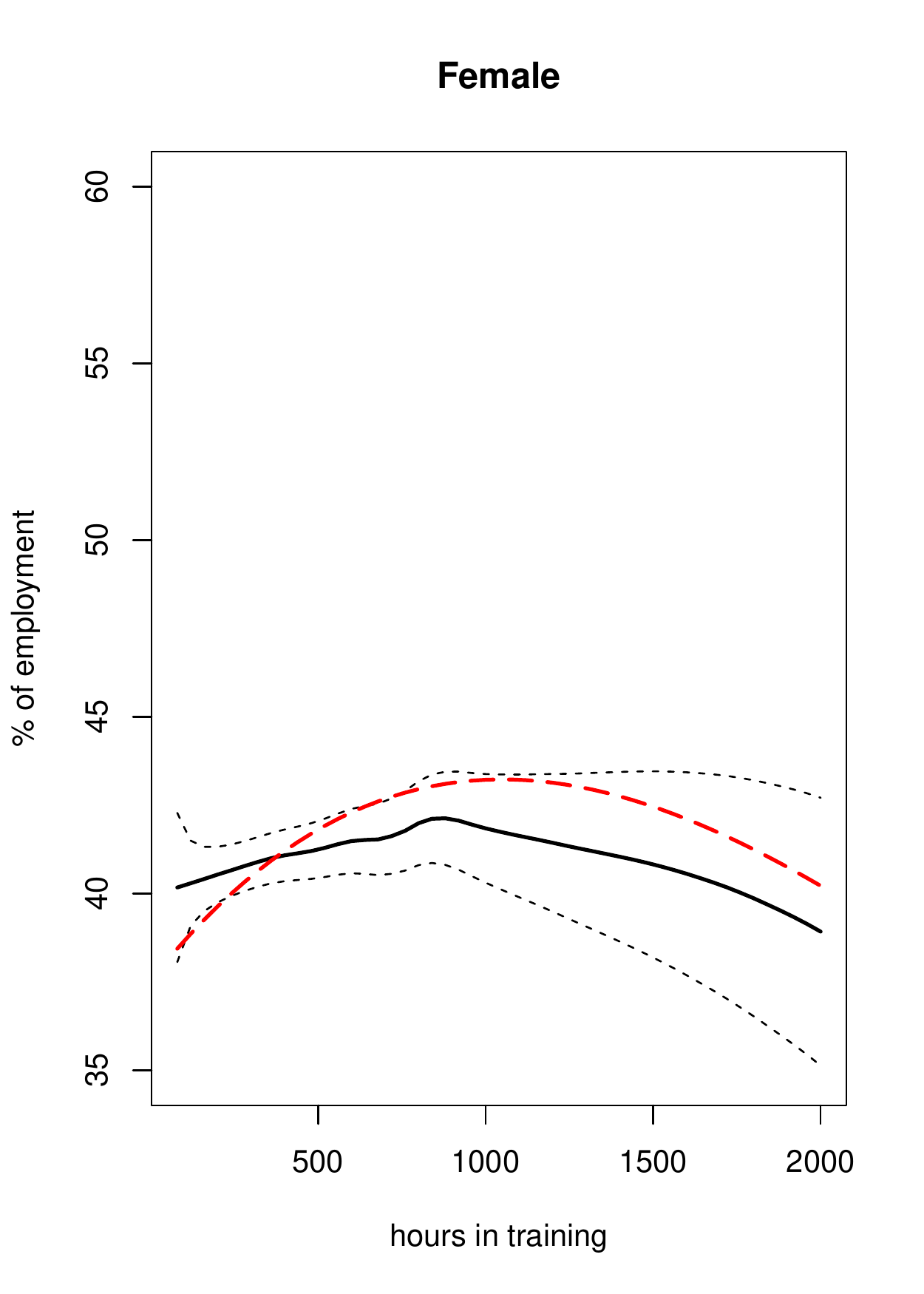}
\includegraphics[width=0.32\textwidth]{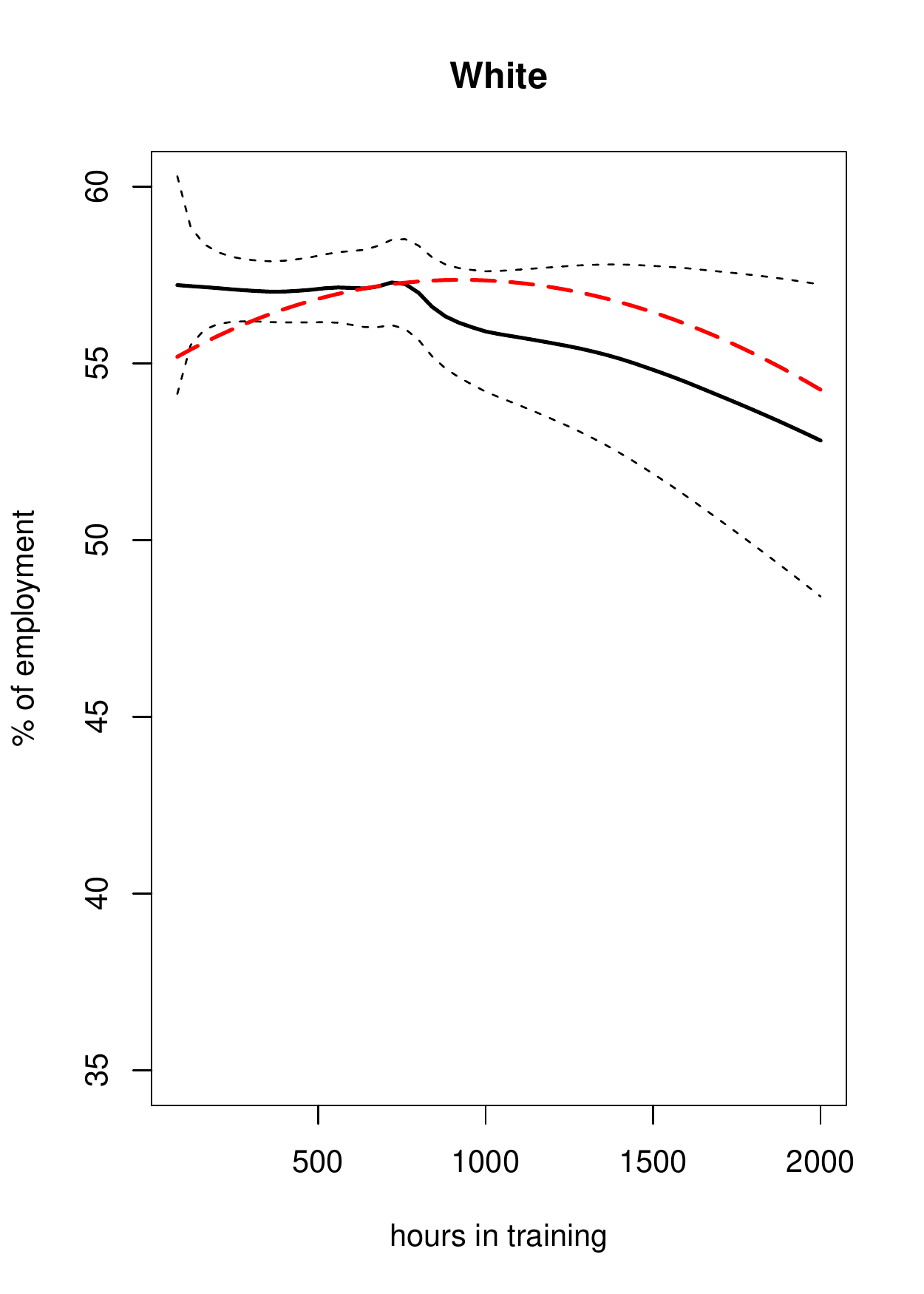}
\includegraphics[width=0.32\textwidth]{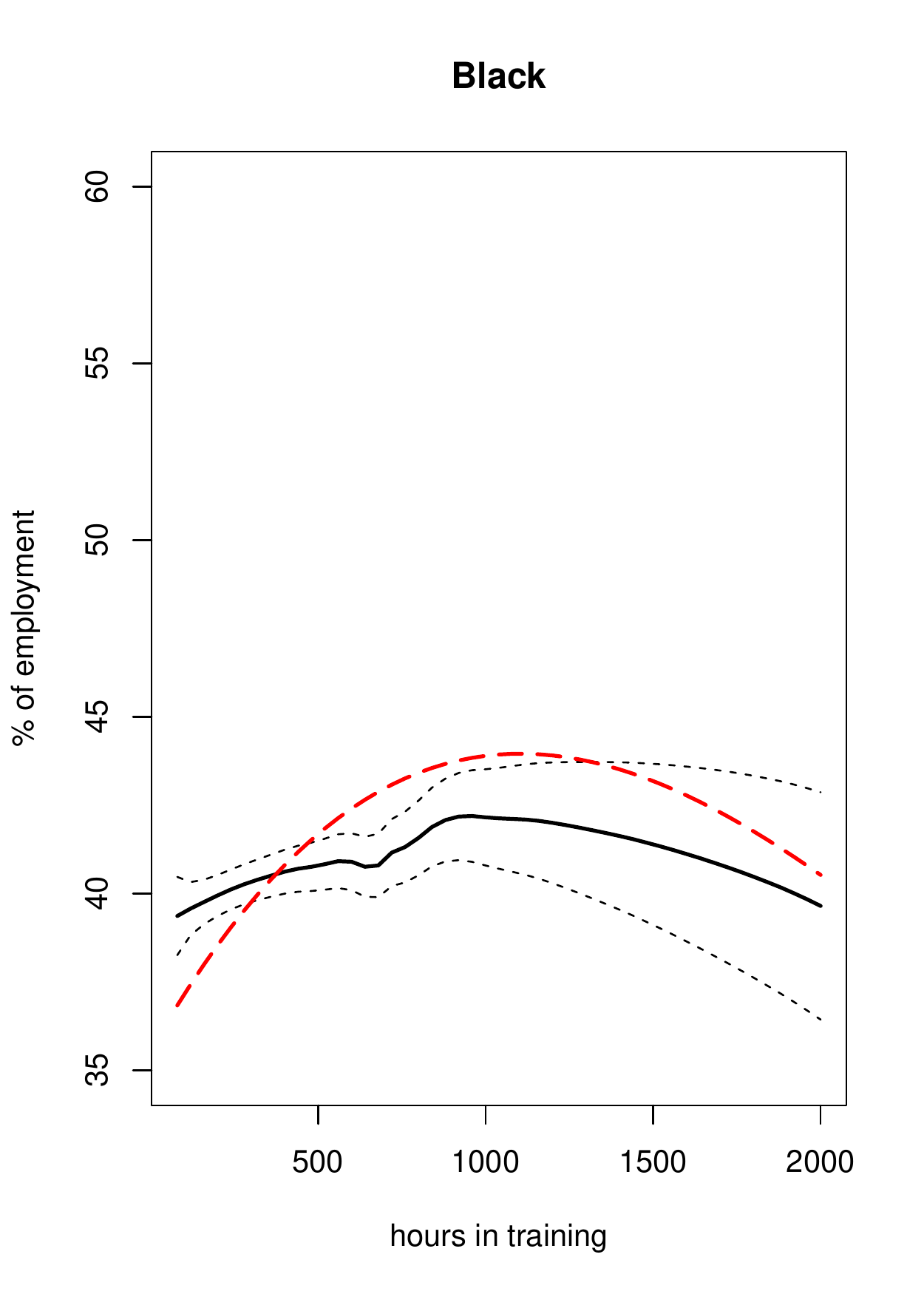}
\includegraphics[width=0.32\textwidth]{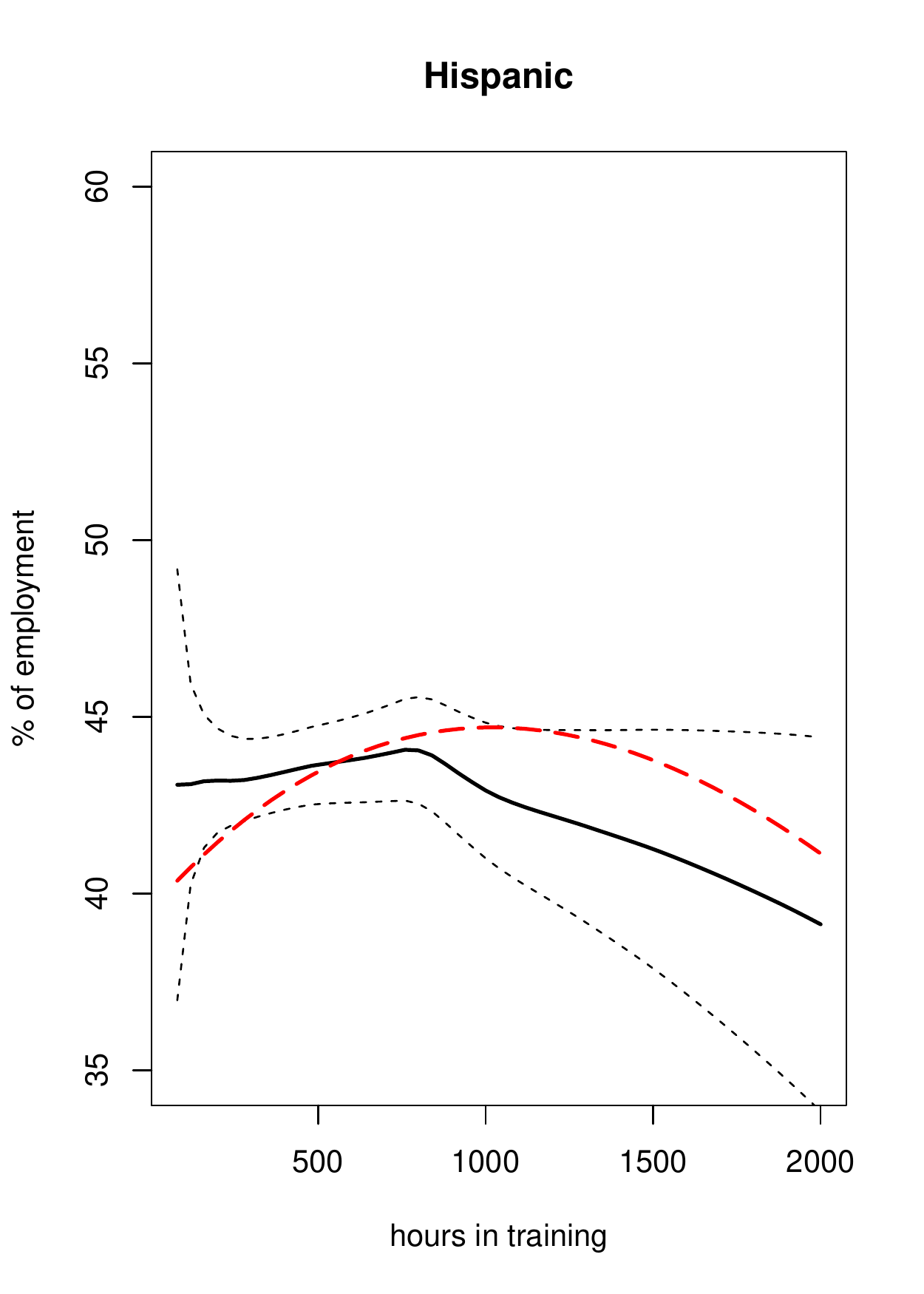}
\end{figure}

\paragraph{Results}
Figure~\ref{FADRF} presents the estimates of the average DRFs $\mathbb{E}[Y(t)]$ for the full sample and the subsamples of different gender and race.
The estimates suggest an inverted-U relationship between the employment and the length of participation for the full sample, the female sample, and the black sample.
For example, the average treatment effect of extending the length from one month to six months is about $2.5\%$ for the full sample, based on the estimate $\hat{\mathbb{E}}[Y(960)] - \hat{\mathbb{E}}[Y(160)] = 2.5$ with standard error $0.61$.
We also see significant differences in the estimates across demographic groups.
In particular, the average employment in the second year for blacks reaches its maximum at around 1000 hours (six mouths) in training.
For example, for blacks, the average treatment effect on employment of extending the length from one month to six months is about 2.43\%, based on the estimate $\widehat{\mathbb{E}}[Y(960)]-\widehat{\mathbb{E}}[Y(160)] = 2.43$ with standard error $0.74$.
In contrast, the average employment for whites decreases over a longer duration in the program, e.g., more than about three months.
The average treatment effect of extending the length from three months to eleven months is about $-3.24\%$ for whites, based on the estimate $\widehat{\mathbb{E}}[Y(1760)]-\widehat{\mathbb{E}}[Y(480)] = -3.24$ with standard error $1.95$.
The average employment for males does not significantly change over the lengths of exposure.
The average employment for females reaches its maximum at around six mouths in the program.
For example, the average treatment effect of extending the length from one month to six months is about 1.54\% for females, based on the estimate $\widehat{\mathbb{E}}[Y(960)]-\widehat{\mathbb{E}}[Y(160)] = 1.54$ with standard error $0.88$.

\begin{figure}[!htp]
\centering

\caption{Quantile dose response functions
\label{FQDRF}}
\includegraphics[width=0.32\textwidth]{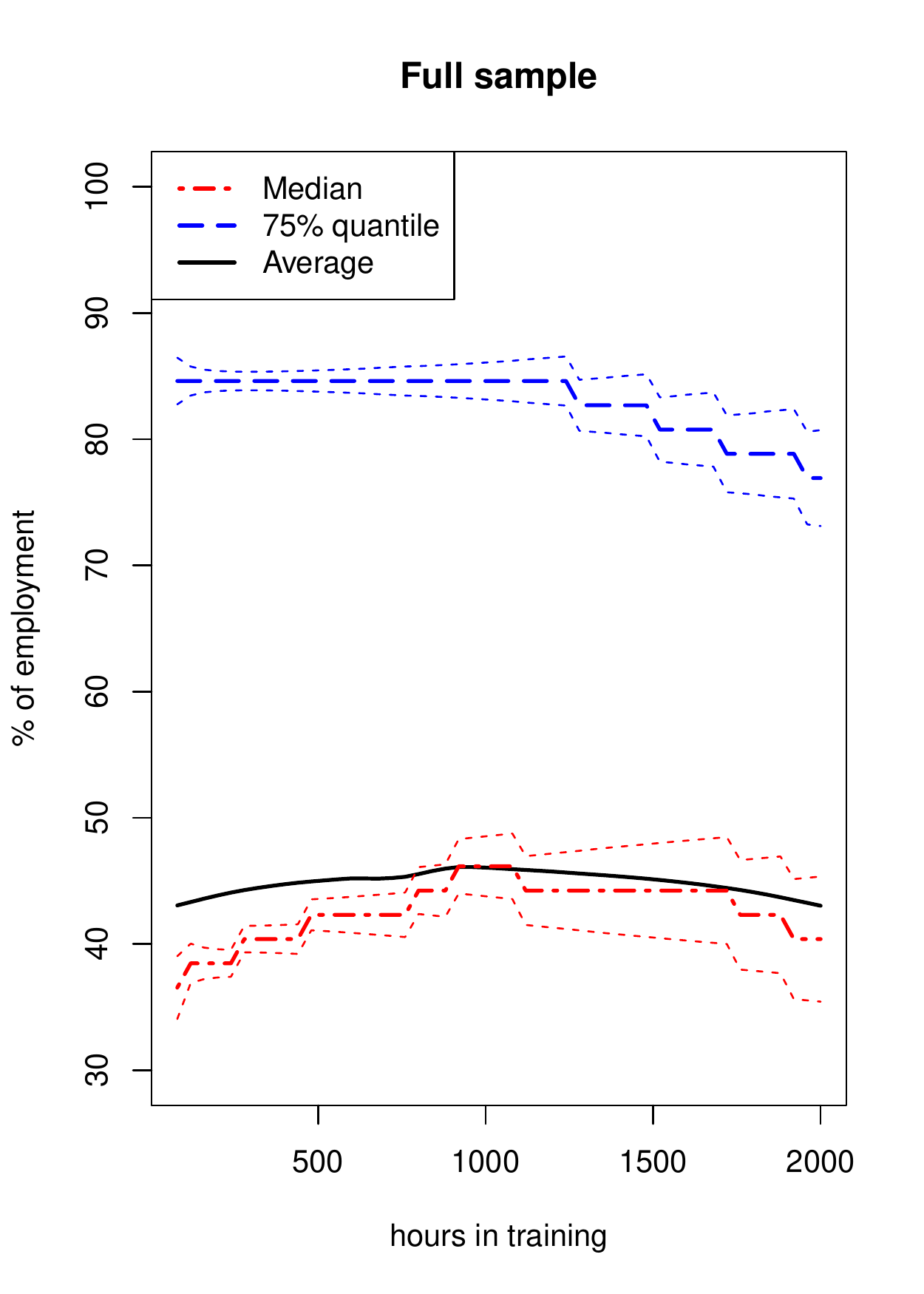}
\includegraphics[width=0.32\textwidth]{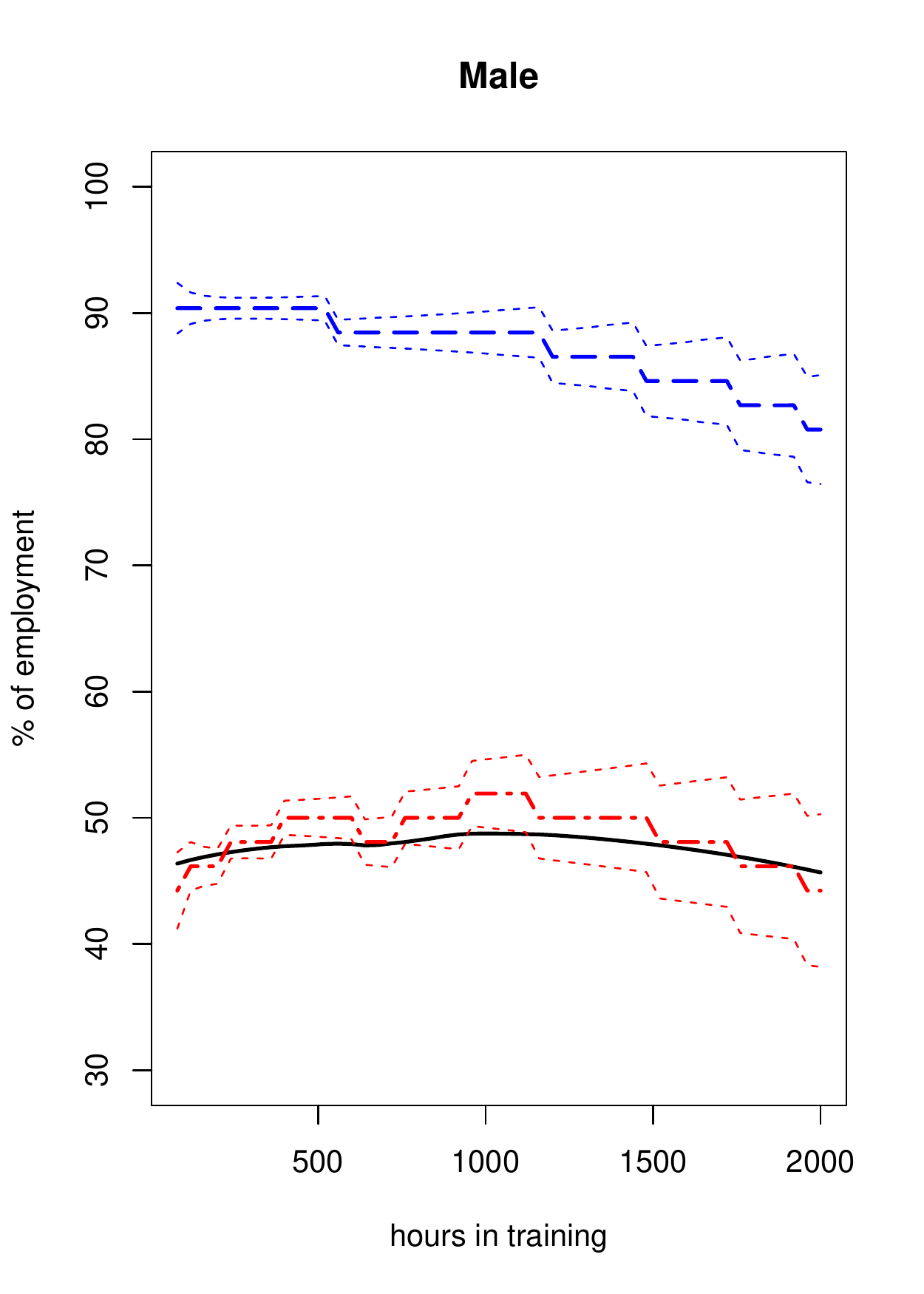}
\includegraphics[width=0.32\textwidth]{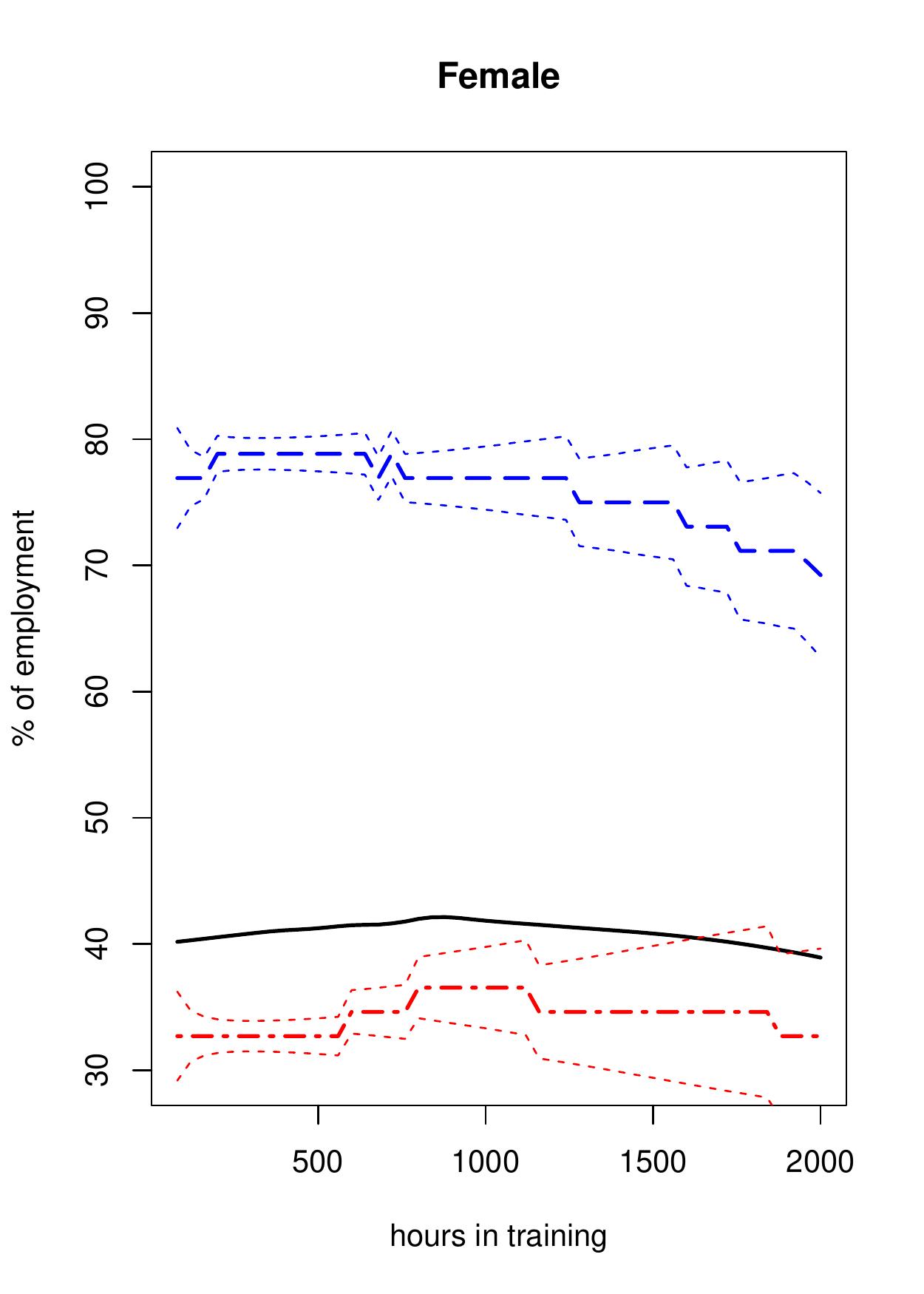}
\includegraphics[width=0.32\textwidth]{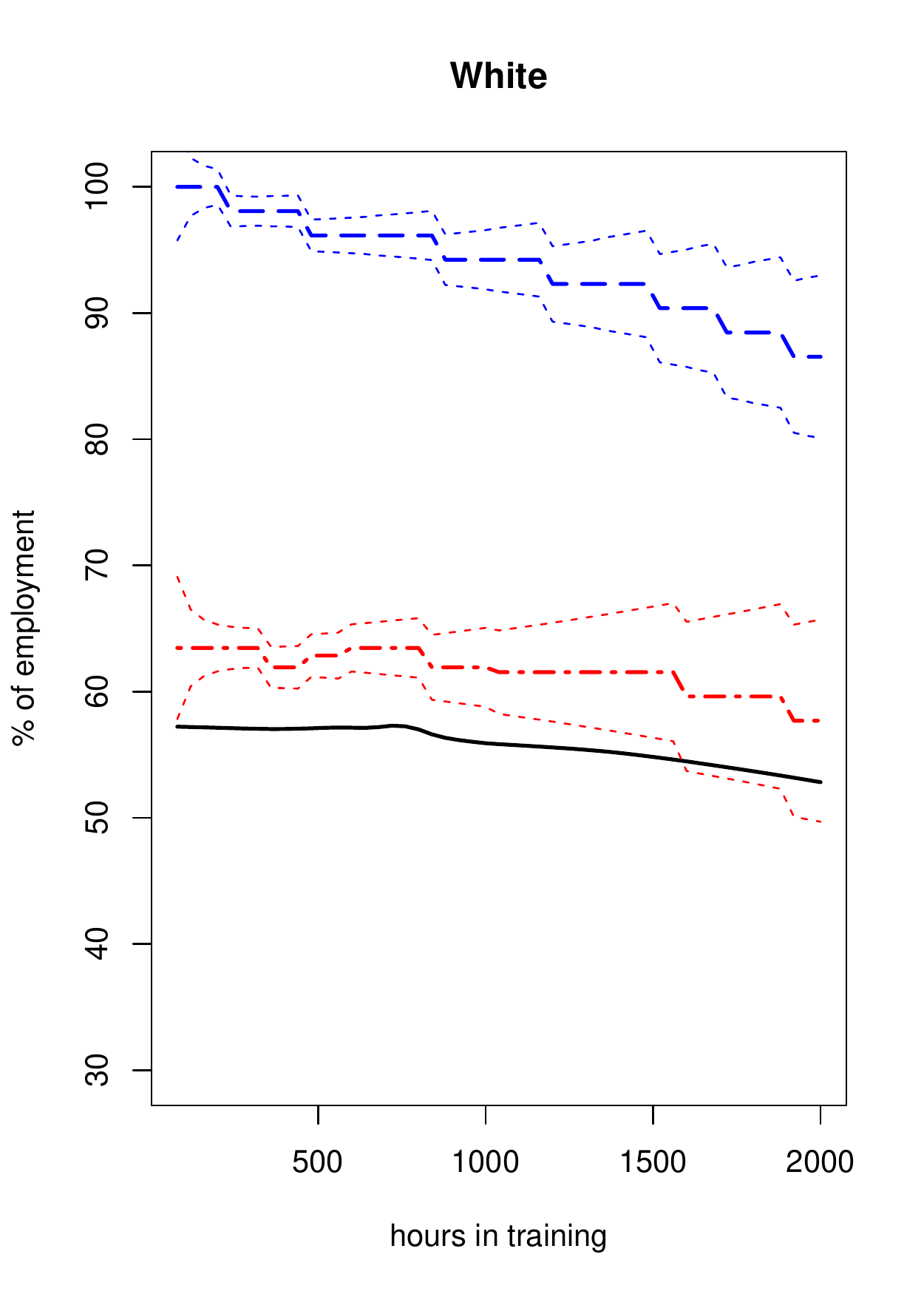}
\includegraphics[width=0.32\textwidth]{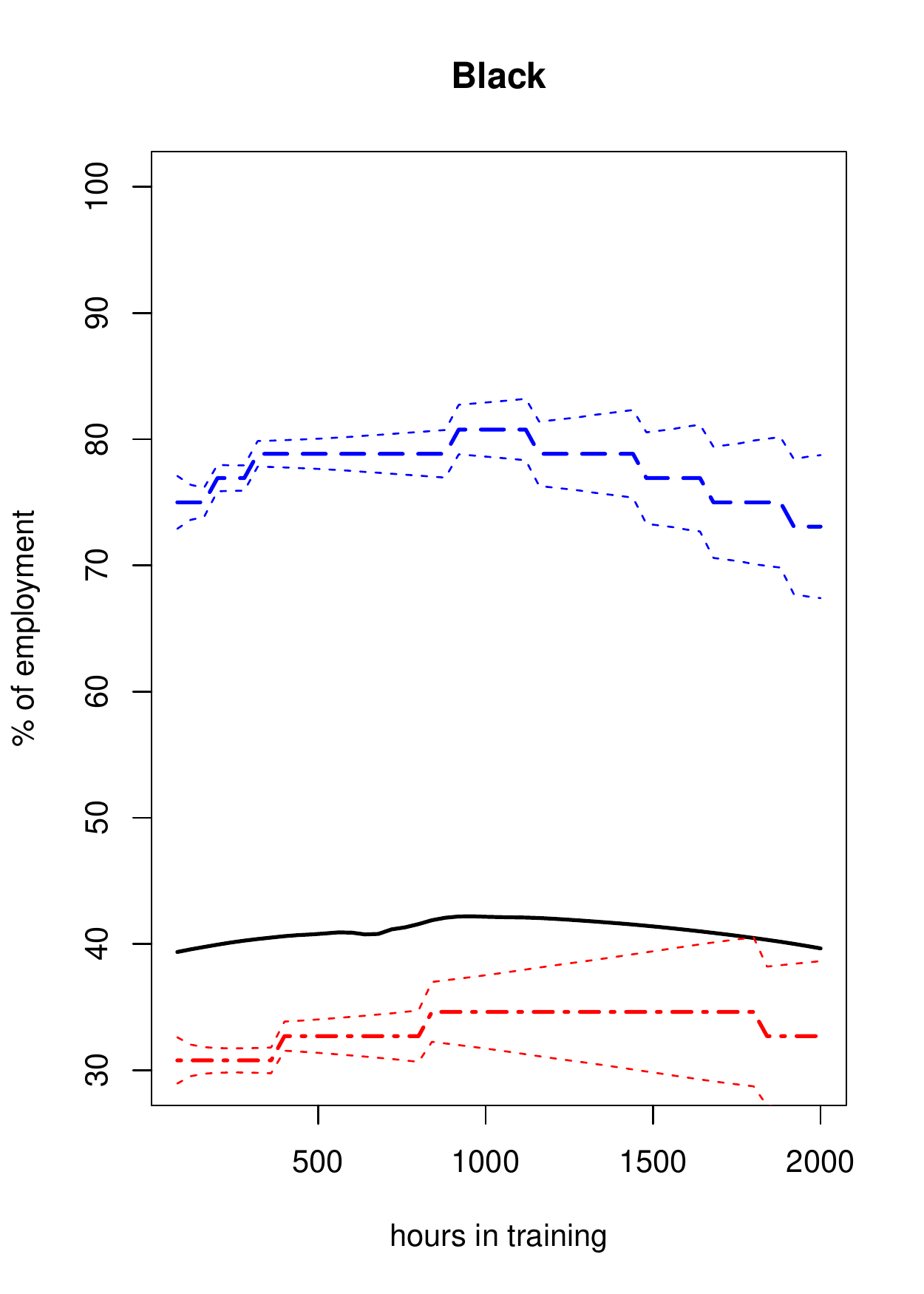}
\includegraphics[width=0.32\textwidth]{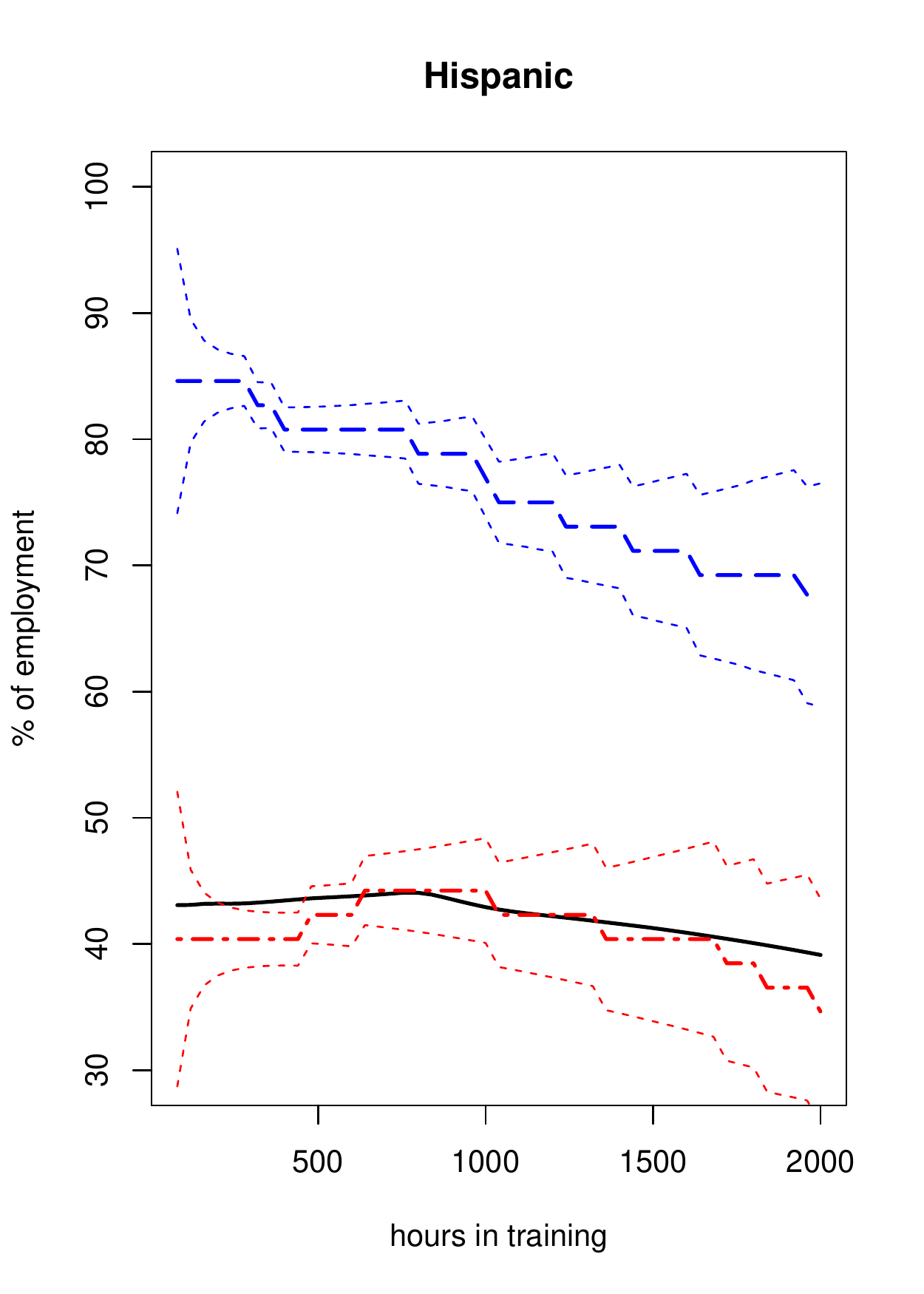}
\end{figure}


We also compute the ordinary least squares (OLS) of employment on a cubic function of the length of exposure and $X$.
The OLS estimates present the same inverted-U relationship for all subsamples that could misspecify and over estimate the effects.
Our semiparametric estimator is more flexible relative to OLS to capture heterogeneities in the effects of length of exposure to academic and vocational training.

Figure~\ref{FQDRF} presents the estimates of the quantile DRFs $Q_{\tau}(Y(t))$ for $\tau = 0.5, 0.75$, for the full sample and the subsamples of different gender and race.
Since the 25th percentile of $Y$ is zero and our asymptotic theory is established for the interior compact support of $Y$, we focus on the median and the 75th percentile.
The median estimates show similar patterns with the average estimates presented in Figure~\ref{FADRF}.
For example, the median treatment effect of extending the length from one month to six months is about $7.69\%$ for the full sample, based on the estimate $\hat Q_{.5}(Y(960)) - \hat Q_{.5}(Y(160)) = 7.69$ with standard error $1.32$.
The $75\%$-quantile DRF seems to decrease over a loner duration in Job Corps across demographic groups.
For example, the $75\%$-quantile treatment effect of extending the length from one month to eleven months is about $-5.77\%$ for the full sample, based on the estimate $\hat Q_{.75}(Y(1760)) - \hat Q_{.75}(Y(160)) = -5.77$ with standard error $1.66$.
For blacks, the average employment  is larger than the median employment, suggesting that the distribution of employment might be skewed to the right.
This is in contrary to whites, whose employment distribution might be skewed to the left.

The estimates for the average and quantile DRFs for the treated $\bar t = 400, 1000, 1760$ (i.e., around 25th, 50th, 75th percentiles of $T$, respectively) are not significantly different from the estimates for the population.
Thus we do not present the results for the DRFs for the treated.

We demonstrate that our analysis uncovers heterogeneities in the effects of training in Job Corps along the different lengths of exposure.
Job Corps staff may suggest participants to take additional or less training.
The decreasing or inverted-U relationship might indicate the lock-in effect, due to the lack of labor market experience in the second year for participants with a longer enrollment.
 \cite*{FFGN12ReStat} have studied the lock-in effect by considering outcome that puts participants on an equal footing in terms of the time they have been in the labor market after exiting the program.
To illustrate our methods and conserve space in this paper, we separate more comprehensive analysis using other labor market outcomes in another project.

\section{Conclusion}
\label{SecCon}
We derive a stochastic expansion showing how the presence of generated regressors affects the limiting behavior of the three-step nonparametric estimator of the partial mean process defined in Eq.$\!$~(\ref{PMGR}).
We allow the generated regressors to be general functions and estimated by a (semi)parametric or nonparametric method. 
The influence of estimating the generated regressors has three important elements: partial mean structure, index, and  projection of the weight.
These elements provide insights on conditions under which the estimation error is ignorable. 
We study continuous treatment effects in nonseparable models.
We provide an inference method for the bounds of the average and quantile structural functions by a fixed trimming approach.
The endogeneity of the continuous treatment is corrected by control variables or the generalized propensity score.

Our stochastic expansion accounting for the estimation error of the generated regressor is uniform over the treatment value $t$ and the distributional threshold value $y$.
Therefore, these results can be applied to more complicated estimation procedure where the partial mean is an intermediate step; for example, test for stochastic dominance in \cite*{LLW09ETA}, 
the average compensating variation in \cite*{DB} and \cite{BL14}, test for overidentifying restriction in \cite{HR16},
and the transformation model in \cite*{VV13}.


\appendix
\appendixpage
\singlespacing

{\footnotesize

We start with a heuristic sketch of the proof for Theorem~\ref{TGR} that gives an outline of the Appendix.
We decompose the estimator to each step in the estimation procedure: 
\begin{align}
&\frac{1}{n} \sum_{i=1}^n \hat F_{Y|T\hat V}(y|t,\hat V_i) W_i \hat \pi_i - \mathbb{E}\big[ F_{Y|T V}(y|t,V) W\pi\big] \notag \\[-4pt]
&=\  \frac{1}{n} \sum_{i=1}^n F_{Y|T V}(y|t,V_i) W_i \pi_i - \mathbb{E}\big[ F_{Y|T V}(y|t,V) W \pi\big] \label{sumT0} \\[-4pt]
&\ \ \ \ + \frac{1}{n}\sum_{i=1}^n \Big( \hat F_{Y|T V}(y|t,V_i) -  F_{Y|T V}(y|t, V_i) \Big) W_i \pi_i\label{sumT1}\\[-4pt]
&\ \ \ \ + \frac{1}{n}\sum_{i=1}^n \Big( \hat F_{Y|T\hat V}(y|t, V_i) - \hat F_{Y|T V}(y|t, V_i) \Big) W_i \pi_i\label{sumTGR}\\[-4pt]
&\ \ \ \ + \frac{1}{n} \sum_{i=1}^n  \nabla_v F_{Y|T V}(y|t, v)'\big|_{v=V_i} (\hat V_i - V_i)W_i \pi_i \label{sumT2}
\\[-4pt]
&\ \ \ \ + \frac{1}{n} \sum_{i=1}^n \hat F_{Y|T\hat V}(y|t,\hat V_i) W_i (\hat \pi_i - \pi_i)+ \mbox{smaller order terms}.  \label{sumTrim}
\end{align}
The term (\ref{sumT1}) is the estimation error of the Step~2 regression with observable regressors, derived in Theorem~\ref{TGaussian} whose proof is in Section~\ref{ASecTGaussian}.
The term (\ref{sumTGR}) is the estimation error from generated regressors for the {\it regressor} role.
The term (\ref{sumT2}) is for the {\it argument} role, i.e., using $\hat V$ as the argument in the regression function.
Section~\ref{ASecTGR} is the proof of Theorem~\ref{TGR}.
Section~\ref{ASecTrim} shows the estimation error of the trimming function in (\ref{sumTrim}) is zero $w.p.a.1$.
We suppress $\pi$ in the proofs for notational ease.
Section~\ref{ASecTLP} is the proof of Theorem~\ref{TLP}. 

\medskip

Section~\ref{SE} presents assumptions and lemmas of stochastic equicontinuity arguments whose proofs are  in Section~\ref{ApfSE}. 
We discuss technical detail in Section~\ref{SecTR}.
The complexity of the function space is measured by the cardinality of the covering sets or the packing number, which can be achieved by assuming smoothness of the functions.
Section~\ref{SPA} provides primitive conditions for the smoothness assumptions and the nonparametric tuning parameters.
The proofs for the inference for the treatment effects in Section~\ref{SecInf} are in Section~\ref{Ainf}. 



\section{Preliminaries}
\label{SE}

\paragraph{Notation.}Let $(Z_1, Z_1,...,Z_n)$ be an $i.i.d.$ sequence of random variables taking values in a probability space $(\mathcal{Z}, \mathcal{B})$ with distribution $P$.
For some measurable function $\phi: \mathcal{Z} \rightarrow \mathbb{R}$, define $\mathbb{E}\phi = \int \phi dP$.
Let $N(\epsilon, \mathcal{V}, \|\cdot\|)$ be the covering number with respect to the semimetric $\|\cdot\|$ and $N_{[\cdot]}(\epsilon, \mathcal{V}, \|\cdot\|)$ be the bracketing number.
Let $\bar O_p(a_n)$ and $\bar o_p(a_n)$ be $O_p(a_n)$ and $o_p(a_n)$ uniformly in $y\in \mathcal{Y}$, $t \in \mathcal{T}_0$ the interior support of $T$.
Let $C$ be a generic constant.

\begin{Assumption}{\bf(Smoothness)}
\begin{enumerate}
\item[(i)] The data $\{Y_i, T_i, X_i, Z_{1i}, W_i\}$, $i=1,...,n$, are independent and identically distributed ($i.i.d.$).
The random vector $V = v_0(S_v)$ is a vector of measurable functions of $S_v$, a subvector of $\{T,X,Z_1\}$.

\item[(ii)] 
For $(t, v) \in \mathcal{T}_0 \times \mathcal{V}_0$ the interior support of $(T, V)$ in $\mathbb{R}^{d_t + d_v}$, $f_{TV}(t, v)$ is bounded away from zero and is $\Delta$-order continuously differentiable with respect to both $t$ and $v$, with uniformly bounded derivatives.

\item[(iii)] 
The unconditional distribution $F_Y(y)$ is continuous on a compact support $\mathcal{Y} = [y_l, y_u] \subset \mathbb{R}$.
The conditional distribution $F_{Y|TV}(y|t,v)$ is $\Delta$-order continuously differentiable with respect to both $t$ and $v$, with uniformly bounded derivatives.

\end{enumerate}
\label{Asm}
\end{Assumption}

\begin{Assumption}[Kernel]
The kernel function $k(u): \mathbb{R} \rightarrow \mathbb{R}$ satisfies the following conditions: 
(i) ($r$-order) $\int k(u)du =1$, $\mu_l \equiv \int u^l k(u) du =0$ for $0 < l < r$, and $\int |u^r k(u)| du < \infty$ for some $r \geq 2$.
(ii) (bounded support) for some $L < \infty$, $k(u) = 0$ for $|u| > L$.
(iii)
$k(u)$ is $r$-times continuously differentiable and the derivatives are uniformly continuous and bounded.
(iv) For an integer $\Delta_k$, the derivatives of the kernel up to order $\Delta_k$ exist and are Lipschitz.
\label{AkernelEJL}
\end{Assumption}

\begin{Assumption}[Complexity - Regression] (i) For each $y \in \mathcal{Y}$ and $t \in \mathcal{T}_0$, $F_{Y|TV}(y|t,\cdot) \in \mathcal{C}^a_M(\mathcal{V})$,\footnote{By \cite{VW96} (P. 154),
$\mathcal{C}_M^\alpha(\mathcal{S})$ is defined on a bounded set $\mathcal{S}$ in $\mathbb{R}^{d_s}$ as follows:
For any vector $q = (q_1, ..., q_d)$ of $d$ integers,
let $D^q$ denote the differential operator
$D^q = \partial^{q.}/(\partial s_1^{q_1} ... \partial s_d^{q_d})$.
Denote $q. = \sum_{l=1}^d q_l$ and $\underline{\alpha}$ to be the greatest integer strictly smaller than $\alpha$.
Let 
$\|g\|_\alpha = \max_{q. \leq \underline{\alpha}} \sup_s |D^qg(s)| + \max_{q. \leq \underline{\alpha}} \sup_{s \neq s'} | D^qg(s) - D^q g(s') |/ \|s-s'\|^{\alpha - \underline{\alpha}}$
where $\max_{q. \leq \underline{\alpha}}$ denotes the maximum over $(q_1, ..., q_d)$ such that $q. \leq \underline{\alpha}$ and the suprema are taken over the interior of $\mathcal{S}$.
Then $\mathcal{C}_M^\alpha(\mathcal{S})$  is the set of all continuous functions $g: \mathcal{S} \subset \mathbb{R}^{d_s} \mapsto \mathbb{R}$ with $\|g\|_\alpha \leq M$.
}
 where  $a > d_v/2$.
(ii) There exists a universal constant $C$ satisfying a H\"{o}lder continuity condition: for any $ y_1, y_2 \in \mathcal{Y}$, $t_1, t_2 \in \mathcal{T}_0$,
$\big\| F_{Y|TV}(y_1|t_1, \cdot) - F_{Y|TV}(y_2| t_2, \cdot) \big\|_\infty \leq C\|(y_1,t_1) - (y_2, t_2)\|^{1/2}$, where $\|\cdot \|_\infty$ is the $\sup$-norm.
\label{AsmSE}
\end{Assumption}
Assumption~\ref{AsmSE} (ii) specifies the complexity of the function space in $y$ and $t$, following \cite{IL10} whose unknown function is a process indexed by the parameter of interest.
By the assumptions on our estimators, Section~\ref{RSE1} in the Appendix shows that the estimator $\hat F_{Y|TV}(y|t, V)$ satisfies Assumption~\ref{AsmSE} with probability approaching one ($w.p.a.1$) for any $y \in \mathcal{Y}$ and $t \in \mathcal{T}_0$.
\begin{Assumption}[Bandwidth]
The bandwidth $h$ satisfies (i) $nh^{2r+d_t} \rightarrow C_B \in [0, \infty)$ and (ii) $nh^{2d - d_t}/\log(n) \rightarrow \infty$, as $n \rightarrow \infty$.
\label{Abw}
\end{Assumption}

Assumption \ref{Acom2} below requires the Step 1 estimator $\hat V = \hat v(T,S)$ to converge fast enough and to take values in a function space that is not too complex $w.p.a.1$.
Assumption \ref{Acom2} is primitive sufficient for Assumption \ref{Acom}, which is modified from Assumptions~2 and 3 in \cite{MRS12A} (MRS12, hereafter).

\begin{Assumption}[Complexity and Accuracy]
The $j$-th component $\hat v_j$ and $v_{0j}$ of vectors $\hat v$ and $v_0$ satisfies the following conditions: for any $j=1,..., d_v$,
\begin{enumerate}
\item[(i)]
$v_{0j} \in \mathcal{C}_M^\alpha(\mathcal{S}_v)$ for $\mathcal{S}_v \subset \mathbb{R}^{d_1}$, $\alpha > d_1/2$, and $M > 0$.

\item[(ii)]
Let the Step~2 bandwidth $h \propto n^{-\eta}$.
$\|\hat v_{j} - v_{0j}\|_\infty = o_p(n^{-\delta})$, for $\delta >  2\eta$.

\item[(iii)]
$\| D^{\alpha} \hat v_j - D^{\alpha} v_{0j} \|_\infty = o_p(n^{\xi^*})$, for some $\xi^* \geq 0$.

\end{enumerate}
\label{Acom2}
\end{Assumption}

Lemma~\ref{LBCLT1} is used in Theorem~\ref{TGaussian} for the partial mean process with observable regressors in (\ref{sumT1}).
We use the bracketing central limit theorem and modify Lemma B.2 in \cite{IL10} for the complexity of the function space.
\begin{Lemma}[Stochastic Equicontinuity - Regression]
\ Define $\mathcal{F}$ to be a class of uniformly bounded functions $f: \mathcal{Y} \times \mathcal{T}\times \mathcal{V} \mapsto \mathbb{R}$ such that 
(i) for each fixed $\bar y \in \mathcal{Y}$ and $\bar t \in \mathcal{T}$, the subclass $\{f(\bar y, \bar t, \cdot) \in \mathcal{F}\}$ is $\mathcal{M}$, where the class $\mathcal{M}$ is a class of functions such that $\log N(\epsilon, \mathcal{M}, \|\cdot\|_\infty) \leq C \epsilon^{-\nu}$ for some $\nu < 2$.
(ii) There exists a universal constant $C$ satisfying a H\"{o}lder continuity condition: for any $f \in \mathcal{F}$,
\begin{align}
\| f(y_1, t_1, \cdot) - f(y_2, t_2,  \cdot) \|_\infty \leq C\|(y_1,t_1) - (y_2, t_2)\|^{1/2}.
\label{HC}
\end{align}
Suppose $F_{Y|TV} \in \mathcal{F}$ and $\hat F_{Y|TV} \in \mathcal{F}$ $w.p.a.1$.
Suppose the weight $W$ is uniformly bounded.
Then
\begin{align*}
\sup_{y \in \mathcal{Y}, t\in \mathcal{T}_0} \bigg|&\frac{1}{\sqrt{n}}\sum_{i=1}^n \Big( \hat F_{Y|TV}(y|t,V_i) - F_{Y|TV}(y|t,V_i) \Big) W_i \\
&- \sqrt{n}\ \mathbb{E} \Big[ \Big( \hat F_{Y|TV}(y|t,V) - F_{Y|TV}(y|t,V) \Big) W \Big] \bigg|= o_p(1).
\end{align*}
\label{LBCLT1}
\end{Lemma}

The preliminary lemmas modify the stochastic equicontinuity argument for the kernel estimator developed in Lemma 1 in MRS12.
The high-level Assumption~\ref{Acom} below is modified from Assumptions~2 and 3 in MRS12. 
In Section~\ref{ASecPCGR}, we show that Assumption~\ref{Acom2} is sufficient for Assumption~\ref{Acom}.

\begin{Assumption}[Complexity - Generated Regressors]
There exists a sequence of sets of functions $\mathcal{M}_{n}$ such that
\begin{enumerate}
\item[(i)]
${\rm Pr}(v_{0j} \in \mathcal{M}_{n}) \rightarrow 1$ and ${\rm Pr}(\hat v_j \in \mathcal{M}_{n}) \rightarrow 1$ as $n \rightarrow \infty$ for all $j=1,..., d_v$.

\item[(ii)] 
For a constant $C > 0$ and a function $v_{nj}$ with $\|v_{nj} - v_{0j}\|_\infty = o(n^{-\delta})$, the set $\mathcal{\bar M}_{n} \equiv \mathcal{M}_{n} \cap \{ u_j: \|v_{nj} - u_{j}\|_\infty \leq n^{-\delta} \} $ can be covered by at most $C\exp(\varrho^{-\beta} n^{\xi} )$ balls with $\|\cdot\|_\infty$-radius $\varrho$ for all $\varrho \leq n^{-\delta}$, where $\delta > 2\eta$, $0 < \beta < 2$, and $\xi > 0$.

\end{enumerate}
\label{Acom}
\end{Assumption}

\begin{Lemma}
Let Assumptions~\ref{Asm}, \ref{AkernelEJL}, and \ref{Acom} hold.
Suppose a function $B_{yt}( W,V)$ is uniformly continuous in $(y,t)$ and continuously differentiable with respect to $(W,V)$, with uniformly bounded derivatives.
Then 
\begin{align*}
\sup_{\stackrel{y \in \mathcal{Y}, t \in \mathcal{T}_0, s\in\mathcal{S}_0}{v_1,v_2 \in \mathcal{\bar M}_n} } \bigg| & \frac{1}{n} \sum_{i=1}^n B_{yt}( W_i,V_i) \big( K_h(v_1(t, s) - V_i) - K_h(v_2(t, s) - V_i)\big) \\
&- \mathbb{E}\Big[ B_{yt}( W,V) \big( K_h(v_1(t, s) - V) - K_h(v_2(t, s) - V)\big)   \Big] \bigg| =  O_p(n^{-\kappa_{11}}),
\end{align*}
where $0 <  \kappa_{11} < \frac{1}{2}(1 - d_v \eta) + (\delta - \eta) - \frac{1}{2} (\delta \beta + \xi)$.
\label{LMRS2}
\end{Lemma}

\begin{Lemma}
Let the conditions of Lemma~\ref{LMRS2} hold.
Suppose a function $B_{t}( V)$ is uniformly continuous in $t$ and 
continuously differentiable with respect to $V$, with uniformly bounded derivatives.
Then 
\begin{align*}
\sup_{\stackrel{y \in \mathcal{Y}, t \in \mathcal{T}_0}{v_1,v_2 \in \mathcal{\bar M}_n} } \bigg|&\frac{1}{n} \sum_{i=1}^n   {\bf 1}{\{Y_i \leq y\}} K_h(T_i - t)  \big( B_{t}( v_1(T_i, S_i)) - B_{t}( v_2(T_i, S_i)) \big) 
\\&
- \mathbb{E}\Big[  {\bf 1}{\{Y \leq y\}} K_h(T - t)    \big( B_{t}( v_1(T, S)) - B_{t}( v_2(T, S)) \big)  \Big] \bigg| = O_p(n^{-\kappa_{12}}),
\end{align*}
where $0< \kappa_{12} < \frac{1}{2}(1-d_t \eta) + \delta - \frac{1}{2} (\delta \beta + \xi)$.
\label{LMRS3}
\end{Lemma}

\begin{Lemma}
Let the conditions of Lemma~\ref{LMRS2} hold.
Suppose a function $B_{yt}( W,V)$ is uniformly bounded in all its arguments and uniformly continuous in $(y,t)$.
Then 
\begin{align*}
\sup_{\stackrel{y \in \mathcal{Y}, t \in \mathcal{T}_0}{v_1,v_2 \in \mathcal{\bar M}_n} } \bigg|&\frac{1}{n} \sum_{i=1}^n   \Big( v_1(T_i, S_i) - v_2(T_i, S_i) \Big)'B_{yt}( W_i,V_i) 
- \mathbb{E}\Big[  \Big(v_1(T, S) - v_2(T, S) ) \Big)' B_{yt}( W,V)\Big] \bigg| 
\end{align*}
$= O_p(n^{-\kappa_{13}})$, where $\kappa_{13} < \frac{1}{2} + \delta - \frac{1}{2} (\delta \beta + \xi)$.\footnote{Lemma~\ref{LMRS4} is for the generated regressors playing the {\it argument} role in (\ref{sumT2}).
When $\xi^* = 0$ ($\xi = 0$), the result coincides with Lemma 1 in \cite{CLK03ETA}.  
Their small order term is controlled by $o_p\left(n^{-1/2}\right)$ using Donsker theorem.
In contrast, we follow MRS12 to use a larger function space $\mathcal{M}_n$ in the sense that for $v \in \mathcal{M}_n$, $v/n^{\xi^*} \in \mathcal{C}_M^\alpha$ in Assumption~\ref{Acom}. 
}
\label{LMRS4}
\end{Lemma}

\begin{Lemma}[Lemma 1 in \cite*{MRS12A}]
Let the conditions of Lemma~\ref{LMRS2} hold.
Then
\begin{align*}
\sup_{\stackrel{t\in \mathcal{T}_0, u  \in \mathcal{V}_0, y\in \mathcal{Y}}{v_1, v_2 \in \mathcal{\bar M}_n}} &\bigg| \frac{1}{n} \sum_{j=1}^n {\bf 1}{\{ Y_j\leq y \}} K_h(T_j-t) \Big( K_h(v_1(T_j, S_{j}) - u) - K_h(v_2(T_j, S_{j}) - u)\Big) \\[-2pt]
&- \mathbb{E}\Big[ {\bf 1}{\{ Y\leq y \}} K_h(T-t) \Big( K_h(v_1(T, S) - u)- K_h(v_2(T, S) - u)  \Big) \Big] \bigg| = O_p(n^{-\kappa_{14}}), \\[7pt]
\sup_{\stackrel{t\in \mathcal{T}_0, u \in \mathcal{V}_0}{ v_1, v_2 \in \mathcal{\bar M}_n}}& \bigg| \frac{1}{n} \sum_{j=1}^n K_h(T_j-t) \Big( K_h(v_1(T_j, S_{j}) - u) - K_h(v_2(T_j, S_{j}) - u)\Big) \\[-2pt]
&- \mathbb{E}\Big[ K_h(T-t) \Big( K_h(v_1(T, S) - u)- K_h(v_2(T, S) - u)  \Big) \Big] \bigg| = O_p(n^{-(\kappa_{14}}),
\end{align*}
where $0 <  \kappa_{14} < \frac{1}{2}(1 - d \eta) + (\delta - \eta) - \frac{1}{2} (\delta \beta + \xi)$.\footnote{Lemma~\ref{LMRS} extends Lemma 1 in MRS12 to the derivative of a kernel function.}
Define the $q$th element of the vector $\nabla K_h(u)$ to be $h^{-(d_u+1)} k'(u_q/h)\Pi_{l\neq q,  l=1}^{d_u} k(u_l/h)$ for $q=1,..., d_v$.
\begin{align*}
\sup_{\stackrel{t\in \mathcal{T}_0, u  \in \mathcal{V}_0, y\in \mathcal{Y}}{v_1, v_2 \in \mathcal{\bar M}_n}} &\bigg| \frac{1}{n} \sum_{j=1}^n {\bf 1}{\{ Y_j\leq y \}} K_h(T_j-t) \Big( \nabla K_h(v_1(T_j, S_{j}) - u) -\nabla K_h(v_2(T_j, S_{j}) - u)\Big) \\[-2pt]
&- \mathbb{E}\Big[ {\bf 1}{\{ Y\leq y \}} K_h(T-t) \Big( \nabla K_h(v_1(T, S) - u)- \nabla K_h(v_2(T, S) - u)  \Big) \Big] \bigg| = O_p\left(n^{-(\kappa_{14}-\eta )}\right). \\[7pt]
\sup_{\stackrel{t\in \mathcal{T}_0, u \in \mathcal{V}_0}{ v_1, v_2 \in \mathcal{\bar M}_n}}& \bigg| \frac{1}{n} \sum_{j=1}^n K_h(T_j-t) \Big( \nabla K_h(v_1(T_j, S_{j}) - u) - \nabla K_h(v_2(T_j, S_{j}) - u)\Big) \\[-2pt]
&- \mathbb{E}\Big[ K_h(T-t) \Big( \nabla K_h(v_1(T, S) - u)- \nabla K_h(v_2(T, S) - u)  \Big) \Big] \bigg| = O_p\left(n^{-(\kappa_{14}-\eta )}\right).
\end{align*}
\label{LMRS}
\end{Lemma}

\begin{Lemma}[Stochastic Equicontinuity - Regression]
Define $\mathcal{F}$ to be a class of uniformly bounded functions $f: \mathcal{Y} \times \mathcal{T} \times \mathbf{\Lambda} \mapsto \mathbb{R}$ such that 
(i) For each fixed $\bar y \in \mathcal{Y}$ and $\bar t \in \mathcal{T}$, the subclass $\{f(\bar y, \bar t, \cdot) \in \mathcal{F}\}$ is $\mathcal{C}_M^\alpha$;
(ii) There exists a universal constant $C$ satisfying a Lipschitz continuity condition: for any $f \in \mathcal{F}$,
$\| f(y_1, t_1, \cdot) - f(y_2, t_2,  \cdot) \|_\infty \leq C\|(y_1,t_1) - (y_2, t_2)\|.$
Define $S \subseteq T$ a subset of the treatment variable with the support $\mathcal{S} \subseteq \mathcal{T}$.
\begin{align*}
\sup_{(y, t, s) \in \mathcal{Y} \times \mathcal{T} \times \mathcal{S}} \sup_{f \in \mathcal{F}} \Big|  \frac{1}{n}\sum_{i=1}^n f(y, t, \Lambda_i) K_h(S_i - s) - \mathbb{E}\Big[ f(y, t, \Lambda) K_h(S - s) \Big]\Big| = O_p\bigg( \sqrt{\frac{\log n}{nh^{d_s}}} \bigg).
\end{align*}
\label{LCLT}
\end{Lemma}

Result \ref{PKunif} below is from Lemma B.3 in \cite{Newey94ET} and  Theorem 3.2 in \cite{HJS88AS}.
Define $g_{YTV}(y,t,v) \equiv F_{Y|TV}(y|t, v) f_{TV}(t,v)$ and $\hat g_{YT\hat V}(y,t,v) \equiv n^{-1} \sum_{j=1}^n {\bf 1}{\{ Y_j\leq y \}}  $\\$\times K_h(T_j-t) K_h(\hat V_j - v)$.
\begin{result}
Suppose the bandwidth $h \rightarrow 0$ and $\log n/ (nh^d) \rightarrow 0$, where the dimension of the regressors is $d = d_t +d_v$.
Suppose Assumptions~\ref{Asm} and \ref{AkernelEJL} hold.
For the first four results below, assume $\Delta \geq r$ and $\Delta_k \geq 0$.
\begin{enumerate}
\item 
$\sup_{(y,t,v) \in \mathcal{Y}\times \mathcal{T}_0 \times \mathcal{V}_0} \Big| \hat g_{YTV}(y, t,v) - g_{YTV}(y, t,v) \Big|  = O_p\Big( \big( \frac{\log n}{nh^d}\big)^{1/2} + h^r \Big)
$

\item
$\sup_{(t,v) \in \mathcal{T}_0 \times \mathcal{V}_0} \Big| \hat f_{TV}(t,v) - f_{TV}(t,v) \Big|  = O_p\Big( \big( \frac{\log n}{nh^d}\big)^{1/2} + h^r \Big)
$

\item
$\sup_{(t,v) \in \mathcal{T}_0 \times \mathcal{V}_0}\Big| \hat f_{T|V}(t|v) - f_{T|V}(t|v) \Big|  = O_p\Big( \big( \frac{\log n}{nh^d}\big)^{1/2} + h^r \Big)
$

\item
$\sup_{(y,t,v) \in \mathcal{Y}\times \mathcal{T}_0 \times \mathcal{V}_0} \Big| \hat F_{Y|TV}(y|t,v) - F_{Y|TV}(y|t,v) \Big|  = O_p\Big( \big( \frac{\log n}{nh^d}\big)^{1/2} + h^r \Big)
$

\item
Now assume $\Delta \geq r + q$ and $\Delta_k \geq q$.
Then $\sup_{(y,t,v) \in \mathcal{Y}\times \mathcal{T}_0\times \mathcal{V}_0 } \Big| \frac{\partial^q}{\partial t^q} \hat F_{Y|TV}(y|t,v) - \frac{\partial^q}{\partial t^q} F_{Y|TV}(y|t,v) \Big|  = O_p\Big( \big( \frac{\log n}{nh^{d+2q}}\big)^{1/2} + h^r \Big)$.
\end{enumerate}
\label{PKunif}
\end{result}

\subsection{Technical remarks}
\label{SecTR}
We discuss technical details and compare our approach with recent seminal work in the literature.
Our estimator can be viewed as a second-order $U$-statistic of a kernel regression function whose regressor and argument are the {\it generated regressor} $V_i = v_0(S_{vi})$.
We specify the function space where the generated regressor $v_0$ belongs.
Our empirical process approach allows a general estimator of $v_0$.
We provide the corresponding low-level conditions that require $v_0(\cdot)$ to be a smooth function and estimated at a particular convergence rate.
In contrast, a third-order $U$-statistics approach requires a specific kernel estimator of the generated regressor that admits an asymptotic linear representation, e.g., \cite{SuUllah} and \cite{VV13}.

For a general estimation problem with nuisance unknown functions, a Lipschitz continuous condition is often imposed on the criterion function or moment function with respect to the infinite-dimensional parameters; for example, \cite{CLK03ETA}. 
Here, the unknown functions involve a regression function and its generated regressor. 
Previous works, such as \cite{HR13ETA}, \cite{Song08ET}, \cite{EJL}, MRS12, \cite{MRS12B} (MRS16, hereafter),  
 impose a high-level smoothness assumption on the regression function with respect to the regressor.
We impose low-level conditions on the estimators.
For example, \cite{EJL} assume a Lipschitz condition: $\sup_v \big| \mathbb{E}[Y|v_1(S_v) = v] - \mathbb{E}[Y|v_0(S_v) = v] \big| \leq C\|v_1- v_0\|_\infty$ for some positive constant $C$.\footnote{\cite{EJL} use this smoothness assumption to obtain the desired covering number for using the proof of Lemma B.2 in \cite{IL10}.
\cite{EJL} derive a stochastic expansion for a full mean by an argument of stochastic equicontinuity in both the regression function and the generated regressor, i.e., to analyze (\ref{sumT0})-(\ref{sumTGR}) together in one term.
To implement the theoretical path-derivative approach, \cite{HR13ETA} assume the derivative of the regression function $\mathbb{E}[Y|T=t, v_0(S_v; \alpha)=v]$ with respective to the scalar $\alpha$ exists.
MRS12 and MRS16 use a Lipschitz continuity assumption based on the index bias to control remainder terms, which we summarize in Assumption~\ref{Acon} in this paper.}
We avoid such high-level assumption at the cost of a slightly stronger estimation accuracy assumption on the first-step estimation. In particular, we use a crude bound to control the reminder terms $\|\hat v - v_0\|_\infty^2/h^2$ rather than $\|\hat v - v_0\|_\infty^2$.\footnote{This estimation accuracy assumption appears in $2(\delta - \eta)$ in $\kappa_2$. 
We use the mean-value expansion similar to Theorem~3 in \cite{GPV00ETCA}.
Thus we bound $\|\hat F_{Y|T\hat V} - F_{Y|TV} \|_{\infty} = O_p(n^{-(\delta-\eta)} + n^{-\kappa_{14}})$ in (\ref{SE0}) and the remainder terms $\|so2\|_\infty = O_p(n^{-\kappa_2})$.  
}
This is a tradeoff  between the high-level smoothness assumption and estimation accuracy assumption.
\cite{Song12SPL} and \cite{HR13ETA} provide sufficient conditions for this type of high-level smoothness assumption.
That is to assume the regression function to be very smooth and insensitive at a local perturbation of the regressor.
\cite{HR16} show that the failure of such smoothness assumption may result in irregular behavior of the estimator. 

A possible alternative to our stochastic equicontinuity argument is 
to bound weighted integrals of the remainder term of (\ref{sumTGR}) as in MRS16,
instead of controlling the remainder terms by supremum norm.
We decide this extension to be out of the scope of this paper for the following reasons:
First, our stochastic equicontinuity argument using a supremum norm serves the purpose to derive the expansion that is uniform over $t$ for the partial mean.  
Second, the remainder terms from the preliminary estimators in a multi-step estimation procedure are often controlled by the supremum norm, i.e., (\ref{sumT1}) and (\ref{sumT2});
see also the examples in MRS16, where the remainders of the Step~2 regression function are controlled by the supremum norm.
Thus we use the supremum norm for all the remainder terms from each step of the estimation procedure. 
Third, the resulting regularity conditions are not unreasonably restrictive for our empirical application of the GPS.

\section{Primitive Conditions}
\label{SPA}
\subsection{Observable regressors}
\label{RSE1}
This section shows Assumptions~\ref{AsmSE}, \ref{Abw}, and the conditions in Theorem~\ref{TGaussian} are sufficient for the high-level conditions in Lemma~\ref{LBCLT1}.
By Theorem 2.7.1 of \cite{VW96}, there exists a constant $C$ depending only on $M, a, diam(\mathcal{V}), d$ such that $\log N(\epsilon, \mathcal{C}_M^a, \|\cdot\|_\infty) \leq C \epsilon^{-d/a}$ for a bounded convex $\mathcal{V}$.
So choosing $\mathcal{C}_M^a(\mathcal{V})$ with $a > d/2$ to be the function space $\mathcal{M}$ satisfies the conditions in Lemma~\ref{LMRS}. 
Assumption~\ref{AsmSE} implies $F_{Y|TV}(y|t,V) \in \mathcal{F}$. 
The condition of $P\big(\forall y\in \mathcal{Y}, \forall t \in \mathcal{T}_0, \hat F_{Y|TV}(y|t, \cdot) \in \mathcal{F}\big) \rightarrow 1$ hold by the following arguments: 
\begin{enumerate}

\item
The condition that $P\big(\forall y\in \mathcal{Y},  \forall t \in \mathcal{T}_0, \hat F_{Y|TV}(y|t, \cdot) \in \mathcal{M} = \mathcal{C}^a_M(\mathcal{V}) \big) \rightarrow 1$ is checked by the uniform convergence of the $q$th derivative of $\hat F_{Y|TV}$ for $q. \leq \underline{a}$
\[
\Big\| D^q \hat F_{Y|TV}(y|t,\cdot) - D^q F_{Y|TV}(y|t,\cdot) \Big\|_\infty = O_p\Big( \sqrt{\frac{\log n}{nh^{d+2q}}}  + h^r \Big) = o_p(1)
\]
by Result \ref{PKunif}, Assumptions~\ref{AsmSE}, and \ref{Abw}.
Similar primitive conditions have been discussed in MRS12, Appendix C of \cite{EJL} and in footnote 11 of \cite{IL10}.

\item
We check the following sufficient high-level assumption modifying Assumption~3.4~(e) in \cite{IL10}:
For any $\epsilon > 0$ and $\delta > 0$, there exists $n_0$ such that for all $n \geq n_0$, for any $ y_1, y_2 \in \mathcal{Y}$, $t_1, t_2 \in \mathcal{T}$,
\begin{align}
{\rm Pr}\bigg( \Big\| & \hat F_{Y|TV}(y_1|t, \cdot) -  \hat F_{Y|TV}(y_2|t, \cdot) - \big( F_{Y|TV}(y_1|t_1, \cdot) - F_{Y|TV}(y_2|t_2, \cdot) \big) \Big\|_\infty \notag \\
&\leq \delta \big\|(y_1,t_1) - (y_2,t_2)\big\|^{1/2} \bigg) \geq 1 -\epsilon.
\label{ILLcon}
\end{align}
(\ref{ILLcon}) is satisfied by Chebyshev's inequality and the mean-square-errors of our kernel estimator for the regressor $\mathbb{E}[{\bf 1}{\{y_2 < Y \leq y_1\}} | T=t, V]$, assuming $y_1 > y_2$.
(\ref{ILLcon}) and the H\"{o}lder continuity of $F_{Y|TV}$ imply $\| \hat F_{Y|TV}(y_1|t, \cdot) - \hat F_{Y|TV}(y_2| t, \cdot) \|_\infty \leq C_L |y_1 - y_2|^{1/2}$, $w.p.a.1$ for $t = t_1, t_2$.\footnote{
Because the estimator $\hat F_{Y|TV}(y|t,V)$ is nonsmooth in $y$, the function space $\mathcal{F}$ is allowed to be less smooth in $y$ by assuming a H\"{o}lder continuity (\ref{HC}).
Alternatively, as discussed in \cite{IL10}, a smoothed CDF estimator is needed if a stronger Lipschitz continuity assumption is made on $\mathcal{F}$.
}

\end{enumerate}

\subsection{Generated regressors}
\label{ASecPCGR}
When the generated regressors are specified and estimated parametrically, $\delta = 1/2$ and Complexity Assumption~\ref{Acom} is satisfied by Example 19.7 in \cite{V00} for a Donsker parametric function. 

Let $\mathcal{M}_{n}$ be the set of functions such that for any $v \in \mathcal{M}_{n}$, $v/n^{\xi^*} \in \mathcal{C}_M^\alpha(\mathcal{S}_v)$, for some $\xi^* \geq 0$. 
By letting $\beta = d_1/\alpha$ and $\xi = \xi^* d_1/\alpha = \xi^*_j\beta$,  Assumption~\ref{Acom}~(ii) is satisfied by Corollary 2.7.2 in \cite{VW96}.
The complexity of the function space is controlled by the uniform bound $\xi^*$ and the differentiability $\alpha$.
Assumption~\ref{Acom2}~(i) and (iii) implies Assumption~\ref{Acom}~(i).
The advantage of the uniform bound $\xi^*$ is that when a stronger smoothness assumption is required (a larger $\alpha$), a larger $\xi^*$ can be a leverage to avoid restrictive assumption on the Step 1 estimation of the generated regressor $\hat v$.

To make the remainder terms smaller order, i.e., $\sqrt{nh^{d_t}} \|R_n\|_\infty = o_p(1)$, assume
$\delta > \max\{ (1+ \eta(4-d_t))/4, (\eta(d+2 -2d_t)\alpha + \xi^*d_1)/(2\alpha-d_1),  \xi^*d_1/(2\alpha-d_1), (\eta(d_v + 4)\alpha + \xi^*d_1)/(4\alpha-d_1) \}$.
This condition implies that if the function space is more restrictive or less complex (smaller $\xi^*$ or larger $\alpha$), the Step~1 estimation can be less accurate (smaller $\delta$).
The tradeoff between the complexity and accuracy assumptions is also discussed in MRS12 and \cite{EJL}.
When the dimension of continuous treatments $d_t$ is larger than four, $\delta$ is allowed to be smaller than $1/4$.
For the cases of full mean, discrete treatment, and semiparametric estimation, the conditions are modified by setting $d_t=0$.   
The regularity conditions are more restrictive in the semiparametric models.

The following primitive bandwidth condition is for the nonparametrically estimated GPS in Section~\ref{SGRGPS}, where $d=2$ and $d_1 = d_x$.
Assumptions~\ref{Abw} and \ref{Acom2} and $\sqrt{nh^{d_t}}\|R_n\|_\infty = o_p(1)$ imply 
$1/(2r+1) < \eta < \min\{1/3, (4\delta-1)/3, \delta/2, (\delta(2\alpha-d_x) - \xi^*d_x)/(2\alpha), (\delta(4\alpha-d_x) - \xi^*d_x)/5\alpha\}$.

\begin{Assumption}[Bandwidth - GPS]
Let $\xi^* = 0$ and $dx(4+d_x)/(d_x+2) < 2\alpha < 4+d_x$.
\begin{enumerate}
\item[(i)]
The bandwidth for the second step regression $h \propto n^{-\eta}$ and the bandwidth for the first step GPS estimation $h_1 \propto n^{-g}$ satisfy 
\begin{align*}
\frac{1}{2r_1+ 1} <  g &<  \frac{1}{2d_x + 5} \\
\max\left\{\frac{1}{2r+ 1}, g\right\} < \eta &< \min\bigg\{  \frac{1-(d_x + 1)g}{4}, \frac{1-2g(d_x + 1)}{3},\\
&\hspace{1.3cm} \frac{\big(1-(d_x + 1)g\big)\big(2-d_x/\alpha\big)}{4}, \frac{\big(1-d_xg\big)\big(4-d_x/\alpha\big) }{10} \bigg\}
\end{align*}
\item[(ii)]
 The second step regression and the first step GPS estimation use the same kernel function with order $r$ and bandwidth $h_1 = h \propto n^{-\eta}$ satisfies $(2r+1)^{-1} < \eta <  (2d_x + 5)^{-1}$.

\end{enumerate}
\label{Abwgps}
\end{Assumption}
\section{Asymptotic Results}
Section~\ref{ASecTGaussian} presents the proof of Theorem~\ref{TGaussian}.
Section~\ref{ASecTGR} presents the proof of Theorem~\ref{TGR}.
Section~\ref{ASecEx} collects the proofs for the results in Section~\ref{SecEx}.

\medskip

For any regressor $V_1 = v_1(T, S)$ with $v_1 \in \bar{\mathcal{M}}_n$, define the regression estimator 
\begin{align*}
& \hat F_{Y|TV_1}(y|t,u) = \frac{ n^{-1}\sum_{i=1}^n {\bf 1}{\{ Y_i\leq y \}} K_h(T_i-t) K_h(v_1(T_i, S_i) - u)}{ n^{-1}\sum_{i=1}^n K_h(T_i-t) K_h(v_1(T_i, S_i) - u)} 
\equiv \frac{\hat g_1(y,t,u)}{\hat f_1t,u)}. 
\end{align*}
Denoting $\hat f_1 = \hat f_1t,u) = \hat f_{TV_1}(t,u)$ and $f_0 = f_{TV}(t,u)$, linearize $\hat F_{Y|TV_1}(y|t,u) = $
\begin{align}
\frac{\hat g_1}{ f_0} + \frac{F_{Y|TV}(y|t,u)}{f_0} \big( f_0 - \hat f_1 \big) + \frac{f_0 - \hat f_1}{f_0} \Big( \hat F_{Y|TV_1}(y|t,u) - F_{Y|TV}(y|t,u)\Big).
\label{linear1}
\end{align}
The first two terms will dominate the first-order asymptotics of $\hat F_{Y|TV_{1}}(y|t,u) - \hat F_{Y|TV_{2}}(y|t,u)$ and the third term will be collected in a smaller-order term $so2$.
That is,  $\hat F_{Y|TV_{1}}(y|t,u) - \hat F_{Y|TV_{2}}(y|t,u) = $
\begin{align}
\frac{1}{n}\sum_{j=1}^n \left( {\bf 1}{\{Y_j \leq y\}} - F_{Y|TV}(y|t, u) \right) \frac{K_h(T_j - t)}{f_{TV}(t,u)} \left( K_h(v_1(T_j, S_j) - u) - K_h(v_2(T_j, S_j) - u) \right) + so2.
\label{eqlinear}
\end{align}

\subsection{Proof of Theorem~\ref{TGaussian} (Observable regressors)}
\label{ASecTGaussian}
By the stochastic equicontinuity in Lemma~\ref{LBCLT1}, the linearization in (\ref{linear1}), and Result \ref{PKunif}, 
\begin{align*}
&\frac{1}{n}\sum_{i=1}^n \left( \hat F_{Y|TV}(y|t, V_i) - F_{Y|TV}(y|t, V_i) \right) W_i \pi_i
\\
&= \int \frac{1}{n}\sum_{j=1}^n \frac{{\bf 1}{\{Y_j \leq y\}} - F_{Y|TV}(y|t, v)}{f_{TV}(t, v)} K_h(T_j - t)K_h(v - V_j)  \mathbb{E}[W|V=v] \pi(v) dF_V(v)   \\
&\ \ \ + O_p\left( \left( \sqrt{\frac{\log n}{nh^{d}}} + h^{r} \right)^2 \right)+ o_p\left(n^{-1/2}\right) \\
&= \frac{1}{n}\sum_{j=1}^n \frac{1}{\sqrt{h^{d_t}}}\psi_{tjn}(y;V, W\pi)  +  O_p\left( \frac{\log n}{nh^{d}} + h^{2r}\right) + o_p\left(n^{-1/2}\right), \mbox{where}
\\
\psi_{tin}(y;V, W\pi) &\equiv \ \frac{\sqrt{h^{d_t}} K_{h}\big(T_i-t\big)}{f_{T|V}(t|V_i)}  \Big( {\bf 1}{\{Y_i \leq y\}} -  F_{Y|TV}(y|t,V_i) \Big)  \mathbb{E}\big[W\big| V = V_i \big] \pi(V_i).
\end{align*}
Then
$\sqrt{nh^{d_t}} \Big( \hat F_{Y(t)}(y;V,W\hat \pi) - F_{Y(t)}(y;V,W\pi) \Big)
= n^{-1/2}\sum_{i=1}^n \psi_{tin}(y;V, W\pi) + o_p(1)$, uniformly in $y \in \mathcal{Y}$ and $t \in \mathcal{T}^*$.

To compute the leading bias, $ \mathbb{E}[\psi_{tjn}(y;V, W\pi) ]/\sqrt{h^{d_t}} = h^r\mathfrak{B}_{LC} + o(h^r)$, 
where 
$\mathfrak{B}_{LC} \equiv \int \mathsf{B}_{LC}(t,v) \mathbb{E}[W|V=v] \pi(v) f_V(v) dv$ with the leading bias of the second-step local constant estimator 
\begin{align}
\mathsf{B}_{LC}(t,v) \equiv\ & \Big\{ \nabla_t^{(r)} \big( F_{Y|TV}(y|t, v) f_{TV}(t,v)  \big) - F_{Y|TV}(y|t, v) \nabla_t^{(r)} f_{TV}(t,v)\notag \\
&+ \nabla_v^{(r)} \big( F_{Y|TV}(y|t, v) f_{TV}(t,v)  \big) - F_{Y|TV}(y|t, v) \nabla_v^{(r)} f_{TV}(t,v)
\Big\} \frac{\mu_r}{r!}\frac{1}{f_{TV}(t, v)},  \label{eBiasLC}
\end{align}
letting $\nabla_v^{(r)}$ be the vector $r^{th}$-order differential operator with respective to $v$.

\begin{Lemma}[Functional Central Limit Theorem]
The process $n^{-1/2}\sum_{i=1}^n \psi_{tin}(\cdot; V,W\pi)$ weakly converges to a Gaussian process $\mathbb{G}_t(\cdot)$ given in Theorem~\ref{TGaussian}, for any $t \in \mathcal{T}^*$.
\label{LDHS}
\end{Lemma}
{\bf Proof of Lemma~\ref{LDHS}}\\
Denote $W_V(v) = \mathbb{E}[W|V = v]$.
For all $\omega \in \Omega$, the triangular array $f_{ni}(\omega, y) = n^{-1/2}\psi_{tin}(y)  = \big( {\bf 1}{\{Y_i(\omega) \leq y\}} - F_{Y|TV}(y|t,V_i(\omega)) \big) (nh^{d_t})^{-1/2} k\big((T_i(\omega)-t)/h\big) W_V(V_i(\omega))/f_{T|V}(t|V_i(\omega))$ are independent within rows.
Define the $n\times 1$ vector $f_n(\omega, y) = \big(f_{n1}(\omega, y),..., f_{nn}(\omega, y) \big)'$ and the random set $\mathcal{F}_{n\omega} = \big\{ f_{n}(\omega, y): y\in \mathcal{Y} \big\}$.   
We skip the subscript $t$ and let $d_t=1$ for notational ease without loss of clarity and generality.
We check the conditions for the functional CLT, Theorem 10.6 in \cite{Pollard}.
\begin{itemize}
\item[(i)] The triangular array processes $\{f_{ni}(\omega, y)\}$ are manageable with respect to the envelopes $F_{ni}(\omega) =  (nh^{d_t})^{-1/2}K\big((T_i(\omega)-t)/h\big)   W_V(V(\omega)/f_{T|V}(t|V_i(\omega))$.
First, $\{ {\bf 1}{\{ Y_i\leq y \}}: y\in \mathcal{Y}, i=1,...,n \}$ and $\{ F_{Y|TV}(y|t,V_i): y\in \mathcal{Y}, i=1,...,n \}$ are manageable by the fact that they are monotone increasing in $y$ (p.221 in \cite{Kosorok}).
And $F_n(\omega)= (F_{n1},..., F_{nn})^\top$ is a $\mathbb{R}^n$-valued function on the underlying probability space.
Then (i) is proved by applying Lemma E1 in \cite{AndrewsShi}.
\end{itemize}
Before we proceed to check the next conditions, it will be convenient to calculate the following expectations.
Define $V(y,T,V) = \big( F_{Y|TV}(y|T,V)- F_{Y|TV}(y|t,V) \big) f_{T|V}(T|V)$.
For notational simplicity, let $F_{y|tV} = F_{Y|TV}(y|t, V)$ and $f_{t|V} = f_{T|V}(t|V)$.
By assumption, $\frac{\partial^r}{\partial T^r} V(y,T,V)$ is bounded uniformly over $y,T,V$, 
and $f_{T|V}$ is uniformly bounded away from zero.
Then
\begin{align*}
\mathbb{E}f_{ni}(y) &= \sqrt{\frac{h^{d_t}}{n}} \mathbb{E}\left[ \int k(u) \Big( F_{Y|TV}(y|t+uh,V)- F_{Y|TV}(y|t,V) \Big) f_{T|V}(t+uh|V)du \ \frac{W_V(V)}{f_{T|V}(t|V)} \right]  \\
&= \sqrt{\frac{h^{d_t}}{n}} \frac{h^r}{r!} \int k(u)u^r du \mathbb{E}\left[ \frac{\partial^r}{\partial T^r} V(y,T,V)\Big|_{T=t}   \frac{W_V(V)}{f_{T|V}(t|V)} \right] = \bar O\Big(h^r  \sqrt{\frac{h^{d_t}}{n}}\Big).
\end{align*}

For any $t,s \in \mathcal{T}$, 
\begin{align}
&n \mathbb{E} \big[ f_{ni}(y_1,t)f_{ni}(y_2,s)\big]\notag \\
&= \mathbb{E} \bigg[ \Big( F_{y_1|TV} - F_{y_1|tV}F_{y_2|TV}- F_{y_1|TV}F_{y_2|sV} + F_{y_1|tV} F_{y_2|sV}\Big) \frac{1}{h^{d_t}} k\Big( \frac{T-t}{h} \Big) k\Big( \frac{T-s}{h} \Big) \frac{W^2_{V}(V)}{f_{t|V} f_{s|V}} \bigg]  \notag\\
&= \mathbb{E}\bigg[ \frac{W^2_{V}(V)}{f_{t|V} f_{s|V}} \int k(u)k\big(u+\frac{t-s}{h}\big) \Big( F_{Y|TV}(y_1|t+uh,V) - F_{Y|TV}(y_1|t,V) F_{Y|TV}(y_2|t+uh,V) \notag \\
&\ \ \ \  - F_{Y|TV}(y_1|t+uh,V)F_{Y|TV}(y_2|s,V) + F_{Y|TV}(y_1|t,V)F_{Y|TV}(y_2|s,V) \Big) f_{TV}(t+uh,V) du  \bigg]  \notag\\
&= \mathbb{E}\bigg[ \Big( F_{Y|TV}(y_1| t,V) - F_{Y|TV}(y_1|t,V) F_{Y|TV}(y_2|t,V) \Big)\frac{W^2_{V}(V)}{f_{T|V}(s|V)}\bigg] \frac{1}{h^{d_t}} \int k(u)k\Big(u - \frac{s - t}{h}\Big) du  + O(h)
\label{acov}
\end{align}
uniformly in $y_1\leq y_2 \in \mathcal{Y}$.

\begin{enumerate}
\item[(ii)] Define $\mathcal{Z}_n(y) = \sum_{i=1}^n \big( f_{ni}(y) - \mathbb{E}f_{ni}(y) \big)$.  Let $y_1 \leq y_2 \in \mathcal{Y}$. 
Using (\ref{acov}), the covariance kernel of the limiting Gaussian process is
\begin{align}
&\lim_{n\rightarrow \infty} \mathbb{E} \mathcal{Z}_n(y_1) \mathcal{Z}_n(y_2) 
= \lim_{n\rightarrow \infty}  \mathbb{E}\big[ \psi_{tin}(y_1)\psi_{tin}(y_2)\big] - \mathbb{E}\big[ \psi_{tin}(y_1) \big] \mathbb{E}\big[ \psi_{tin}(y_2) \big] \notag \\
&=\mathbb{E}\bigg[ \Big( F_{Y|TV}(y_1| t,V) - F_{Y|TV}(y_1|t,V) F_{Y|TV}(y_2|t,V) \Big) \frac{W_V^2(V)}{f_{T|V}(t|V)} \bigg] \int k^2(u) du. 
\label{PTGacov}  
\end{align}
\item[(iii)] Using (\ref{acov}), 
\begin{align*}
\sum_{i=1}^n  \mathbb{E}F_{ni}^2 = \mathbb{E}\bigg[ \frac{W_V^2(V)}{f_{T|V}(t|V)}  \bigg] \int k^2(u) du + O(h).
\end{align*}

\item[(iv)] For any $\epsilon > 0$, $\sum_{i=1}^n  \mathbb{E}F_{ni}^2\ {\bf 1}\big\{ F_{ni} > \epsilon \big\} \rightarrow 0$ holds.
This is because ${\bf 1}\big\{ F_{ni} > \epsilon\big\} =0$ for $n$ large enough, by assuming $k$ is bounded, $f_{T|V}$ is bounded away from zero, and $\sqrt{nh^{d_t}} \rightarrow \infty$.
\item[(v)] Uniform in $y_1, y_2$, $n \mathbb{E}\big| f_{ni}(y_1) - f_{ni}(y_2) \big|^2 \longrightarrow $
\begin{align*} 
\rho(y_1, y_2)^2  =  \int k^2(u)du\mathbb{E}\left[  \frac{W_V^2(V)}{f_{T|V}(t|V)} \Big( F_{y_2|tV} - F_{y_1|tV} - \big(F_{y_2|tV} - F_{y_1|tV}\big)^2 \Big) \right] .
\end{align*}
Therefore, uniformly in $y_1, y_2$,
$\rho_n(y_1, y_2) = \big( \sum_{i=1}^n  \mathbb{E}\big| f_{ni}(y_1) - f_{ni}(y_2) \big|^2 \big)^{1/2}\rightarrow \rho(y_1, y_2)$.
\end{enumerate}

\subsection{Proof of Theorem~\ref{TGR} (Generated regressors)}
\label{ASecTGR}
Decompose \begin{align*}
&\frac{1}{n} \sum_{i=1}^n \big( \hat F_{Y|T\hat V}(y|t, \hat v(T_i, S_i)) - F_{Y|TV}(y|t, v_0(T_i, S_i)) \big) W_i\pi_i \\
&= 
\frac{1}{n} \sum_{i=1}^n \big( \hat F_{Y|T\hat V}(y|t, \hat v(T_i, S_i)) - \hat F_{Y|T\hat V}(y|t, v_0(T_i, S_i))+  F_{Y|T V}(y|t, v_0(T_i, S_i)) - F_{Y|TV}(y|t, \hat v(T_i, S_i))
\\
&\ \ \ + \hat F_{Y|T\hat V}(y|t, v_0(T_i, S_i)) - F_{Y|TV}(y|t, v_0(T_i, S_i)) 
 +  F_{Y|TV}(y|t, \hat v(T_i, S_i)) - F_{Y|TV}(y|t, v_0(T_i, S_i)) \big) W_i \pi_i\\
&=  T_{1,n} + T_{2,n} + O_p(\|\hat v - v_0\|^2) + so1,\ \mbox{where}  \\[3pt]
&T_{1,n}  \equiv  \frac{1}{n} \sum_{i=1}^n \big( \hat F_{Y|T\hat V}(y|t, v_0(T_i, S_i)) - F_{Y|TV}(y|t, v_0(T_i, S_i)) \big) W_i\pi_i \\
&T_{2,n} \equiv \frac{1}{n} \sum_{i=1}^n \nabla_v F_{Y|TV}(y|t, v_0(T_i, S_i))' \big( \hat v(T_i, S_i) - v_0(T_i, S_i) \big) W_i\pi_i\\
& so1 \equiv \frac{1}{n} \sum_{i=1}^n \big( \hat F_{Y|T\hat V}(y|t, \hat v(T_i, S_i)) - \hat F_{Y|T\hat V}(y|t, v_0(T_i, S_i)) \\
&\ \ \ \ \ \ \ +  F_{Y|T V}(y|t, v_0(T_i, S_i)) - F_{Y|TV}(y|t, \hat v(T_i, S_i)) \big) W_i\pi_i. 
\end{align*}
$T_{1,n}$ is the estimation error from the Step~2 nonparametric regression in (\ref{sumT1}) and the first-step generated regressor in  (\ref{sumTGR}).  
$T_{2,n}$ contributes (\ref{sumT2}).
By Lemma~\ref{LMRS4} with $B_{yt}(w, v) = \nabla_v F_{Y|TV}(y|t, v) w$,
\begin{align}
T_{2,n} &= \mathbb{E}\big[ \nabla_v F_{Y|TV}(y|t, v_0(T, S))' \big( \hat v(T, S) - v_0(T, S) \big) W \pi\big] + O_p\big(n^{-\kappa_{13}}\big).  \label{T2n} 
\end{align}

The first part of $T_{1,n}$ is $n^{-1} \sum_{i=1}^n W_i\pi_i\big( \hat F_{Y|T\hat V}(y|t, V_i) - \hat F_{Y|TV}(y|t,V_i) \big)$ in (\ref{sumTGR}).
In the following calculation, we use Fubini's theorem to interchanges the order of integration and extract the generated regressor from its {\it regressor} role to the argument of the kernel estimator.
The idea is to make use of the marginal integration in Step 3 that averages over $V_i$ and view the generated regressor $\hat V$ as arguments of the kernel functions.
Finally, a standard Taylor series expansion linearizes the partial mean to be a linear functional of the estimation error of the generated regressor $\hat V - V$.
By (\ref{eqlinear}), Lemma~\ref{LMRS2} and Lemma~\ref{LMRS3}, 
\begin{align}
\frac{1}{n} &\sum_{i=1}^n W_i\pi_i \Big( \hat F_{Y|TV_1}(y|t, V_i) - \hat F_{Y|TV_2}(y|t,V_i) \Big)  \notag \\
=&\ \frac{1}{n} \sum_{i=1}^n \frac{W_{i}\pi_i }{f_{TV}(t,V_i)} \frac{1}{n} \sum_{j=1}^n \Big( {\bf 1}{\{Y_j \leq y\}} - F_{Y|TV}(y|t,V_i)\Big)\notag\\
&\times K_h(T_j- t) \Big( K_h(v_1(T_j, S_j) - V_i) - K_h(v_2(T_j, S_j) - V_i)\Big) + so2 \notag \\
=&\ \frac{1}{n} \sum_{j=1}^n K_h(T_j- t) \mathbb{E}\bigg[ \frac{W {\bf 1}{\{f_{TV}(t, V) \geq c\}}}{f_{TV}(t, V)}\Big( {\bf 1}{\{Y_j \leq y\}}  - F_{Y|TV}(y|t, V)\Big)\notag\\
&\times \Big( K_h(v_1(T_j, S_j) - V) - K_h(v_2(T_j, S_j) - V)\Big) \bigg] + O_p(n^{-\kappa_{11}}) + so2 \notag\\
=&\ \frac{1}{n} \sum_{j=1}^n \bigg\{ \frac{\mathbb{E}[W|V=v_1(T_j, S_j)]}{f_{T|V}(t| v_1(T_j, S_j))}\Big( {\bf 1}{\{Y_j \leq y\}} - F_{Y|TV}(y|t, v_1(T_j, S_j))\Big) {\bf 1}{\{f_{TV}(t,v_1(T_j, S_j)) \geq c\}} \notag\\
&- \frac{\mathbb{E}[W|V=v_2(T_j, S_j)]}{f_{T|V}(t| v_2(T_j, S_j))}\Big( {\bf 1}{\{Y_j \leq y\}}- F_{Y|TV}(y|t, v_2(T_j, S_j))\Big) {\bf 1}{\{f_{TV}(t,v_2(T_j, S_j)) \geq c\}}  \bigg\} \notag \\
& \times K_h(T_j- t)  + O_p(n^{-\kappa_{11}} + h^{r}\|v_1-v_2\|_\infty/\sqrt{nh^{d_t}}) + so2 \notag\\
=&\ \mathbb{E}\bigg[ \bigg\{ \frac{\mathbb{E}[W|V=v_1(t, S)]}{f_{T|V}(t| v_1(t, S))}\Big( F_{Y|TS}(y|t,S)- F_{Y|TV}(y|t, v_1(t, S))\Big)\notag\\
&- \frac{\mathbb{E}[W|V=v_2(t, S)]}{f_{T|V}(t| v_2(t, S))}\Big( F_{Y|TS}(y|t,S)- F_{Y|TV}(y|t, v_2(t, S))\Big) \bigg\} \notag \\
& \times f_{T|S}(t|S)   {\bf 1}{\{f_{TV}(t,v_1(t, S) \geq c, f_{TV}(t,v_2(t, S)\geq c \}}\bigg] + O_p(n^{-\kappa_{11}} + n^{-\kappa_{12}} + h^{r}\|v_1-v_2\|_\infty) + so2 \notag\\
=&\  \mathbb{E}\bigg[ \Big( v_1(t, S) - v_2(t, S) \Big)' \bigg\{ F_{Y|TS}(y|t,S)  \Big( -\frac{\mathbb{E}[W|V=v_2(t, S)]}{f_{T|V}(t|v_2(t, S))} \nabla_v f_{T|V}(t| v) + \nabla_v \mathbb{E}[W|V=v] \Big)\notag\\ 
& -  \mathbb{E}[W|V=v_2(t, S)] \nabla_v F_{Y|TV}(y|t, v)  \notag\\
&+ \mathbb{E}[W|V=v_2(t, S)] F_{Y|TV}(y|t,v_2(t, S)) \frac{\nabla_v f_{T|V}(t|v)}{f_{T|V}(t|v_2(t, S))}  \notag\\
&- F_{Y|TV}(y|t, v_2(t, S))\nabla_v \mathbb{E}[W|V=v] \bigg\}\bigg|_{v=v_2(t, S)} \frac{f_{T|S}(t|S)}{f_{T|V}(t|v_2(t, S))} \pi(v_2(t,S))\bigg]\notag\\
&+ O\big( \|v_1 - v_2 \|^2_\infty \big)  +O_p(n^{-\kappa_{11}} + n^{-\kappa_{12}} + h^{r} \|v_1-v_2\|_\infty ) + so2, \notag
\end{align}
where the second equality uses Lemma~\ref{LMRS2} and the fourth equality uses Lemma~\ref{LMRS3}.
In the third equality, the event $\{ f_{TV}(t,v_1(t,S)) \geq c, f_{TV}(t,v_1(t+hu,S)) < c \}$ has asymptotic probability zero. 
The fourth equality is because the events $\{f_{TV}(t,V_{1j}) \geq c, f_{TV}(t,V_{2j}) < c \}$
and $\{f_{TV}(t,V_{1j}) \geq c, f_{TV}(t,V_{2j}) < c \}$ have asymptotic probability zero.
The last equality is by a Taylor expansion for any $v_1, v_2 \in \bar{\mathcal{M}}_n$.

Together with the other term in $T_{1,n}$ $n^{-1} \sum_{i=1}^n \hat F_{Y|TV}(y|t, V_i) - F_{Y|TV}(y|t, V_i)$ by Theorem~\ref{TGaussian} and $T_{2,n}$ in (\ref{T2n}), Theorem~\ref{TGR} (I) are derived.
For (I\!I), the above results hold by dropping $T$ in $v_0(S)$. 
The smaller order terms are controlled in the following.
By Lemma~\ref{LMRS}, uniformly in $v_1, v_2 \in \bar{\mathcal{M}}_n$, $\big\| \hat g_1 - \hat g_{2}\big\|_\infty =$
\begin{align}
& \sup_{y,t,u}  \big|\mathbb{E}\big[  {\bf 1}{\{ Y\leq y \}} K_h(T-t)  \big( K_h(v_1(T, S) - u) - K_h(v_2(T, S) - u)\big) \big] \big| + O_p(n^{-\kappa_{14}}) \notag \\
&\leq \sup_{y,t,u}\mathbb{E}\Big[ \Big| {\bf 1}{\{ Y\leq y \}} K_h(T-t)  \sum_{l=1}^{d_v}  \frac{1}{h_l}k'\Big(\frac{V_{2l}-u_l}{h_l}\Big)\prod_{j=1,j \neq l}^{d_v} \frac{1}{h}k\Big(\frac{V_{2j}-u_j}{h}\Big) \Big| \Big]  \max_{j\in \{1,..., d_v\}}\Big\|\frac{v_{1j} - v_{2j}}{h}\Big\|_\infty 
\notag \\
&\ \ \  + \sup_{y,t,u} \mathbb{E}\bigg[ \bigg| {\bf 1}{\{ Y\leq y \}} K_h(T-t)  \frac{1}{2}\bigg( \sum_{l=1}^{d_v} \frac{\|k''\|_\infty}{h_l} \ \prod_{j\neq l} \frac{1}{h_l}k\Big(\frac{V_{2j}-u_j}{h_l}\Big)  \notag \\
&\ \ \ +  \sum_{l,m=1, l \neq m}^{d_v} \frac{\|k'\|_\infty}{h_m} \frac{1}{h_l}k'\Big(\frac{V_{2l}-u_l}{h_l}\Big)\prod_{j \neq l,m} \frac{1}{h}k\Big(\frac{V_{2j}-u_j}{h}\Big) \bigg) \bigg| \bigg]  \max_{j\in \{1,..., d_v\}}\Big\|\frac{v_{1j} - v_{2j}}{h}\Big\|_\infty^2 + O_p(n^{-\kappa_{14}}) \notag\\
&= O_p\Big(\max_{j\in \{1,..., d_v\}}\Big\|\frac{v_{1j} - v_{2j}}{h}\Big\|_\infty  \Big) + O_p(n^{-\kappa_{14}}) = O_p\big( n^{- (\delta-\eta)_{\min}} + n^{-\kappa_{14}} \big)
\label{SE0}
\end{align} 
given 
$\|\nabla^2 K\|_\infty < \infty$ and $\delta > 2\eta $.

The remainder term of order $n^{-\kappa_1}$ is from the stochastic equicontinuity argument.
The smaller-order terms from linearizing the estimator are governed by $O_p(n^{-\kappa_2})$.
By Result \ref{PKunif} and (\ref{SE0}),
the smaller order term in $\hat F_{Y|TV_{1}} - \hat F_{Y|TV_2}$ is
\begin{align*}
\big\| so2 \big\|_\infty &\leq \Big\| \frac{1}{f} \big( f - \hat f_1 \big) \big( \hat F_{Y|TV_1} - F_{Y|TV}\big) \Big\|_\infty + \Big\| \frac{1}{f} \big( f - \hat f_{2} \big) \big( \hat F_{Y|TV_{2}} - F_{Y|TV}\big) \Big\|_\infty  \\
&= O_p\Big( \Big\| \frac{1}{f} \big( f - \hat f+ \hat f - \hat f_1 \big) \big( \hat F_{Y|TV_1} - \hat F_{Y|TV}  + \hat F_{Y|TV}  - F_{Y|TV}\big) \Big\|_\infty \Big) \\
&= O_p\bigg( \Big( \sqrt{\frac{\log n}{nh^{d}}} + h^r + \frac{\|v_1 - v_0\|_\infty}{h} + n^{-\kappa_{14}}\Big)^2 \bigg) = O_p(n^{-\kappa_2}). 
\end{align*}

Decompose 
$so1 = so11 + so12$, where 
\begin{align*}
so11 \equiv &\  \frac{1}{n} \sum_{i=1}^n \big( \hat F_{Y|T\hat V}(y|t, \hat v(T_i, S_i)) - \hat F_{Y|T\hat V}(y|t, v_0(T_i, S_i)) \\
&+  \hat F_{Y|T V}(y|t, v_0(T_i, S_i)) - \hat F_{Y|TV}(y|t, \hat v(T_i, S_i)) \big) W_i\pi_i 
\\
so12 \equiv &\  \frac{1}{n} \sum_{i=1}^n \big( \hat F_{Y|TV}(y|t, \hat v(T_i, S_i)) - \hat F_{Y|TV}(y|t, v_0(T_i, S_i)) \\
&+  F_{Y|T V}(y|t, v_0(T_i, S_i)) - F_{Y|TV}(y|t, \hat v(T_i, S_i)) \big) W_i\pi_i \\
=&\  \frac{1}{n} \sum_{i=1}^n \left( \nabla_v\hat F_{Y|TV}(y|t, v_0(T_i, S_i))  - \nabla_v F_{Y|TV}(y|t, v_0(T_i, S_i))\right) \\
&\times \left( \hat v(T_i, S_i)- v_0(T_i, S_i) \right)  W_i\pi_i + O_p(\|\hat v-v_0\|^2_\infty) \\
=&\  O_p\left( \left( \sqrt{\frac{\log n}{nh^{d+2}}} + h^r \right) n^{-\delta} + n^{-2\delta}\right)  = O_p(n^{-\kappa_2}),
\end{align*}
uniformly in $(y,t)$, by Result \ref{PKunif}. 

For any function $f(u) = f_A(u) + f_B(u)$, $f(u_1) - f(u_2) = f_A(u_1) - f_A(u_2) + O_p(\|f_B\|_\infty) = f_A'(u_2)(u_1-u_2) + O_p(\|u_1-u_2\|_\infty^2 + \|f_B\|_\infty)$. 
Let $so2 = f_B$ in (\ref{eqlinear}).
So by (\ref{eqlinear}) and Lemma~\ref{LMRS}, 
\begin{align*}
&so11\\
&= 
\frac{1}{n}\sum_{i=1}^n \frac{1}{n}\sum_{j=1}^n \bigg\{ \left( {\bf 1}{\{Y_j \leq y\}} - F_{Y|TV}(y|t, V_i) \right) \frac{K_h(T_j - t)}{f_{TV}(t,V_i) } \bigg( - \nabla K_h(v_1(T_j, S_j) - V_i) \\
&\ \ \ + \nabla K_h(v_2(T_j, S_j) - V_i) \bigg) + \nabla_u\bigg(\left( {\bf 1}{\{Y_j \leq y\}} - F_{Y|TV}(y|t, u) \right) \frac{K_h(T_j - t)}{f_{TV}(t,u)}\bigg) \bigg|_{u=V_i} \big(K_h(v_1(T_j, S_j) - V_i) \\
&\ \ \ - K_h(v_2(T_j, S_j) - V_i) \big) \bigg\} (\hat V_i - V_i) W_i\pi_i+ O_p(\|\hat v-v_0\|^2_\infty + \| so2\|_\infty ) \\
&= O_p\bigg( \bigg\{ \sup_{y,t,v}\left\| \mathbb{E}\left[  \left( {\bf 1}{\{Y \leq y\}} - F_{Y|TV}(y|t, v) \right) \frac{K_h(T - t)}{f_{TV}(t,v)} \big( - \nabla K_h(\hat v(T, S) - v) + \nabla K_h(v_0(T, S) - v) \big)  \right] \right\| \\
&\ \ \ +  \sup_{y,t,v} \left\| \mathbb{E}\left[ \nabla_u \left(\left( {\bf 1}{\{Y \leq y\}} - F_{Y|TV}(y|t, u) \right) \frac{K_h(T - t)}{f_{TV}(t,u)}\right)\bigg|_{u=v} \big( K_h(\hat v(T, S) - v) - K_h(v_0(T, S) - v) \big)  \right] \right\| \\
&\ \ \ + n^{-\kappa_{14} + \eta} + n^{-\kappa_{14}} \bigg\} \|\hat v - v_0 \|_\infty \bigg) + O_p( \|\hat v-v_0\|^2_\infty + \| so2\|_\infty )\\
 &= O_p\left( \frac{\|\hat v-v_0\|^2_\infty}{h^2} + n^{-\kappa_{14} + \eta - \delta} + \| so2\|_\infty \right)  = O_p(n^{-\kappa_{14} + \eta -\delta} + n^{-\kappa_2}).
\end{align*}

\subsection{Trimming}
\label{ASecTrim}
We show (\ref{sumTrim}) $n^{-1} \sum_{i=1}^n \hat F_{Y|T\hat V}(y|t, \hat V_i) W_i (\hat \pi_i - \pi_i) = o_p\left(n^{-1/2}\right)$.
Let $f$ be a generic function of $V$.
Suppose the infeasible trimming function $\pi_i = {\bf 1}\{f(V_i) \geq c\} = {\bf 1}\{V_i \in \mathcal{V}^*\}$.
Let the estimated trimming function $\hat \pi_i = {\bf 1}\{\hat f(\hat V_i) \geq c\} = {\bf 1}\{\hat V_i \in \hat{\mathcal{V}}^*\}$. 
Rewrite 
\[
\hat \pi_i - \pi_i = {\bf 1}\{\hat f(\hat V_i) \geq c, f(V_i) < c\} - {\bf 1}\{\hat f(\hat V_i) < c, f(V_i) \geq c\}.
\]
We study the first term and the second term follows the same reasoning.
Let $a_n$ be a positive sequence converging to zero satisfying  $a_n^{-1}  \sup_{v \in \mathcal{V}^*} |\hat f(v) - f(v)|= o_p(1)$ and $ a_n^{-1}\|\hat v -v_0\|_\infty = o_p(1)$.
\begin{align*}
{\bf 1}\{\hat f(\hat V_i) \geq c, f(V_i) < c\} &= {\bf 1}\{\hat f(\hat V_i) \geq c + a_n > c > f(V_i)\} + {\bf 1}\{\sigma + a_n > \hat f(\hat V_i) \geq c > c - a_n > f(V_i) \} 
\\
&+ {\bf 1}\{c + a_n > \hat f(\hat V_i) \geq c > f(V_i) \geq c - a_n \} 
\\
&\leq 2\  {\bf 1}\{| \hat f(\hat V_i) - f(V_i) | \geq a_n\} + {\bf 1}\{c >  f(V_i) \geq c- a_n\}.
\end{align*}
By the chosen $a_n$, the first term equals zero $w.p.a.1$.
Because $c$ is a fixed constant, the event $\{c >  f(V_i) \geq c- a_n\}$ has asymptotic probability zero.
The second term is zero $w.p.a.1$.
Therefore, $\hat \pi_i - \pi_i$ equals zero $w.p.a.1$.

\subsection{Proof of Theorem~\ref{TLP}}
\label{ASecTLP}

\begin{Assumption}[Continuity Assumption 4(i)(ii) in MRS16]
For any $v_1 \in \mathcal{C}_M^\alpha(\mathcal{S}_v)$, the conditional expectation $\tau^B(v,y, v_1) = \mathbb{E}[\rho(T,S)|v_1(T,S) = v]$
with $\rho(T,S) = F_{Y|TS}(y|T,S) - F_{Y|TV}(y|T, v_0(T,S)) $ exists and is $q+1$ times differentiable with respect to its first argument, with derivatives that are uniformly bounded in absolute value over $(v,y,v_1)$, and satisfies 
$|\tau^B(v,y, v_1) - \tau^B(v,y, v_2) | \leq C_B^\ast \|v_1 - v_2|_\infty$ a.s. for all $v_1, v_2 \in \mathcal{C}_M^\alpha(\mathcal{S}_v)$, $y\in \mathcal{Y}$, and a constant $C_B^\ast > 0$.
\label{Acon}
\end{Assumption}

\paragraph{Partial mean}

Let $w(t) \equiv (1, t, ..., t^p)^\top$ and $N_h(t,v)\equiv \mathbb{E}\big[w( (T_i - t)/h, (V_i -v)/h) w( (T_i - t)/h, (V_i -v)/h)^\top K_h(T_i - t) K_h(V_i - v)  \big]$.
It is well-known in the literature, e.g., (A.25) in MRS12, that the stochastic expansion uniformly over $t,v$ 
\begin{align*}
&\hat F_{Y|TV}(y|t,v) - F_{Y|TV}(y|t,v) \\
= \ & \frac{1}{n}\sum_{i=1}^n e_1^\top
 N_h(t,v)^{-1} w( (T_i - t)/h, (V_i -v)/h) K_h(T_i - t) K_h(V_i - v) \big({\bf 1}\{Y_i \leq y\} - F_{Y|TV}(y|T_i,V_i)\big)
\\&+ h^{p+1}\mathsf{B}_{LP}(t,v) (1+ o(1) ) + O_p(\log(n)/(nh^d)). 
\end{align*}
To conserve space, we do not copy the expression for the leading bias $h^{p+1}\mathsf{B}_{LP}(t,v)$ for a odd $p$.

The partial mean of a local polynomial estimator 
\begin{align*}
&\int ( \hat F_{Y|TV}(y|t,v) - F_{Y|TV}(y|t,v)) f_V(v) dv \\
=\ & \frac{1}{n}\sum_{i=1}^n 
\int f_{TV}(t,v)^{-1} K_h(T_i - t) K_h(V_i - v) ({\bf 1}\{Y_i \leq y\} - F_{Y|TV}(y|T_i,V_i)) f_V(v) dv
\\&+  h^{p+1}\int \mathsf{B}_{LP}(t,v) f_V(v) dv (1+o(1)) + O_p(\log(n)/(nh^d)) \\
=\ &
\frac{1}{n}\sum_{i=1}^n  K_h(T_i - t) f_{T|V}(t|V_i)^{-1}   ({\bf 1}\{Y_i \leq y\} - F_{Y|TV}(y|T_i,V_i))
\\&+ h^{p+1}\int \mathsf{B}_{LP}(t,v) f_V(v) dv (1+o(1)) + O_p(\log(n)/(nh^d)). 
\end{align*}

Following the proof of Theorem~\ref{TGaussian}, we obtain the same first-order asymptotic.

\paragraph{Generated regressors}

We modify the proof for the local constant estimator in Section~\ref{ASecTGR}.
We use the stochastic expansion in Theorem 2 in MRS16 to analyze the influence of $\hat V$ as a regressor in $T_{1,n}$,
where the remainder terms is $O_p(n^{-\kappa^\ast})$ with 
$\kappa^\ast \equiv \min\{(1-d\eta)/2 + \delta - \eta - (\delta + \xi^*)d_1/(2\alpha), p \eta + \delta, 2\delta -\eta\}$. 
That is, $\sup_{t,v,y} \big| n^{-1}\sum_{i=1}^n \hat F_{Y|T\hat V}(y|t, V_i) - \hat F_{Y|TV}(y|t, V_i) - (\varphi_n^A(t,v,y, \hat v)  + \varphi_n^B(t,v,y, \hat v)  ) \big| = O_p(n^{-\kappa^\ast})$.
Then we can obtain $REG_{yt}$ by computing the following.
\begin{align*}
&\int_\mathcal{V} \varphi_n^A(t,v,y, v_1) \mathbb{E}[W|V=v ]  f_V(v) dv \\
=\ & \int_\mathcal{V}  e_1^\top N_h(t,v)^{-1} \mathbb{E}[  w( (T - t)/h, (V -v)/h)   K_h(T-t) K_h(V-v) \nabla_vF_{Y|TV}(y|t,V) \\
&
(v_1(T,S) - v_0(T,S))] \mathbb{E}[W|V=v ] f_V(v) dv\\
=\ & \int_\mathcal{V} e_1^\top
 N_h(t,v)^{-1}  \int_\mathcal{S} \int_\mathcal{T} w( (T - t)/h, (v_0(T,S) -v)/h)   
K_h(T-t) K_h(v_0(T,S)-v)  \nabla_vF_{Y|TV}(y|t,v_0(T,S)) \\
&
(v_1(T,S) - v_0(T,S)) f_{TS}(T,S)dTdS  \mathbb{E}[W|V=v ] f_V(v) dv
\\
=\ &   \int_\mathcal{S} \nabla_vF_{Y|TV}(y|t,v_0(t,S)) f_{TV}(t,v_0(t,S))^{-1} 
   (v_1(t,S) - v_0(t,S)) \mathbb{E}[W|V=v_0(t,S) ] f_V(v_0(t,S)) f_{TS}(t, S)dS  
\\
&+ O_p(h^r \|v_1(t,\cdot) - v_0(t,\cdot) \|_\infty)
\\
=\ &   \int_\mathcal{S} \frac{f_{T|S}(t|S)}{f_{T|V}(t|v_0(t,S))} 
\nabla_vF_{Y|TV}(y|t,v_0(t,S))    \mathbb{E}[W|V=v_0(t,S) ] (v_1(t,S) - v_0(t,S)) f_S(S)dS  
 +  O_p( n^{-r \eta - \delta}).
\end{align*}

\medskip

\begin{align*}
&\int_\mathcal{V} \varphi_n^B(t,v,y, v_1) \mathbb{E}[W|V=v ]  f_V(v) dv \\
=\ & \int_\mathcal{V} 
e_1^\top N_h(t,v)^{-1} \mathbb{E}[ w( (T - t)/h, (V -v)/h)  K_h(T - t) \nabla K_h(V - v) (v_1(T,S) - v_0(T,S)) \rho(T,S)]
\mathbb{E}[W|V=v ]  f_V(v) dv 
\\
=\ & \int_\mathcal{V} 
 f_{TV}(t,v)^{-1} \int_\mathcal{S}
 \nabla K_h(v_0(t,S) - v) (v_1(t,S) - v_0(t,S)) \rho(t,S) f_{TS}(t,S)dS
\mathbb{E}[W|V=v ]  f_V(v) dv   ( 1 +  O_p(h^r) )
\\
=\ & \int_\mathcal{S} 
 \int_\mathcal{V}
  K_h(v_0(t,S) - v) 
\nabla_v\left(\frac{\mathbb{E}[W|V=v ]}{f_{T|V}(t|v)} \right) dv  (v_1(t,S) - v_0(t,S)) \rho(t,S) f_{TS}(t,S) dS  ( 1 +  O_p(h^r) )
\\
=\ & \int_\mathcal{S} 
\nabla_v\left(\frac{\mathbb{E}[W|V=v ]}{f_{T|V}(t|v)} \right) \Big|_{v = v_0(t,S)}  (v_1(t,S) - v_0(t,S)) \rho(t,S) f_{T|S}(t|S)f_S(S) dS  +  O_p( n^{-r \eta - \delta}),
\end{align*}
where the third equality is from integration by parts.
Thus we obtain $REG_{yt}(s,v)$.

$ARG_{yt}(w,v)$ is from (\ref{T2n}).
Under Assumption~\ref{Acon}, the remainder terms 
$so11 = O_p(n^{-2\delta} + n^{-\kappa_{14} + \eta - \delta} + n^{-\kappa^\ast})$
and 
\begin{align*}
so12 
=&\  \frac{1}{n} \sum_{i=1}^n \left( \nabla_v\hat F_{Y|TV}(y|t, v_0(T_i, S_i))  - \nabla_v F_{Y|TV}(y|t, v_0(T_i, S_i))\right) \\
&\times \left( \hat v(T_i, S_i)- v_0(T_i, S_i) \right)  W_i\pi_i + O_p(\|\hat v-v_0\|^2_\infty) \\
=&\  O_p\left( \left( \sqrt{\frac{\log n}{nh^{d+2}}} + h^p \right) n^{-\delta} + n^{-2\delta}\right).
\end{align*}

Therefore we obtain the results.

\subsection{Proof in Section \ref{SecEx}}
\subsubsection{Proof of Corollary~\ref{GRGPS}}
\label{ASecEx}
Note that $V = v_0(X) = f_{T|X}(t|X)$ is not a function of $T$.
To compute the influence of estimating the GPS in Theorem~\ref{TGR}, 
\begin{align*}
\Delta_{yt}(\hat v(X)) =&\  \int_{\mathcal{X}^\dagger} \big(\hat v(x) - v_0(x)\big)' REG_{yt}\big(x,v_0(x)\big)\ dF_{X}(x) \\
&+
\int_{\mathcal{W}} \!\int_{\mathcal{X}^\dagger}
\big(\hat v(x) - v_0(x)\big)' ARG_{yt}\big(w, v_0(x)\big)  \ dF_{XW}(x,w) \\
=&\ \mathbb{E}\big[ \big( \hat f_{T|X}(t|X) - f_{T|X}(t|X) \big) A_{yt}(X, W\pi) \big], 
\end{align*}
where $A_{yt}(X, W\pi) \equiv \big\{ \nabla_v F_{Y|TV}(y|t, v) \big( W - \mathbb{E}[W|V=v] \big) + \big( F_{Y|TV}(y|t,v) - F_{Y|TX}(y|t,X) \big)\big(\nabla_vf_{T|V}(t|v) \mathbb{E}[W|V=v] \big/ f_{T|X}(t|X) -   \nabla_v \mathbb{E}[W|V=v] \big)\big\}\pi(X)\big|_{v=v_0(X)}$.
\begin{itemize}
\item[(I)]
\begin{itemize}
\item[(a)]
The linear representation of the kernel estimator is
\begin{align*}
&\hat f_{T|X}(t|x) - f_{T|X}(t|x) = \frac{1}{n}\sum_{i=1}^n \big( K_{h_1}(T_i - t) - f_{T|X}(t|x) \big) K_{h_1}(X_i - x) /  f_X(x) + R_n^v(t,x).\\
&\| R_n^v \|_\infty = O_p\Big( \|\hat f_X - f_X\|_\infty \| \hat f_{T|X} - f_{T|X} \|_\infty \Big) = O_p\bigg(\bigg( \sqrt{\frac{\log n}{nh_1^{d_x}}} + h_1^{r_1}\bigg)\bigg( \sqrt{\frac{\log n}{nh_1^{d_x+d_t}}} + h_1^{r_1}\bigg)\bigg).
\end{align*}
Then 
\begin{align*}
\Delta_{yt}(\hat v(X)) &= \mathbb{E}\big[\big( \hat f_{T|X}(t|X) - f_{T|X}(t|X) \big) A_{yt}(X, W\pi) \big] \\
&= \frac{1}{n}\sum_{i=1}^n \Big( K_{h_1}(T_i - t) - f_{T|X}(t|X_i) \Big)\mathbb{E}\big[A_{yt}(X_i, W\pi)\big|X=X_i\big] +  O_p\left( \|R_n^v\|_\infty \right) \\
&= O_p\big((nh_1^{d_t})^{-1/2} + h_1^{r_1} + \|R_n^v\|_\infty \big).
\end{align*}

The bias $\mathbb{E}\Big[ \Big( K_{h_1}(T - t) - f_{T|X}(t|X) \Big)\mathbb{E}\big[A_{yt}(X, W\pi)\big|X\big] \Big]
= h_1^{r_1} \mathfrak{B}_{GPS} + o(h_1^{r_1})$,
where
\begin{align}\mathfrak{B}_{GPS} \equiv \frac{\mu_r}{r!} \mathbb{E}\big[ \nabla^{(r)}_t f_{T|X}(t|X) \mathbb{E}[A_{yt}(X, W\pi) | X] \big]. 
\label{EBiasGPS}
\end{align}

\item[(b)] The linear representation of the MLE is
\begin{align*}
\hat f_{T|X}(t|x) - f_{T|X}(t|x) = &\
 \Big((\hat \beta - \beta_0)', \hat \sigma^2 - \sigma_0^2 \Big)\left( x \frac{\partial}{\partial \mu} \zeta(t, \mu, \sigma_0^2)\Big|_{\mu=x'\beta_0},\frac{\partial}{\partial \sigma^2_0}  \zeta(t, x'\beta_0, \sigma_0^2) \right)' \\
&+ O_p\left(\Big\|\left( (\hat \beta - \beta_0)', \hat \sigma^2 - \sigma_0^2 \right) \Big\|^2\right).
\end{align*}

\end{itemize}
\item[(I\!I)]
Assumption~\ref{Abwgps} is sufficient for $O_p \big( h_1^{r_1} + \|R_n^v\|_\infty \big) = o_p\big((nh_1^{d_t})^{-1/2}\big)$ and $n^{-\xi^*} \sqrt{\log n/(nh_1^{d_1 + 2\alpha})} = o(1)$ for Assumption~\ref{Acom2} (iii) by Result \ref{PKunif}. 
\begin{align}
&\sqrt{nh}  \frac{1}{n}\sum_{i=1}^n \Big( \hat F_{Y|T\hat V}(y|t,\hat v(X_i)) W_i\hat \pi_i - \mathbb{E}[ F_{Y|TV}(y|t, v_0(X)) W \pi(X)]  \Big) \notag \\
=\ & \sqrt{\frac{h}{n}}  \sum_{i=1}^n \bigg\{
 F_{Y|TV}(y|t, v_0(X_i)) W_i \pi_i - \mathbb{E}[ F_{Y|TV}(y|t, v_0(X)) W \pi(X)]  
+ \frac{1}{\sqrt{h}}\psi_{tin}(y;V,W\pi) \notag \\
&+   \big( K_{h_1}( T_i - t ) - f_{T|X}(t|X_i) \big)\bigg\{ \nabla_v F_{Y|TV}(Y|t,v_0(X_i)) \big( \mathbb{E}[W|X_i] - \mathbb{E}[W|V=v_0(X_i)] \big) \notag  \\
&+ \Big( F_{Y|TV}(y|t,v_0(X_i)) - F_{Y|TX}(y|t, X_i) \Big)\Big( \nabla_v f_{T|V}(t|v)|_{v=v_0(X_i)}\frac{\mathbb{E}[W|V=v_0(X_i)]}{f_{T|X}(t|X_i)} \notag\\
& - \nabla_v \mathbb{E}[W|V=v] |_{v=v_0(X_i)}\Big)  \bigg\} \pi_i\bigg\}
+ o_{p}(1).
\label{CTEnoW}
\end{align}
When $h_1=h$ and $W=1$, the influence function is reduced to $n^{-1/2}\sum_{i=1}^n \psi_{tin}(y;X,\pi)$.
\end{itemize}
The rest of the results follow Theorem~\ref{TGR}.

\begin{Lemma}
Suppose $A(X)$ is a positive function of $X$. 
Let $B$ be any measurable function of $Y$ such that the following moments exist.  Then
\begin{align*}
\mathbb{E}\big[ var\big(B(Y)\big|T=t, X\big)  A(X) \big] \leq \mathbb{E}\big[ var\big(B(Y)\big|T=t, V(X)\big)  A(X)\big].
\end{align*}
Equality holds if and only if $\mathbb{E}[B(Y)|T=t, V(X)] =\mathbb{E}[B(Y)|T=t, X]$ almost surely, when $A(X) > 0$ almost surely.
\label{LvarGPS}
\end{Lemma}
{\bf Proof of Lemma~\ref{LvarGPS}} \\
First note that $\mathbb{E}[ var(B(Y)|T=t, X)  A(X)] = \mathbb{E}[ var(B(Y)|T=t, X)  A(X)  f_T(t)/f_{T|X}(t|X) | T=t ]$.
And $f_T(t)/f_{T|X}(t|X)$ is a function of $X$.
So we could abuse the notation of $A(X)$ and prove $\mathbb{E}[ var(B(Y)|T=t, V(X)) A(X)| T=t] \geq \mathbb{E}[ var(B(Y)\big|T=t, X) A(X) | T=t ]$.

By the law of iterated expectations,
\begin{align*}
&\mathbb{E}[ var(B(Y)|T=t, V(X))  A(X)| T=t] = \mathbb{E}[  \mathbb{E}[ (B(Y) - \mathbb{E}[B(Y)|T=t, V(X)] )^2| T=t, X]  A(X)| T=t ]. 
\end{align*}
We could skip conditioning on $T=t$ for notational ease and observe that
\begin{align*}
&\mathbb{E}\big[ \big(B(Y) - \mathbb{E}[B(Y)|V(X)] \big)^2\big| X \big] - \mathbb{E}\big[ \big(B(Y) - \mathbb{E}[B(Y)|X] \big)^2\big| X \big]\\
&= \mathbb{E}[B^2(Y)|X] - 2\mathbb{E}[B(Y)|X]  \mathbb{E}[B(Y)|V(X)] + \mathbb{E}[B(Y)|V(X)]^2 
- \mathbb{E}[B^2(Y)|X] + \mathbb{E}[B(Y)|X]^2\\
&= \big( \mathbb{E}[B(Y)|V(X)] - \mathbb{E}[B(Y)|X] \big)^2 \geq 0. 
\end{align*}

\subsubsection{Proof of Theorem \ref{CIAGPS}}
We follow the proofs of Theorems 2.1 and 3.1 in \cite{HI04}.
\begin{enumerate}
\item[(i)]
\begin{align*}
f_{T|Y(t), V}(s|y, v, \bar v) &= \int f_{T|X, Y(t), V}(s|x, y, v, \bar v)\ dF_{X|Y(t), V}(x|y, v, \bar v)\\
&= \int f_{T|X, V}(s|x, v, \bar v)\ dF_{X|Y(t), V}(x|y, v, \bar v) \\
&= \int f_{T|X}(s|x)\ dF_{X|Y(t), V}(x|y,v,\bar v) \\
& = f_{T|X}(s|x)
\end{align*}
for $s \in \{t, \bar t\}$, where the second equality is by the unconfoundedness assumption.
By the same argument, 
$f_{T|V}(s|v,\bar v) = \int f_{T|X, V}(s|x, v, \bar v) dF_{X|V}(x|v,\bar v) = f_{T|X}(s|x)$
for $s \in \{t, \bar t\}$. 
\item[(ii)] 
Using the first result, for $s \in \{t, \bar t\}$, 
$f_{Y(t)|T,V}(y|s, v, \bar v) = f_{T|Y(t), V}(s|y, v, \bar v)\ f_{Y(t)|V}(y|v, \bar v) / f_{T|V}(s|v, \bar v) = f_{Y(t)|V}(y|v, \bar v).$ 
\end{enumerate}

\subsubsection{Proof of Theorem \ref{TTT}}
Denote $F_i \equiv F_{Y|TV}(y|t, V_i)$, $\hat F_i \equiv \hat F_{Y|T\hat V}(y|t, \hat V_i)$, and $\theta \equiv \mathbb{E}[F_{Y|TV}(y|t, V) W]$.
We decompose 
\begin{align}
&\frac{1}{n}\sum_{i=1}^n \hat F_i \hat W_i - \theta = \frac{1}{n}\sum_{i=1}^n \hat F_i W_i  - \theta + \frac{1}{n}\sum_{i=1}^n  F_i \big( \hat W_i-W_i \big) + \frac{1}{n}\sum_{i=1}^n \big( \hat F_i - F_i\big) \big( \hat W_i-W_i \big).  \label{dcomTT}
\end{align}

For the second term in (\ref{dcomTT}) associated with estimating the weight of the form $W = A/B$, linearize
$\hat W_i - W_i = \hat A_i/\hat B - A_i/ B = (\hat A_i - A_i)/B - (\hat B - B)A/B^2 + O(|\hat B - B|^2 + |\hat B - B|\|\hat A_i-A_i\|_\infty)$.
So
\begin{align}
\frac{1}{n}\sum_{i=1}^n F_i  \big( \hat W_i - W_i \big) =\ & \frac{1}{n}\sum_{i=1}^n F_i
\frac{\hat A_i - A_i}{B} -  \frac{1}{n}\sum_{i=1}^n F_i W_i  \frac{\hat B - B}{B} \label{dcomTT2} \\
&+ O_p\big((|\hat B - B|^2 + |\hat B - B|\|\hat A_i-A_i\|_\infty) \big). \notag
\end{align}
For the second term in (\ref{dcomTT2}) associated with $B=f_T(\bar t)$, 
\begin{align}
\frac{1}{n}\sum_{i=1}^n F_i  W_i  \frac{\hat B - B}{B} =\ & \theta  \frac{\hat B - B}{B}+ \left(\frac{1}{n}\sum_{i=1}^n F_i  W_i -  \theta\right)  \frac{\hat B - B}{B}\notag\\
= \ &\theta \frac{1}{n}\sum_{i=1}^n \frac{K_{h}(T_i - \bar t)}{f_T(\bar t)} - \theta + R_n^W,
\label{Wde}
\end{align}
where $\sup_{y, t, \bar t}|R_n^W| = O_p((nh)^{-1} + h^{2r})$ for $W_1$ and $\sup_{y, t, \bar t}|R_n^W| = O_p(n^{-1}h^{-1/2} + h^{r}n^{-1/2})$ for $W_2$.
\\[10pt]
{\bf (a) (Kernel)}\ \ We first show the proof for $W_1 = K_h(T-\bar t)/f_T(\bar t)$.
The first term in (\ref{dcomTT}) is zero.
For the third term in (\ref{dcomTT}), $n^{-1}\sum_{i=1}^n \big( \hat F_i - F_i\big) \big( \hat W_i-W_i \big) = O_p\big( (1/\sqrt{nh} + h^r) (\sqrt{\log n/(nh)} + h^r)\big) = o_p((nh)^{-1})$.

To use Theorem \ref{TGR} for the first term in (\ref{dcomTT}), 
we need Lemma~\ref{LCLT} to replace the Lemma~\ref{LBCLT1} for the stochastic equicontinuity argument.
Then Lemma~\ref{LCLT} implies
\begin{align*}
\sup_{(y, t, \bar t) \in \mathcal{Y} \times \mathcal{T} \times \mathcal{T}} \Big|  \frac{1}{n}\sum_{i=1}^n \Big( \hat F_{Y|TV}(y|t, v_0(X_i)) &- F_{Y|TV}(y|t, v_0(X_i)) \Big) K_h(T_i - \bar t) \\
- \mathbb{E}\Big[ \Big(\hat F_{Y|TV}(y|t, v_0(X)) &- F_{Y|TV}(y|t, v_0(X))\Big) K_h(T - \bar t) \Big]\Big| = O_p\bigg( \sqrt{\frac{\log n}{nh}} \bigg) = o_p(1).
\end{align*}
Then we use a similar argument for (\ref{CTEnoW}) to obtain 
\begin{align}
& \frac{1}{n}\sum_{i=1}^n \Big( \hat F_i W_i - \theta \Big) \notag \\
=\ & \frac{1}{n} \sum_{i=1}^n \bigg\{
 F_i W_i  - \theta 
+ \frac{1}{\sqrt{h}}\psi_{tin}(y;V,W\pi) \notag \\
&+   \big( K_{h}( T_i - t ) - f_{T|X}(t|X_i) , K_{h}( T_i - \bar t ) - f_{T|X}(\bar t|X_i) \big)^\top   \Big( F_{Y|TV}(y|t,v_0(X_i)) - F_{Y|TX}(y|t, X_i) \Big)\notag \\
&\times \Big( \nabla_v f_{T|V}(t|v)|_{v=v_0(X_i)}\frac{\mathbb{E}[W|V=v_0(X_i)]}{f_{T|X}(t|X_i)}  - \nabla_v \mathbb{E}[W|V=v] |_{v=v_0(X_i)}\Big)   \bigg\}
+ o_{p}((nh)^{-1/2}) \notag \\
=\ & \frac{1}{n} \sum_{i=1}^n \bigg\{
 F_i W_i  - \theta 
+ \frac{K_{h}\big(T_i-t\big)}{f_{T|X}(t|X_i)}  \Big( {\bf 1}{\{Y_i \leq y\}} -  F_{Y|TV}(y|t,V_i) \Big)
\frac{f_{T|X}(\bar t|X)}{f_T(\bar t)}
\notag \\
&+   \big(F_{Y|TV}(y|t,V_i) - F_{Y|TX}(y|t,X_i)\big)\Big( K_{h}( T_i - t )\frac{f_{T|X}(\bar t|X_i)}{f_{T|X}(t|X_i)}  - K_{h}( T_i - \bar t )  \Big)  \frac{1}{f_T(\bar t)}
   \bigg\}
+ o_{p}((nh)^{-1/2}) 
\notag \\
=\ &\frac{1}{n} \sum_{i=1}^n \bigg\{ 
\frac{K_h(T_i - t)}{f_{T|X}(t|X_i)}\big(  {\bf 1}{\{Y_i \leq y\}} - F_{Y|TX}(y|t, X_i) \big) \frac{f_{T|X}(\bar t|X_i)}{f_T(\bar t)}
 \notag \\
 &- \frac{K_h(T_i - \bar t)}{f_T(\bar t)} \big( F_{Y|TV}(y|t, V_i)  - F_{Y|TX}(y|t, X_i)  \big)
+ F_i W_i  - \theta \bigg\}  + o_{p}((nh)^{-1/2}). 
\label{noW}
\end{align}

Combining  (\ref{Wde}) and (\ref{noW}) gives the result.

\bigskip
The result for $\hat F_{Y(t)}(y; X,  \hat W_1 \hat \pi)$ is implied by the stochastic expansion in Theorem~\ref{TGaussian} with $\Delta_{yt}=0$ and (\ref{Wde}).
\\[10pt]
{\bf (b) (MLE)}\ \ 
The terms in (\ref{dcomTT2}) are $O_p(n^{-1/2})$.
The result follows Theorem~\ref{TGaussian}.

\section{Inference for the Treatment Effects}
\label{Ainf}
\subsection{Proof of Theorem~\ref{TFDM}}
By the functional delta method (e.g., Theorem 3.9.4 in \cite{VW96}) and the linearity of the Hadamard derivative, the weak convergence to a Gaussian process is implied.
Using the results in (\ref{acov}), for the diagonal term $t \neq s$,
$\lim_{n \rightarrow \infty} \mathbb{E}\big[ \psi_{tin}(y_1) \psi_{si}(y_2)\big] = 0$.
It follows $Cov = \lim_{n \rightarrow \infty}\mathbb{E}\big[ \big(\Gamma'_\theta(\psi_{tin}) - \Gamma'_\theta(\psi_{\bar tin})\big)^2\big] = \lim_{n \rightarrow \infty} \mathbb{E}\big[ \big(\Gamma'_\theta(\psi_{tin})^2 \big]  + \mathbb{E}\big[  \Gamma'_\theta(\psi_{\bar tin})^2\big]$.
\\
{\bf Mean}\ \
Using integration by parts, $\int_{\mathcal Y} y d{\bf 1}{\{ Y \leq y \}} = Y$.
So $\Gamma'\big( {\bf 1}{\{Y \leq y\}} - F_{Y|TX}(y|t,X) \big) = Y - \mathbb{E}[Y|t,X]$.
\\
{\bf Quantile processes}\ \
The Hadamard derivative is shown in Example 3.9.24 in \cite{VW96}.

\subsection{Proof of Theorem~\ref{TMCLT} (Multiplier method)}
Decompose 
\begin{align*}
\frac{1}{\sqrt{n}}\sum_{i=1}^n U_i \hat \psi_{tin}(y) &= \frac{1}{\sqrt{n}}\sum_{i=1}^n U_i \psi_{tin}(y) + \frac{1}{\sqrt{n}}\sum_{i=1}^n U_i \big(\hat \psi_{tin}(y) - \psi_{tin}(y)  \big)
\end{align*}
We use the functional CLT by checking the conditions of Theorem 10.6 in \cite{Pollard}, as in the proof of Lemma~\ref{LDHS}.
We show the first term $n^{-1/2} \sum_{i=1}^n U_i \psi_{tin}(\cdot) \Longrightarrow \mathbb{G}_t(\cdot)$.
The second empirical process is $\bar o_p(1)$ by showing its weak convergence to a Gaussian process with zero covariance kernel.

Define $f^u_{ni}(y) = U_i f_{ni}(y) =  U_i  n^{-1/2} \psi_{tin}(y)$ whose envelope is $F^u_{ni} = U_iF_{ni} = U_i n^{-1} h^{d_t/2} K_h\big( T_i - t\big) W_V(V_i)/ f_{t|V_i}$.
Then (i) holds. $\mathbb{E}f_{ni}^u(y)=0$ and $\mathcal{Z}^u_n(y) = \sum_{i=1}^n f_{ni}^u(y)$.
\begin{enumerate}
\item[(ii)] $ \mathbb{E}[\mathcal{Z}^u_n(y_1) \mathcal{Z}^u_n(y_2)] = \mathbb{E}\big[ \sum_{i=1}^n  f^u_{ni}(y_1)f^u_{ni}(y_2)\big] = n^{-1}\sum_{i=1}^n \psi_{tin}(y_1) \psi_{tin}(y_2) \mathbb{E}U_i^2 \stackrel{p}{\rightarrow} H(y_1, y_2)$
defined in (\ref{PTGacov}), by the weak law of large number.

\item[(iii)]
$\sum_{i=1}^n  \mathbb{E}{F_{ni}^{u}}^2 = n^{-1} \sum_{i=1}^n h^{d_t} K_h^2\big( T_i - t \big) W_V(V_i)^2/f^2_{t|V_i} \rightarrow \int k^2(u)du \mathbb{E}[ W_V(V)^2f_{t|V}^{-1}].$

\item[(iv)] $B \equiv  \inf_{(V_i, T_i)} f_{t|V_i}\big(  W_V(V_i)^2 h^{d_t} K_h\big(T_i-t\big) \big)^{-1}$ exists, because $f_{T|V}$ is bounded away from zero, the weight and the kernel are uniformly bounded.
For any $\epsilon > 0$, 
\begin{align*}
&\sum_{i=1}^n  \mathbb{E}{F_{ni}^u}^2\ {\bf 1}\big\{ F_{ni} > \epsilon \big\} = \frac{h^{d_t}}{n} \sum_{i=1}^n K_h^2\Big(T_i-t\Big)   \frac{W_V(V_i)^2}{f^2_{t|V_i}} \cdot \mathbb{E}\Big[ U_i^2\ {\bf 1}\Big\{ U_i > \frac{\sqrt{n} \epsilon f_{t|V_i}}{\sqrt{h^{d_t}}K_h\big(T_i-t\big)W_V(V_i)}  \Big\} \Big]  \\
&\leq  \frac{h^{d_t}}{n} \sum_{i=1}^n K_h^2\Big(T_i-t\Big)   \frac{W_V(V_i)^2}{f^2_{t|V_i}} \cdot \mathbb{E}\Big[ U_i^2\ {\bf 1}\Big\{ U_i > \sqrt{nh^{d_t}} \epsilon B \Big\} \Big] \rightarrow \int k^2(u)du \cdot \mathbb{E}\big[W_V(V_i)^2 f_{t|V}^{-1} \big] \cdot 0.
\end{align*}

\item[(v)] Denote $F_{Y|TV}(y|t,V_i) = F_{yi}$ and ${\bf 1}{\{ Y_i \leq y \}} = {\bf 1}_{yi}$.
Then for any $y_1 \leq y_2$, 
\begin{align}
&\rho^u_n(y_1,y_2)^2 \\
&= \sum_{i=1}^n \mathbb{E}\Big( f_{ni}^u(y_1) - f_{ni}^u(y_2) \Big)^2 
= \frac{1}{n}\sum_{i=1}^n \Big( \psi_{tin}^2(y_1) + \psi_{tin}^2(y_2) - 2 \psi_{tin}(y_1) \psi_{tin}(y_2) \Big) \notag \\
&= \frac{h^{d_t}}{n}\sum_{i=1}^n  K_h^2\big( T_i - t \big)W_V(V_i)^2 \Big( {\bf 1}_{y_1i} - 2 {\bf 1}_{y_1i} F_{y_1i} + F_{y_1i}^2
+ {\bf 1}_{y_2i} - 2{\bf 1}_{y_2i} F_{y_2i} + F_{y_2i}^2 \label{unify1}\\
&\hspace{135pt} -2 {\bf 1}_{y_1i}  + 2 {\bf 1}_{y_1i} F_{y_2i} + 2{\bf 1}_{y_2i} F_{y_1i} - 2F_{y_1i} F_{y_2i} \Big) \Big/ f_{t|V_i}^2 \label{unify2} \\
&= \frac{h^{d_t}}{n}\sum_{i=1}^n\frac{K_h^2\big(T_i - t \big)}{f_{t|V_i}^2} W_V(V_i)^2 \Big(  {\bf 1}_{y_1i}\big( -1-2F_{y_1i} + 2F_{y_2i} \big) + {\bf 1}_{y_2i}\big( 1-2F_{y_2i} +2 F_{y_1i} \big) + \big(F_{y_1i} - F_{y_2i}\big)^2 \Big) \notag\\
&\rightarrow \int k^2(u)du \cdot \mathbb{E}\Big[ \frac{W_V(V_i)^2}{f_{t|V}} \Big( F_{y_1} ( -1  -2F_{y_1} + 2F_{y_2}) + F_{y_2} ( 1  -2F_{y_2} + 2F_{y_1}) + ( F_{y_1} - F_{y_2})^2 \Big) \Big] \notag\\
&= \int k^2(u)du \cdot \mathbb{E}\Big[ \frac{W_V(V)^2}{f_{t|V}} (F_{y_2} - F_{y_1})(1 - F_{y_2} + F_{y_1})  \Big] 
\equiv \rho^u(y_1, y_2)^2. \notag
\end{align}
It remains to show that for all deterministic sequences $\{y_{1n}\}$ and $\{ y_{2n} \}$ such that $\rho^u(y_{1n}, y_{2n}) \rightarrow 0$, $\rho^u_n(y_{1n}, y_{2n}) \rightarrow 0$.
Using the same argument in Lemma~\ref{LDHS}, $\sqrt{nh^{d_t}} \big\{ n^{-1}\sum_{i=1}^n \psi_{tin}^2(y_1)  -  \int k^2(u)du \cdot \mathbb{E}\big[ W_V(V)^2 \big( F_{y_1} - F_{y_1}^2\big)\big/ f_{t|V} \big]  \big\}$ converges to a Gaussian process of $y_1$.
So the first part $n^{-1}\sum_{i=1}^n \big( \psi_{tin}^2(y_1) + \psi_{tin}^2(y_2) \big)$ in (\ref{unify1}) converges uniformly in $y_1, y_2$. 

For the second part $-2n^{-1}\sum_{i=1}^n \psi_{tin}(y_1) \psi_{tin}(y_2)$ in (\ref{unify2}) indexed by both $y_1$ and $y_2$, we focus on one of the terms, defining $A_n(y_1, y_2)\equiv n^{-1}\sum_{i=1}^n h^{d_t}K_h^2\big( T_i - t\big) W_V(V_i)^2 {\bf 1}_{y_1i} F_{y_2i}/f_{t|V_i}^2$ and  $A(y_1, y_2) \equiv \int k^2(u)du \cdot \mathbb{E}\big[ F_{y_1}F_{y_2} W_V(V)^2\big/f_{t|V} \big]$.
It suffices to show that for all deterministic sequences $\{y_{1n}\}$ and $\{ y_{2n} \}$ such that $A(y_{1n}, y_{2n}) \rightarrow 0$, $A_n(y_{1n}, y_{2n}) \rightarrow 0$.
By assumption, $W_V(V)^2/f_{t|V} < \delta < \infty$.
$A(y_{1n}, y_{2n}) \rightarrow 0$ means that for any $\epsilon > 0$, there exists an integer $N_0$ such that for $n > N_0$,
\begin{align}
A(y_{1n}, y_{2n}) &\leq \int k^2(u)du\cdot \delta \cdot \mathbb{E}\big[F_{Y|TV}(y_{1n}|t,V) F_{Y|TV}(y_{2n}|t,V)\big] \notag \\
&\leq C \cdot \mathbb{E}\big[F_{Y|TV}\min\{y_{1n}, y_{2n}\}|t,V) \big] < \epsilon \label{FCLTv}
\end{align}
defining $C = \int k^2(u)du\cdot \delta$ for notational ease.
Actually, $\min\{y_{1n}, y_{2n}\}$ can be either $y_{1n}$ or $y_{2n}$.
It is not required both the deterministic sequence to converge.

Since $F_{Y|TV}$ is increasing in $y$, there exists $y_0$ such that $C \mathbb{E}[F_{Y|TV}(y_0|t,V)] = \epsilon$.  
Then for any $y < y_0$, $C \mathbb{E}[F_{Y|TV}(y|t,V)] \leq \epsilon$.
Then (\ref{FCLTv}) implies either $y_{1n} < y_0$ or $y_{1n} < y_0$ or both for $n > N_0$.


First, note that $n^{-1}\sum_{i=1}^n h^{d_t} K_h^2\big( T_i - t \big) {\bf 1}{\{Y_i \leq y_0\}} \big/f_{t|V_i} \rightarrow \int k^2(u)du \mathbb{E}\big[ F_{y_0|tV} \big]$, i.e., for any $\epsilon_1> 0$, there exists an integer $N_1$ such that $\big| n^{-1}\sum_{i=1}^n h^{d_t}K_h^2\big(T_i - t\big) {\bf 1}{\{Y_i \leq y_0\}} \big/f_{t|V_i} - \int k^2(u)du$\\ $\times \mathbb{E}\big[ F_{y_0|tV} \big]\big| < \epsilon_1$, for $n > N_1$.
For the case $y_{1n} < y_0$, for $n > \max\{N_1, N_0\}$, 
\begin{align*}
A_n(y_{1n}, y_{2n}) &\leq \delta \frac{1}{n}\sum_{i=1}^n h^{d_t}K_h^2\big(T_i - t \big) {\bf 1}_{y_{1n}i} F_{y_{2ni}}\big/ f_{t|V_i}
\leq \delta \frac{1}{n}\sum_{i=1}^n h^{d_t}K_h^2\big(T_i - t\big) {\bf 1}{\{Y_i \leq y_0\}} \big/f_{t|V_i}  \\
&\leq \delta \int k^2(u)du \mathbb{E}\big[ F_{y_0|tV} \big] + \delta\epsilon_1 = \epsilon + \delta\epsilon_1. 
\end{align*}
For the other case $y_{2n} < y_0$, use the similar argument by $ n^{-1}\sum_{i=1}^n h^{d_t} K_h^2\big(T_i - t \big) F_{y_0i} \big/f_{t|V_i} \rightarrow \int k^2(u)du \mathbb{E}\big[ F_{y_0|tV} \big]$.
Then it is shown $A_n(y_{1n}, y_{2n})\rightarrow 0$.
The same argument applies to other terms in (\ref{unify2}).
\end{enumerate}
Therefore, the FCLT implies $n^{-1/2}\sum_{i=1}^n U_i \psi_{tin}(\cdot) \Longrightarrow \mathbb{G}_t(\cdot)$.  
Next we need to show 
\begin{align*}
\frac{1}{\sqrt{n}}\sum_{i=1}^n U_i \Big( \hat \psi_{tin}(y) - \psi_{tin}(y) \Big) \equiv \frac{1}{\sqrt{n}}\sum_{i=1}^n U_i \sqrt{h^{d_t}}K_h\big( T_i - t\big)  \Big( \hat \varphi_{tin}(y) - \varphi_{tin}(y) \Big) = \bar o_p(1)
\end{align*}
where $\varphi_{tin}(y) = \big({\bf 1}{\{Y_i \leq y\}} - F_{Y|TV}(y|t, V_i)\big) \mathbb{E}[W|V = V_i]/f_{T|V}(t|V_i)$ and a consistent estimator $\hat \varphi_{tin}(y) = \big({\bf 1}{\{Y_i \leq y\}} - \hat F_{Y|TV}(y|t, V_i)\big) \hat{ \mathbb{E}}[W|V = V_i]/\hat f_{T|V}(t|V_i)$.
\begin{enumerate}
\item[(i)] Given the sample, $\hat F_{y|tV_i}$ is monotone increasing in $y$ by construction, so $\big\{ f_{ni}(y) \equiv U_i n^{-1/2}h^{d_t/2}$\\$\times K_h\big( T_i - t\big) \big( \hat \varphi_{tin}(y) - \varphi_{tin}(y) \big) \big\}$ are manageable.
Note $\mathbb{E}f_{ni}(y)=0$, $\mathbb{E}f_{ni}^2(y)=  n^{-1}h^{d_t}K_h^2\big(T_i - t \big) \big( \hat \varphi_{tin}(y) - \varphi_{tin}(y) \big)^2$, and $\mathcal{Z}_n(y) = \sum_{i=1}^n f_{ni}(y)$.
Assuming $f_{T|V}(t|V)$ and $W$ are uniformly bounded away from zero and above, define the envelope $F_{ni} = U_i \sqrt{h^{d_t}/n}K_h\big( T_i - t \big) C$.

\item[(ii)] $\big| \mathbb{E}\mathcal{Z}_n(y_1) \mathcal{Z}_n(y_2)\big| = \big|\mathbb{E} \sum_{i=1}^n f_{ni}(y_1) f_{ni}(y_2) \big| \leq n^{-1} \sum_{i=1}^nh^{d_t}K_h^2\big( T_i - t\big) \big| \hat \varphi_{tin}(y_1) - \varphi_{tin}(y_1)\big|$\\
$\times \big| \hat \varphi_{tin}(y_2) - \varphi_{tin}(y_2)\big| \leq  n^{-1}\sum_{i=1}^n h^{d_t}K_h^2\big( T_i - t \big) \big\| \hat \varphi_{tin} - \varphi_{tin}\big\|_\infty^2 = O_p(1) o_p(1) = o_p(1).$

\item[(iii)] and (iv) are the same as the previous calculation for the first dominating term.
\item[(v)] $0 \leq \sum_{i=1}^n \mathbb{E}\big[ f_{ni}(y_1) - f_{ni}(y_2) \big]^2 \leq n^{-1}\sum_{i=1}^n h^{d_t}K_h^2\big( T_i - t \big)  \big\| \hat \varphi_{tin}(y_1) - \varphi_{tin}(y_1) -\big( \hat \varphi_{tin}(y_2) - \varphi_{tin}(y_2) \big) \big\|_\infty^2 \rightarrow 0$.
\end{enumerate}

\section{Proofs of Lemmas for Stochastic Equicontinuity}
\label{ApfSE}
\subsection{Proof of Lemma~\ref{LBCLT1}}
Define $Z_{ni}(\upsilon) = n^{-1/2} f(y, t, V_i)W_i$, indexed by $\upsilon = (y, t, f) \in \Upsilon = \mathcal{Y} \times \mathcal{T}_0\times \mathcal{F}$.
The bracketing CLT will imply $\sum_{i=1}^n \big( Z_{ni}(\upsilon) - \mathbb{E}Z_{ni}(\upsilon) \big)$ is asymptotic stochastic equicontinuous in $\upsilon$ with respect to the pseudo-metric $\rho(\upsilon_1, \upsilon_2) = \max\{ |y_1 - y_2|, \|t_1 - t_2\|,\| f_1 - f_2\|_\infty \}$.
It suffices to check the conditions for Theorem 2.11.9 in \cite{VW96}: 
\begin{enumerate}
\item[(i)] Since the functions are assumed to be uniformly bounded above and below, ${\bf 1}{\{\| Z_{ni} \|_\Upsilon > \eta\}} = 0$ for $n$ large enough.
So for any $\eta > 0$,
$\sum_{i=1}^n \mathbb{E}\big[ \| Z_{ni} \|_\Upsilon {\bf 1}{\{\| Z_{ni} \|_\Upsilon > \eta\}} \big] = o_p(1)$.

\item[(ii)] It is straightforward to modify Lemma B.2 in \cite{IL10} to replace their Lipschitz continuity with H\"{o}lder continuity, 
\[
N(\epsilon_1^{1/2} C_L + \epsilon_2, \mathcal{F}, \| \cdot \|_\infty ) \leq N\Big(\epsilon_1, \mathcal{Y}\times \mathcal{T}, |\cdot| \Big) \times \sup_{y\in\mathcal{Y}, t\in \mathcal{T}_0} N\Big( \epsilon_2, \mathcal{M}, \|\cdot\|_\infty \Big).
\]
Since $\mathcal{Y}\times \mathcal{T}_0$ is a compact set, the result remains.  
\begin{align*}
N(\epsilon, \mathcal{F}, \| \cdot \|_\infty ) &\leq N\Big(\big(\epsilon/(2C_L)\big)^2, \mathcal{Y}\times \mathcal{T}_0, |\cdot| \Big) \times \sup_{y\in\mathcal{Y}, t \in \mathcal{T}_0} N\Big( \epsilon/2, \mathcal{M}, \|\cdot\|_\infty \Big) \\
N_{[\ ]}(\epsilon, \Upsilon, L_2 ) &\leq N\Big(\frac{\epsilon}{2C}, \mathcal{Y}\times \mathcal{T}_0, |\cdot| \Big) \times N\Big( \frac{\epsilon}{2C}, \mathcal{F}, \|\cdot\|_\infty \Big).
\end{align*}
Therefore, $\int_0^{\delta_n} \sqrt{\log N_{[\ ]} (\epsilon, \Gamma, L_2) } d\epsilon \rightarrow 0$, $\forall \delta_n \rightarrow 0$.

\item[(iii)] 
By the H\"{o}lder continuity assumption, for any $\rho(\upsilon_1, \upsilon_2) = o(1)$, 
\begin{align*}
&\sum_{i=1}^n \mathbb{E}\left[ \big( Z_{ni}(\upsilon_1) - Z_{ni}(\upsilon_2) \big)^2\right] = \mathbb{E}\left[\Big( f_1(y_1, t_1, V) W - f_2(y_2, t_2, V) W \Big)^2 \right] \\
&= \mathbb{E}\left[ \Big( f_1(y_1, t_1, V) - f_2(y_1, t_1, V) + f_2(y_1, t_1, V)  - f_2(y_2, t_2, V) \Big)^2 W^2 \right] = o(1).
\end{align*}
\end{enumerate}

\subsection{Proof of Lemma~\ref{LMRS2}}
The proof modifies the proof of Lemma 1 in MRS12. 
Define $\Delta_i(v_1, v_2) = B_{yt}( W_i,V_i)  \big( K_h(V_i - v_1) - K_h(V_i - v_2)\big) - \mathbb{E}\big[B_{yt}( W,V) \big( K_h(V-v_1)- K_h(V-v_2)  \big) \big]$.
The following observation is useful in the proof (i) $|n^{-1}\sum_{i=1}^n \Delta_i(v_1, v_2) | \leq C n^{d_v\eta} \max_j \| v_{1j} - v_{2j}\|_\infty/h$, 
(ii) $\mathbb{E}\Delta_i( v_1, v_2)^2 \leq C n^{d_v \eta} \big(\max_j \| v_{1j} - v_{2j}\|_\infty /h\big)^2$, 
(iii) $|\Delta_i( v_1, v_2)| \leq C n^{d_v \eta} \max_j \| v_{1j} - v_{2j}\|_\infty/h$.
The bound (ii) is the key to determine the rate $\kappa_{11}$.

When $\kappa_{11} \leq (\delta - \eta)$, the results hold from a direct bound.  Consider the case $\kappa_{11} > (\delta - \eta)$.
For $s \geq 0$, let $\mathcal{\bar M}^*_{s,n,j}$ be a set of functions chosen such that for each $v_j \in \mathcal{\bar M}_{n,j}$, there exists $v_j^*\in \mathcal{\bar M}^*_{s,n,j}$ such that $\| v_j-v_j^* \|_\infty \leq 2^{-s} n^{-\delta}$.
Define $\mathcal{\bar M}^*_{s,n} = \mathcal{\bar M}^*_{s,n,1} \times ... \times \mathcal{\bar M}^*_{s,n,d}$.
For $v_1, v_2 \in \mathcal{\bar M}_{n}$, choose $v_1^s, v_2^s \in \mathcal{\bar M}^*_{s,n}$ such that $\| v_{1,j}^s - v_{1,j} \|_\infty \leq 2^{-s} n^{-\delta}$ and $\| v_{2,j}^s - v_{2,j} \|_\infty \leq C 2^{-s} n^{-\delta}$ for all $j, s \geq 0$.
The functions in $ \mathcal{\bar M}^*_{s,n,j}$ are the midpoints of a ($2^{-s} n^{-\delta}$)-covering of $ \mathcal{\bar M}_{n,j}$.
So the cardinality $\# \mathcal{\bar M}_{s,n,j}^*$ is at most $C\cdot \exp\big( \big(2^{-s} n^{-\delta}\big)^{-\beta} n^{\xi} \big)$.

Consider the chain $\Delta_i( v_1, v_2) = \Delta_i(v_1^0, v_2^0) - \sum_{s=1}^{G_n} \Delta_i(v_1^{s-1}, v_1^s) + \sum_{s=1}^{G_n} \Delta_i(v_2^{s-1}, v_2^s) - \Delta_i( v_1^{G_n}, v_1) +  \Delta_i(v_2^{G_n}, v_2)$, where $G_n$ is chosen to be the smallest integer that satisfies $G_n > (1+ c_G)(\kappa_{11} + d_v\eta - (\delta-\eta)) \log n/\log 2$ for a constant $c_G > 0$.
So for $l=1,2$, by (i),
\[
T_1 = \Big| \frac{1}{n}\sum_{i=1}^n  \Delta_i(v_l^{G_n}, v_l) \Big| \leq C 2^{-G_n} n^{d_v\eta -(\delta - \eta)} \leq Cn^{-\kappa_{11}}.
\]

For any $a \geq c_G$, define the constant $c_a = (\sum_{s=1}^\infty 2^{-as} )^{-1}$.
\begin{align}
&{\rm Pr}\Big( \sup_{v_1 \in \mathcal{\bar M}_n} \big| \frac{1}{n}\sum_{i=1}^n \sum_{s=1}^{G_n} \Delta_i(v_1^{s-1}, v_1^s)  \big| > n^{-\kappa_{11}} \Big)  \notag \\
&\leq  \sum_{s=1}^{G_n} {\rm Pr}\Big(  \sup_{v_1 \in \mathcal{\bar M}_n} \Big| \frac{1}{n}\sum_{i=1}^n \Delta_i(v_1^{s-1}, v_1^s)  \Big|  \geq c_a 2^{-as} n^{-\kappa_{11}} \Big)  \label{eSEMRS1} \\
&\leq \sum_{s=1}^{G_n}  \# \mathcal{\bar M}_{s-1,n}^* \# \mathcal{\bar M}_{s,n}^*{\rm Pr}\Big( \frac{1}{n}\sum_{i=1}^n \Delta_i(v_1^{*,s}, v_1^{**,s})  \geq c_a 2^{-as} n^{-\kappa_{11}}\Big) \label{eSEMRS2} \\
&\ \ + \sum_{s=1}^{G_n} \# \mathcal{\bar M}_{s-1,n}^* \# \mathcal{\bar M}_{s,n}^* {\rm Pr}\Big( \frac{1}{n}\sum_{i=1}^n \Delta_i(\tilde v_1^{*,s}, \tilde v_1^{**,s})  < - c_a 2^{-as} n^{-\kappa_{11}}\Big)  \equiv T_2 + T_3. 
\label{eSEMRS3}
\end{align}
In (\ref{eSEMRS2}) and (\ref{eSEMRS3}), the functions $v_1^{*,s}, \tilde v_1^{*,s} \in \mathcal{\bar M}_{s-1,n}^*$ and $v_1^{**,s}, \tilde v_1^{**,s}\in \mathcal{\bar M}_{s,n}^*$ are chosen such that 
 \begin{align*}
{\rm Pr}\Big( \frac{1}{n}\sum_{i=1}^n \Delta_i(v_1^{*,s}, v_1^{**,s})  \geq c_a 2^{-as} n^{-\kappa_{11}}\Big)  &= \max_{v_1^{s-1}, v_1^s} {\rm Pr}\Big( \frac{1}{n}\sum_{i=1}^n \Delta_i(v_1^{s-1}, v_1^{s})  \geq c_a 2^{-as} n^{-\kappa_{11}}\Big) \\
{\rm Pr}\Big( \frac{1}{n}\sum_{i=1}^n \Delta_i(\tilde v_1^{*,s}, \tilde v_1^{**,s})  < -c_a 2^{-as} n^{-\kappa_{11}}\Big) &= \max_{v_1^{s-1}, v_1^s} {\rm Pr}\Big( \frac{1}{n}\sum_{i=1}^n \Delta_i(v_1^{s-1}, v_1^{s})  < - c_a 2^{-as} n^{-\kappa_{11}}\Big). 
 \end{align*}
To show $T_2$ and $T_3 \leq \exp(-cn^c)$ converging to zero at an exponential rate, we use the Markov inequality and $\mathbb{E}[e^X] \leq 1 + \mathbb{E}[X] + \mathbb{E}[X^2] \leq 1+ \mathbb{E}[X^2] \leq \exp(\mathbb{E}X^2)$ by $\mathbb{E}[X] = 0$ and $|X| \leq C = 1/2$.
\begin{align}
\mathbb{E}\Big[ \exp\big( \gamma_{n,s} n^{-1} \Delta_i(v_1^{**,s}, v_1^{**,s}) \big) \Big] &\leq \exp\big( \gamma_{n,s}^2 n^{-2} \mathbb{E}[ \Delta^2_i(v_1^{**,s}, v_1^{**,s})] \big) \label{LMRSii} \\
&\leq \exp\big( C\gamma^2_{n,s} n^{-2} n^{d_v\eta - 2(\delta - \eta)} 2^{-2s} \big) \notag
\end{align}
by (ii).  
To satisfy $|X| \leq C$,
\begin{align*}
|X| &= \big| \gamma_{n,s}n^{-1}\Delta_i( v_1^{**,s}, v_1^{**,s}) \big| \leq C \gamma_{n,s} n^{-1} n^{d_v\eta -(\delta - \eta)} 2^{-s} \\
&\leq c_\gamma n^{(\delta - \eta) - \kappa_{11}} 2^{-as+s} \leq c_\gamma n^{(c_G - a)(\kappa_{11} - (\delta - \eta))} \leq C,
\end{align*}
where the first inequality is  by (iii) and the second inequality comes from $\gamma_{n,s}$ chosen below.
When $a < 1$, $C n^{(\delta - \eta) - \kappa_{11}} 2^{-as+s} \leq C n^{(\delta - \eta) - \kappa_{11}} 2^{G_n(1-a)}$.
The above inequality holds by the chosen $G_n$.
When $a \geq 1$, the above inequality holds for $n$ large enough.
Therefore, 
\begin{align*}
T_2 &\leq C \sum_{s=1}^{G_n} \exp\Big( d_v(1+2^{-\beta}) 2^{s\beta} n^{\delta \beta + \xi} - \gamma_{n,s} c_a 2^{-as} n^{-\kappa_{11}} + C \gamma^2_{n,s} n^{-1+d_v\eta - 2(\delta-\eta) } 2^{-2s} \Big) \\
&= C \sum_{s=1}^{G_n} \exp \Big( d_v(1+2^{-\beta}) 2^{s\beta} n^{\delta \beta+ \xi}  - c_\gamma 2^{2(1-a)s} n^{1-2\kappa_{11} - d_v\eta + 2(\delta - \eta)} \Big) \\
&\leq C\sum_{s=1}^{G_n} \exp(-c^s n^c) \leq \exp(-cn^c).
\end{align*}
The equality comes from choosing $\gamma_{n,s} = c_\gamma 2^{-as+2s} n^{-\kappa_{11} +1-d_v \eta + 2(\delta -\eta)}$ so that the last two terms in the first line is of the same order.
Then $\kappa_{11}$ is chosen such that the second term dominates.
And choose $a$ and $c_\gamma$ such that the sum of the last two terms is negative.
Similarly, $T_3 \leq \exp(-cn^c)$. 

Because $ \mathcal{\bar M}_{0,n}^*$ can always be chosen such that it contains only a single element and use (i),
\[
T_4 = {\rm Pr}( \sup_{v_1, v_2 \in \mathcal{\bar M}_n} \big| \frac{1}{n}\sum_{i=1}^n \Delta_i(v_1^0, v_2^0) \big| > n^{-\kappa_{11}}  ) \leq \exp(-cn^c).
\]

Therefore, 
\begin{align}
\sup_{(y,t) \in \mathcal{Y}\times\mathcal{T}_0} {\rm Pr}\Big( \sup_{ v_1, v_2 \in \mathcal{\bar M}_n} \Big| \frac{1}{n}\sum_{i=1}^n B_{yt}( W_i, V_i)  \Big( K_h(V_i - v_1) - K_h(V_i-v_2)\Big) \notag \\
- \mathbb{E}\Big[ B_{yt}( W,V)   \Big( K_h(V-v_1)- K_h(V-v_2 )  \Big) \Big] \Big| \geq Cn^{-\kappa_{11}}   \Big) \leq \exp(-cn^c).
\label{unif1}
\end{align}
{\bf Uniformity in $(y, t) \in \mathcal{Y} \times \mathcal{T}_0$.}\\
\ \ For $C_t > 0$, choose a grid $\mathcal{Y}_n \times  \mathcal{T}_n$ with $O(n^{C_t})$ points, such that for each $(y, t) \in\mathcal{Y} \times \mathcal{T}_0$, there exists a grid point $(y^*, t^*) = (y^*(y), t^*(t)) \in \mathcal{Y}_n \times \mathcal{T}_n$ such that $\|y - y^*\| \leq n^{-c C_t}$ and $\|t - t^*\| \leq n^{-c C_t}$.

Define $D_i(y, t, v_1) = B_{yt}( W_i, V_i) K_h(V_i - v_1)$.
Choosing $C_t$ large enough implies
\begin{align*}
&\sup_{\stackrel{y \in \mathcal{Y}, t \in \mathcal{T}_0}{v_1 \in \mathcal{\bar M}_n}} \Big| \frac{1}{n} \sum_{i=1}^n D_i( y^*, t^*, v_1) - D_i(y, t,v_1)  - \mathbb{E}\big[ D_i(y^*, t^*, v_1) - D_i(y, t, v_1)\big] \Big|  \leq C n^{-cC_t}n^{d_v \eta}  \leq n^{-\kappa_{11}}.
\end{align*}
The triangle inequality implies the statement in this lemma.

\subsection{Proof of Lemma~\ref{LMRS3}}
The proof is implied by the proof of Lemma~\ref{LMRS2}.
For any $(y,t) \in \mathcal{Y} \times \mathcal{T}_0$, $\Delta_i(y, v_1, v_2) = {\bf 1}{\{Y_i \leq y\}} K_h(T_i - t)  \big( B_t(v_1(T_i, S_i)) - B_t(v_2(T_i, S_i)) \big) $.
(i) $|n^{-1} \sum_{i=1}^n \Delta_i(y, v_1, v_2) | \leq C n^{-d_t\eta}\max_j \| v_{1j} - v_{2j}\|_\infty$.
(ii) $\mathbb{E}\Delta_i(y, v_1, v_2)^2 \leq C n^{-d_t\eta}\max_j \| v_{1j} - v_{2j}\|_\infty^2$.
(iii) $|\Delta_i(y, v_1, v_2)| \leq C n^{-d_t\eta} \max_j \| v_{1j} - v_{2j}\|_\infty$.

\begin{itemize}
\item[(I)] 
{\bf Uniformity in $(y, t) \in \mathcal{Y} \times \mathcal{T}_0$.}\\
\ \ Following the proof of Lemma~\ref{LMRS2}, 
define $D_i(y,t,v_1) \equiv {\bf 1}{\{ Y_i\leq y \}} K_h(T_i-t) B_t(v_1(T_i, S_i))$.
For $C_t > 0$, choose a grid $\mathcal{T}_n$ with $O(n^{C_t})$ points such that for each $t \in \mathcal{T}_0$, there exists a grid point $t^*= t^*(t) \in \mathcal{T}_n$ such that $\|t - t^*\| \leq n^{-c C_t}$.
And there exists a partition $\mathcal{Y}_n \equiv \{y_0, y_1, ..., y_L\} \subset \mathcal{Y}$ such that $\mathbb{E}[D_i(y_l, t, v_1)] - \mathbb{E}[D_i(y_{l-1}, t, v_1)] < n^{-c C_t}$.
Thus for $y_{l-1} \leq y< y_l$, 
$ n^{-1}\sum_{i=1}^n D_i(y_{l-1}, t, v_1) - \mathbb{E}[D_i(y_{l-1}, t, v_1)] -  n^{-c C_t}
\leq
n^{-1}\sum_{i=1}^n D_i(y, t, v_1) - \mathbb{E}[D_i(y, t, v_1)] \leq n^{-1}\sum_{i=1}^n D_i(y_l, t, v_1) - \mathbb{E}[D_i(y_l, t, v_1)] +  n^{-c C_t}$.

For a grid point $(y^*, t^*) \in \mathcal{Y}_n \times \mathcal{T}_n$,
choosing $C_t$ large enough and the triangle inequality imply
\begin{align*}
&\sup_{\stackrel{y \in \mathcal{Y}, t \in \mathcal{T}_0}{v_1 \in \mathcal{\bar M}_n}} \Big| \frac{1}{n} \sum_{i=1}^n D_i( y^*, t^*, v_1) - D_i(y, t,v_1)  - \mathbb{E}\big[ D_i(y^*, t^*, v_1) - D_i(y, t, v_1)\big] \Big|  \\
&\leq C n^{-cC_t}n^{(d_t+1) \eta}  \leq n^{-\kappa_{12}}.
\end{align*}

\item[(I\!I)]
Consider the dependent variable to be $A_y(Y)$ where $\{ A_y: y \in \mathcal{A}\}$ is a class of uniformly bounded functions, $A_y$ is Lipschitz continuous in $y$, and $\mathcal{A}$ is a compact set.
The result in (\ref{unif1}) holds by replacing the dependent variable ${\bf 1}{\{Y \leq y\}}$ with $A_y(Y)$.

Next we show the uniformity over $y \in \mathcal{A}$.
Define $D_i(y, t, v_1) = A_y(Y_i)K_h(T_i-t) B_{t}( v_1(T_i, S_i))$.
Following (I), similarly choose a grid $\mathcal{A}_n$ with $O(n^{C_t})$ points such that for each $y \in \mathcal{A}$, there exists a grid point $y^* = y^*(y) \in \mathcal{A}_n$ such that $\|y - y^*\| \leq n^{cC_t}$.  Then by the Lipschitz continuity of $A_y$, 
\begin{align*}
&\sup_{\stackrel{t \in \mathcal{T}_0}{ y \in \mathcal{A}, v_1 \in \mathcal{\bar M}_n}} \Big| \frac{1}{n} \sum_{i=1}^n D_i(y^*, t^*, u^*, v_1) - D_i(y, t^*, u^*, v_1)  - \mathbb{E}\big[ D_i(y^*, t^*, u^*, v_1) - D_i(y, t^*, u^*, v_1)\big] \Big| \\
&\leq C |y^*-y| n^{d_t \eta} \leq C n^{-cC_t}n^{d_t \eta} \leq n^{-\kappa_{12}}.
\end{align*}
We therefore prove 
\begin{align*}
\sup_{\stackrel{u \in \mathcal{V}, t \in \mathcal{T}_0}{ y \in \mathcal{A}, v_1, v_2 \in \mathcal{\bar M}_n}}\bigg| & \frac{1}{n}\sum_{i=1}^n A_y(Y_i)K_h(T_i-t) \Big( B_{t}( v_1(T_i, S_i)) -B_{t}( v_2(T_i, S_i))\Big) \\
&- \mathbb{E}\Big[ A_y(Y) K_h(T-t) \Big( B_{t}( v_1(T, S)) -B_{t}( v_2(T, S)) \Big) \Big] \bigg| \leq Cn^{-\kappa_{12}}
 \end{align*}
for $n$ large enough $w.p.a.1$.\footnote{\cite*{Song08ET} derives the uniform convergence rate of series estimators  for the regression $\mathbb{E}[A_y(Y)|v_0(T,S) = v]$ over an infinite-dimensional $y$, $v_0$, and $v$.  The extension of our result to an infinite-dimensional space $\mathcal{A}$ is possible but beyond the scope of this paper.
}
\end{itemize}

\subsection{Proof of Lemma~\ref{LMRS4}}
The proof is implied by the proof of Lemma~\ref{LMRS2}, where $\Delta_i(v_1, v_2) = B_{yt}( V_i, W_i) \big( v_1(T_i, S_i) - v_2(T_i, S_i) \big) - \mathbb{E}\big[ B_{yt}( V, W) \big( v_1(T, S) - v_2(T, S) \big) \big]$.
Note that the following still holds the same as the proof of Lemma~\ref{LMRS2}: (i) $|n^{-1} \sum_{i=1}^n \Delta_i(v_1, v_2) | \leq C \max_j \| v_{1j} - v_{2j}\|_\infty$.
(ii) $\mathbb{E}\Delta_i(v_1, v_2)^2 \leq C \max_j \| v_{1j} - v_{2j}\|_\infty^2$.
(iii) $|\Delta_i(v_1, v_2)| \leq C \max_j \| v_{1j} - v_{2j}\|_\infty$.
That is, the proof is essentially the same for the case $\eta = 0$, $h=1$, and $K_h(x) = x$.

\subsection{Proof of Lemma~\ref{LMRS}}
The proof is implied by the proof of Lemma~\ref{LMRS2}.
For (I) define $\Delta_i(y, v_1, v_2) \equiv {\bf 1}{\{ Y_i\leq y \}} K_h(T_i-t) \big( K_h(v_1(T_i, S_i) - u) - K_h(v_2(T_i, S_i) - u)\big) - \mathbb{E}\big[ {\bf 1}{\{ Y\leq y \}} K_h(T-t) \big( K_h(v_1(T, S) - u)- K_h(v_2(T,S) - u)  \big) \big]$.
The following observation is useful in the proof (i) $|n^{-1}\sum_{i=1}^n \Delta_i(y, v_1, v_2) | \leq C n^{d\eta} \max_j \| v_{1j} - v_{2j}\|_\infty/h$, 
(ii) $\mathbb{E}\Delta_i(y, v_1, v_2)^2 \leq C n^{d \eta} \big(\max_j \| v_{1j} - v_{2j}\|_\infty /h\big)^2$, 
(iii) $|\Delta_i(y, v_1, v_2)| \leq C n^{d \eta} \max_j \| v_{1j} - v_{2j}\|_\infty/h$.
The bound (ii) is the key to determine the rate $\kappa_1$.

For (I\!I), (ii) is modified to $\mathbb{E}\Delta_i(y, v_1, v_2)^2 \leq C n^{(d+2) \eta} \big(\max_j \| v_{1j} - v_{2j}\|_\infty /h\big)^2$.

\subsection{Proof of Lemma \ref{LCLT}}
Let $\nu \equiv (y, t, s) \in \Upsilon \equiv \mathcal{Y} \times \mathcal{T} \times \mathcal{S}$.
The pseudo-metric $\rho((\nu_1, f_1),(\nu_2, f_2)) = \max\{ |y_1 - y_2|, \|t_1 - t_2\|, \|s_1-s_2\|, \| f_1 - f_2\|_\infty \}$.
We first show $\Upsilon \times \mathcal{F}$ can be covered by a finite number $L_n$ of cubes $I_k$ with centers $(\nu_k, f_k)$ and length $l_n$ for $k = 1, 2,..., L(n)$.
The cardinality of the $\epsilon$-covering set for the function space $\mathcal{C}_M^\alpha(\Lambda)$ is at most $N\big( \epsilon, \mathcal{C}_M^\alpha, \|\cdot\|_\infty \big) = C \exp(\epsilon^{-2})$ by choosing $\alpha > d_\lambda/2$.
In the proof of Lemma~\ref{LBCLT1}, we have shown that  the covering number 
\begin{align*}
N(\epsilon, \mathcal{F}, \| \cdot \|_\infty ) &\leq N\Big(\big(\epsilon/(2C)\big)^2, \mathcal{Y}\times \mathcal{T}, |\cdot| \Big) \times \sup_{y\in\mathcal{Y}, t \in \mathcal{T}} N\Big( \epsilon/2, \mathcal{C}_M^\alpha, \|\cdot\|_\infty \Big) 
\end{align*}
and $N_{[\ ]}(\epsilon, \Upsilon \times \mathcal{F}, L_2 ) \leq N\Big(\frac{\epsilon}{2C}, \Upsilon, \|\cdot\| \Big) \times N\Big( \frac{\epsilon}{2C}, \mathcal{F}, \|\cdot\|_\infty \Big)$.
The covering number of the compact set is $N\Big(\frac{\epsilon}{2C}, \Upsilon, \|\cdot\| \Big) = C\cdot (\epsilon/(2C))^{-(1+d_t + d_s)}$.

\begin{align*}
&\sup_{\nu \in \Upsilon} \sup_{f(y,t,\cdot) \in \mathcal{F}} \Big| \frac{1}{n}\sum_{i=1}^n f(y, t, \Lambda_i) K_h(S_i - s) - \mathbb{E}\big[f(y, t, \Lambda) K_h(S- s)\big]  \Big|\\
\leq\ & \max_{k} \sup_{(\nu, f) \in \Upsilon \times \mathcal{F} \bigcap I_k} \Big| \frac{1}{n}\sum_{i=1}^n f(y, t, \Lambda_i) K_h(S_i - s) - f_k(y_k, t_k, \Lambda_i) K_h(S_i- s_k)  \Big|\\
&+ \max_{k} \Big| \frac{1}{n}\sum_{i=1}^n f_k(y_k, t_k, \Lambda_i) K_h(S_i - s_k) - \mathbb{E}\big[f_k(y_k, t_k, \Lambda) K_h(S- s_k)\big]  \Big|\\
&+ \max_{k} \sup_{(\nu, f) \in \Upsilon \times \mathcal{F} \bigcap I_k} \Big| \mathbb{E}\big[ f_k(y_k, t_k, \Lambda) K_h(S- s_k)\big] - \mathbb{E}\big[f(y, t, \Lambda) K_h(S- s)\big]  \Big|\\
\equiv\ & Q_1 + Q_2 + Q_3
\end{align*}
Because $f$ is uniformly bounded between $(0,1)$, showing $Q_2 = O_p(\sqrt{\log n/(nh^{d_s})})$ is the same as the proof of the uniform convergence rate of $\hat f_S(s) = n^{-1} \sum_{i=1}^n K_h(S_i - s)$ in \cite{Masry96} or Chapter 1.12 in \cite{LiRacine}.
We do not repeat.

For $Q_1$, 
\begin{align*}
&\Big| \frac{1}{n}\sum_{i=1}^n f(y, t, \Lambda_i) K_h(S_i - s) - f_k(y_k, t_k, \Lambda_i) K_h(S_i- s_k)  \Big| \\
\leq\ & 
\Big| \frac{1}{n}\sum_{i=1}^n f(y, t, \Lambda_i) K_h(S_i - s) - f(y_k, t_k, \Lambda_i) K_h(S_i- s)  \Big|\\
&+ \Big| \frac{1}{n}\sum_{i=1}^n f(y_k, t_k, \Lambda_i) K_h(S_i - s) - f_k(y_k, t_k, \Lambda_i) K_h(S_i- s)  \Big|\\
&+ \Big| \frac{1}{n}\sum_{i=1}^n f_k(y_k, t_k, \Lambda_i) K_h(S_i - s) - f_k(y_k, t_k, \Lambda_i) K_h(S_i- s_k)  \Big| \\
\equiv\ & Q_{11} + Q_{12} + Q_{13}.
\end{align*}
By Lipschitz continuity, $\max_{k} \sup_{(\nu, f) \in \Upsilon \times \mathcal{F} \bigcap I_k}  Q_{11} = O_p(l_n  h^{-d_s})$ and  $\max_{k} \sup_{(\nu, f) \in \Upsilon \times \mathcal{F} \bigcap I_k}  Q_{13} = O_p\big(l_n h^{-(d_s+1)}\big)$.
For $\max_{k} \sup_{(\nu, f) \in \Upsilon \times \mathcal{F} \bigcap I_k}  Q_{12} = O_p(l_n h^{-d_s})$.
By choosing $l_n = \sqrt{\log n\ h^{d_s +1}/n}$, we have $\max_{k} \sup_{(\nu, f) \in \Upsilon \times \mathcal{F} \bigcap I_k}  Q_1$\\ $= O_p(\sqrt{\log n/(nh^{d_s})})$.
By exactly the same argument, we can show $\max_{k} \sup_{(\nu, f) \in \Upsilon \times \mathcal{F} \bigcap I_k}  Q_3 = O_p(\sqrt{\log n/(nh^{d_s})})$.
}

\bibliographystyle{chicagoa}	
\bibliography{database}		

\end{document}